\documentclass[twocolumn, twocolappendix]{aastex631}

\usepackage{rotating}
\usepackage{tikz}
\usetikzlibrary{backgrounds}
\usetikzlibrary{shapes, arrows.meta, positioning, fit}
\usepackage{hyperref}
\usepackage{amsmath}

\begin{document}

\title{The GammaTPC Gamma-Ray Telescope Concept}

\correspondingauthor{Tom Shutt}
\email{tshutt@slac.stanford.edu}

\author[0000-0002-4974-409X]{Tom Shutt}
\affiliation{SLAC National Accelerator Laboratory, Menlo Park, CA 94025, USA}
\affiliation{Kavli Institute for Particle Astrophysics and Cosmology, Stanford University, Stanford, CA 94305-4085 USA}

\author[0009-0002-4698-3631]{Bahrudin Trbalic}
\affiliation{SLAC National Accelerator Laboratory, Menlo Park, CA 94025, USA}
\affiliation{Kavli Institute for Particle Astrophysics and Cosmology, Stanford University, Stanford, CA 94305-4085 USA} 

\author{Eric Charles}
\affiliation{SLAC National Accelerator Laboratory, Menlo Park, CA 94025, USA}
\affiliation{Kavli Institute for Particle Astrophysics and Cosmology, Stanford University, Stanford, CA 94305-4085 USA} 


\author[0000-0002-7574-1298]{Niccolo Di Lalla}
\affiliation{Kavli Institute for Particle Astrophysics and Cosmology, Stanford University, Stanford, CA 94305-4085 USA}

\author{Oliver Hitchcock}
\affiliation{Kavli Institute for Particle Astrophysics and Cosmology, Stanford University, Stanford, CA 94305-4085 USA}

\author[0009-0001-3418-8254]{Sam Jett}
\affiliation{Kavli Institute for Particle Astrophysics and Cosmology, Stanford University, Stanford, CA 94305-4085 USA}

\author[0000-0003-2785-018X]{Ryan Linehan}
\affiliation{Fermi National Accelerator Laboratory, Batavia, Illinois 60510-0500}

\author{Steffen Luitz}
\affiliation{SLAC National Accelerator Laboratory, Menlo Park, CA 94025, USA}
\affiliation{Kavli Institute for Particle Astrophysics and Cosmology, Stanford University, Stanford, CA 94305-4085 USA} 

\author{Greg Madejski}
\affiliation{SLAC National Accelerator Laboratory, Menlo Park, CA 94025, USA}
\affiliation{Kavli Institute for Particle Astrophysics and Cosmology, Stanford University, Stanford, CA 94305-4085 USA} 

\author{Aldo Pe\~na-Perez}
\affiliation{SLAC National Accelerator Laboratory, Menlo Park, CA 94025, USA}

\author{Yun-Tse Tsai}
\affiliation{SLAC National Accelerator Laboratory, Menlo Park, CA 94025, USA}

\begin{abstract}
We present GammaTPC, a transformative 0.1-10\,MeV $\gamma$-ray instrument concept featuring a tracker using a liquid argon time projection chamber (LAr TPC) technology with the novel GAMPix high spatial resolution and ultra low power charge readout. These enable an economical instrument with unprecedented effective area and sensitivity. We discuss the design and technology in some detail, including how a LAr TPC can be staged in space. Finally, we present a first study of the sensitivity of the instrument in the Compton regime using a new framework for analyzing Compton telescope data.
\end{abstract}



\section{Introduction}\label{sec:introduction}

Of all the bands of the electromagnetic spectra of astrophysical sources, the least well explored is the mid-range $\gamma$-ray regime, $\sim0.1$--50\,MeV. While the soft-to-medium energy X-rays are well-covered via various imaging X-ray facilities, and the high-energy $\gamma$-ray band is observed by the \emph{Fermi} Large Area Telescope~\citep{Ajello2009} and ground-based detectors, the sensitivity of existing observations has limited them to observing the very brightest sources. Previous experiments, such as the COMPTEL and OSSE instruments on-board the Compton $\gamma$-ray Observatory (CGRO, 1991--2000), and the instruments flown on INTEGRAL (2002--), had a small effective area or were background-limited for faint sources~\citep{COMPTEL:1993,OSSE:1997}.  The upcoming COSI mission is very exciting and features excellent energy resolution, but the instrument is relatively small with modest effective area~\citep{cosi_2019}. 

For this reason, we are developing a new $\gamma$-ray instrument concept, GammaTPC, for the energy range from below 100\,keV to as high as 1\,GeV. Featuring a liquid argon (LAr) time projection chamber (TPC) tracker which takes advantage of the economic scalability of TPC technology, it will enable an all-sky survey and transient instrument with sensitivity in the Compton regime substantially better than both COSI and the proposed AMEGO-X~\citep{McEnery:2019tcm} MIDEX mission, pointing and energy resolution roughly comparable to AMEGO-X, and good polarization sensitivity. This paper describes the instrument concept in detail, and presents a first study of its sensitivity in the Compton regime, for which we have developed new Compton telescope analysis techniques.

\subsection{MeV \texorpdfstring{$\gamma$}{gamma}-ray Science}

The discovery potential in the MeV band is tremendous, because the number of sources detected in any band scales roughly as the $-3/2$ power of sensitivity (for an isotropic population). COMPTEL detected $\sim$ 30 celestial point-like sources~\citep{Schoenfelder:2000bu}, thus an instrument that is 1000 times more sensitive would allow for a 30000-fold increase of sources, permitting population studies in addition to studies of individual sources. 

The $\gamma$-ray emission from celestial sources is due to non-thermal processes, namely energetic particles interacting with ambient magnetic fields (synchrotron emission), ambient photons (inverse Compton)~\citep{RybickiLightman}, or with other particles (proton-nucleon collisions leading to $\gamma$-ray emission from pion decays). Proton-proton collisions resulting in $\pi^0$ production also result in $\gamma$-ray emission. Distinct features are also present in celestial $\gamma$-ray spectra, such as lines due to radioactive decay of nuclei or features associated with $e^+-e^-$ annihilation at 511\,keV. Understanding which of those processes operate in astrophysical sources is essential in discerning their natures and how they accelerate particles, but this requires sensitive broad-band spectra. Furthermore, since production of very energetic photons requires comparably energetic radiating particle, the $\gamma$-ray studies probe the processes involving particles with energies not accessible in terrestrial laboratories. 

Many classes of $\gamma$-ray emitting sources are powered by relativistic jets~\citep{Blandford:2018iot} produced in active galactic nuclei, supernovae, pulsars including millisecond pulsars, and some Galactic binary sources. Synchrotron radiation dominates the emission in most frequency bands (radio to X-ray), while inverse Compton produces energetic $\gamma$-rays. The soft MeV band, located at the crossroad of dominance of the two processes, contains information regarding the particle content of the jet. Specifically, the ``tail end'' of the synchrotron range reveals the energies of the {\sl most energetic} particles in the jet, while the onset of the inverse Compton emission, being produced by the {\sl least energetic} yet {\sl most numerous} particles, sheds light on the total energy content of the jet. Measuring {\sl both} should lead to inferences about poorly understood processes of jet launching and particle acceleration. 

Further, since synchrotron radiation is polarized while inverse Compton radiation generally is not, our proposed instrument will be able to differentiate between the two. Comparison of $\gamma$-ray vs. optical polarization should provide important insight into both the particle acceleration processes and geometry of such sources. Even more fundamental is the question whether such synchro-Compton models are appropriate, or whether hadronic processes are required. An entire class of extremely luminous flat-spectrum radio quasars (FSRQs) formed in the early universe (redshifts $z >$ 1) has its inverse Compton peak in the MeV range; these MeV blazars \citep{Bloemen1995} are important to the radiation budget of the universe yet are observed only in the tip-of-the-iceberg sense with current instruments.

The last few years have witnessed very rapid growth of multi-messenger astrophysics. We have evidence that neutron star mergers can produce gravitational waves (GW), and at least in one definite case, a merger with a GW signal was associated with a short $\gamma$-ray burst~\citep{Monitor:2017mdv}. Moreover, the first source of very high-energy neutrinos was observed in 2018, and identified with a flaring blazar~\citep{IceCube:2018dnn}. Going back a few decades, $\gamma$-ray lines were observed from SN1987A, a nearby neutrino source~\citep{SN1987A} associated with a supernova explosion. In all these cases, $\gamma$-ray observations were critical to understanding the underlying physical phenomena driving these extremely energetic sources. Given this rapid evolution in the field, it is no longer sufficient to simply find and identify electromagnetic counterparts; it is crucial to leverage joint electromagnetic and gravitational wave/neutrino/cosmic-ray observations to address compelling science questions on the nature of these extreme sources using the unique multi-messenger data as a probe of fundamental physics~\citep{Burns:2019zzo}. 

With its large effective area and high pointing accuracy, the envisioned instrument will detect prompt $\gamma$-ray flashes coincident with gravitational wave (GW) events in real time. By providing relatively precise localization, it will enable follow-up observations across multiple wavelengths, ultimately allowing for the determination of the event's origin and redshift~\citep{GBM:2017lvd}. Moreover, medium-energy $\gamma$-rays provide a unique probe of astrophysical nuclear processes, directly measuring radioactive decay, nuclear de-excitation, and positron annihilation: our instrument will illuminate the processes of element formation in extreme environments such as supernovae and kilonovae~\citep{Abbott:2017dke}. This concept aligns with the priorities identified in the Astro2020 Decadal Survey~\citep{Decadal2020}, to maintain capabilities across the multiwavelength spectrum and eventually develop a probe-class mission (see L.4.3.1) in the context of the ``New Windows on the Dynamic Universe'' theme. In addition, the production of very high (energy?) neutrinos in an AGN requires the presence of very energetic protons, and an instrument with MeV sensitivity could clarify the underlying jet composition. 
This observatory will provide essential capabilities in the $\gamma$-ray band to enable multi-messenger astrophysics with the next generation of gravitational waves and neutrino observatories.

In young supernova remnants, soft X-rays can arise either as thermal (bremsstrahlung) emission from hot plasma, energized by the explosion, or from synchrotron emission arising from strong shocks colliding with the interstellar medium, or both. The situation is much less clear regarding their $\gamma$-ray emission; $\gamma$-rays can be produced by energetic electrons via the inverse Compton process, or via decay of pions produced via proton-proton collisions~\citep[see, e.g.,][]{Funk2017}. The ``smoking gun'' is the shape of the broad-band spectrum in the soft $\gamma$-ray regime, where the proposed instrument is most sensitive.

An exciting possibility afforded by a large instrument with high MeV sensitivity is to sharply improve upon the catalog of millisecond $\gamma$-ray pulsars found by \emph{Fermi}~\citep{Smith_2023} as many of these objects are significantly brighter (in $F_\nu$) at 10-100 MeV than GeV. This could prove powerful for pulsar timing array (PTA) sensitivity, as $\gamma$-rays have no time delay from intervening matter, in contrast to radio, and also will yield a uniformly sampled set of sources which provide a better array for timing. Note that \emph{Fermi} is already within a factor of $\sim$~3 of the sensitivity of the NANOGrav nano-Hz gravity wave background~\citep{agazie2023nanograv}. More speculatively, this enhanced pulsar sensitivity could provide a path to a new, independent measurement of the Hubble constant~\citep{McGrath_2022}.

Finally, an MeV telescope with large area and good angular resolution will be a powerful probe of dark matter in the increasingly interesting mass range below the GeV scale~\citep{Coogan_2021, berlin2023}, and will provide a unique test of dark matter as evaporating primordial black holes.~\citep{Coogan_2021b} In addition, by resolving the contribution to the Galactic center $\gamma$-ray excess from millisecond pulsars in the Milky Way halo in the energy range below 100\,MeV, such an instrument could rule out or strengthen the dark matter interpretation of the Galactic center excess seen by \emph{Fermi}~\citep{Ajello:2015kwa}. 


\subsection{The GammaTPC Instrument Concept}\label{sec:instrument_concept}

\begin{figure*}
    \centering
    \includegraphics[width=0.5\textwidth]{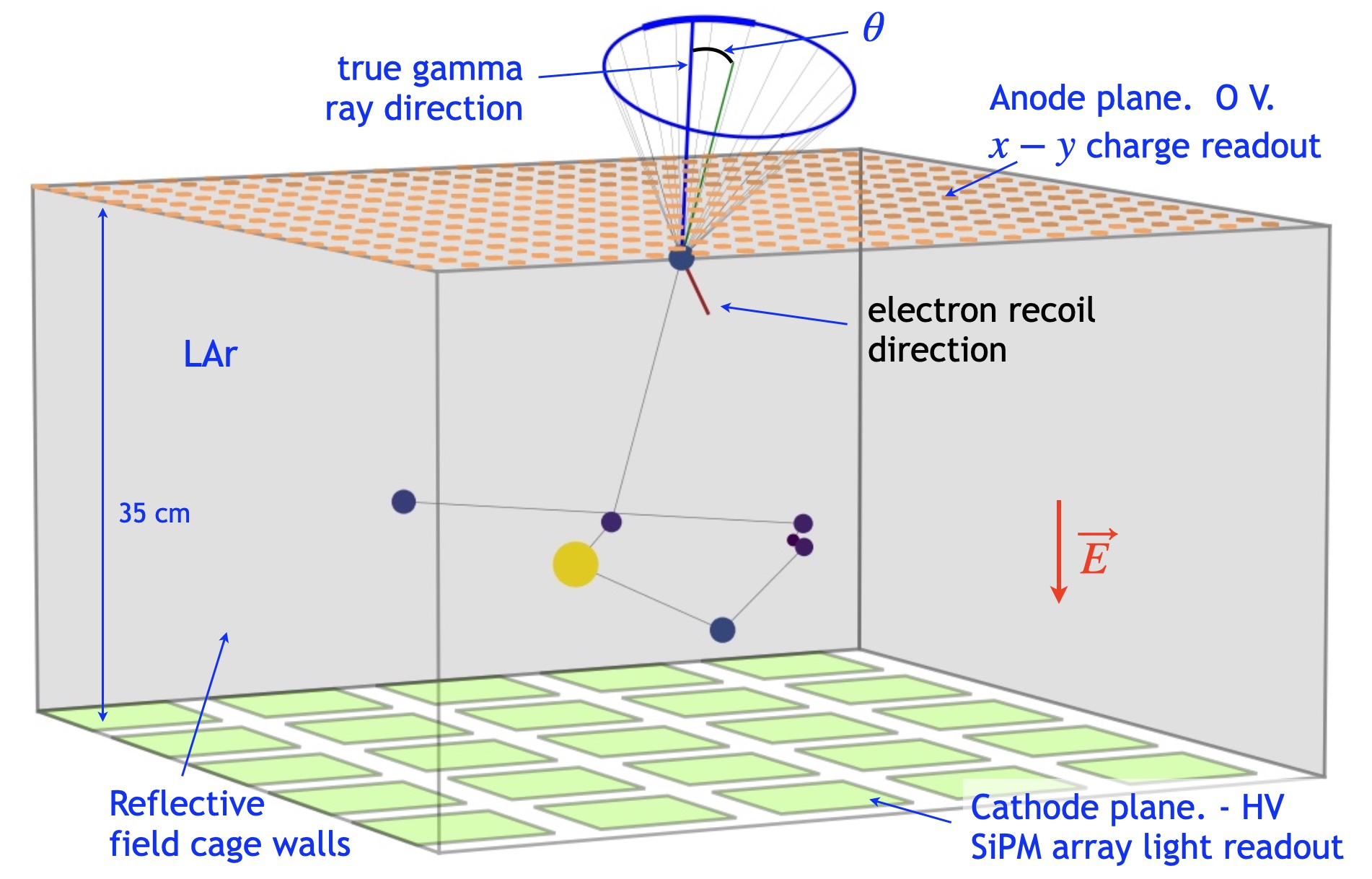}
    \includegraphics[width=0.4\textwidth]{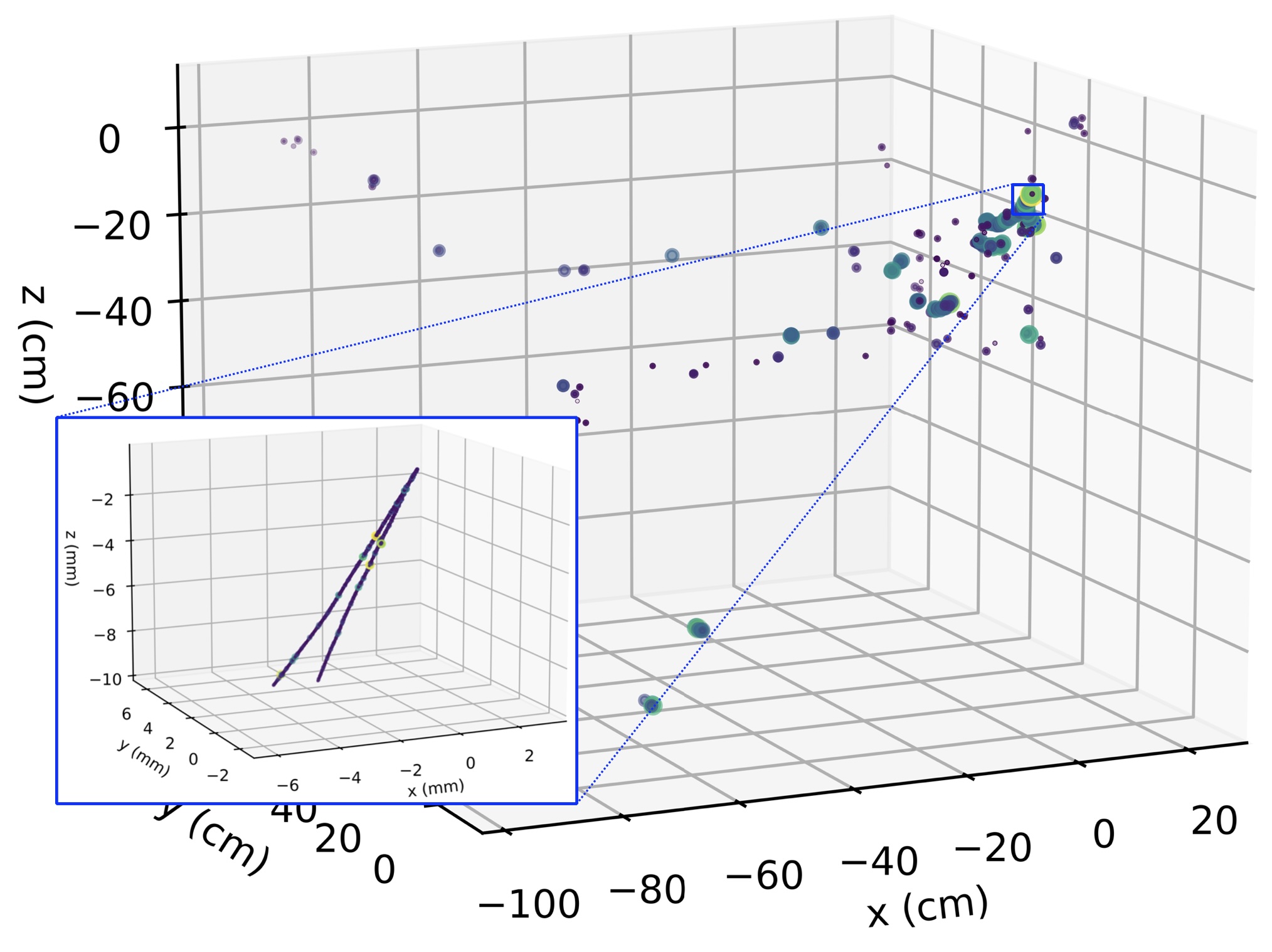}
    \caption{(left) A 1.0 MeV gamma-ray undergoing multiple Compton scatters and final photoabsorption in a uniform LAr volume configured as a TPC, along with the reconstructed sequence of interactions and event cone. The initial direction of the first scatter's electron recoil, if measured, reduces the cone to an arc. (right) The full set of interactions of a 100 MeV gamma-ray in LAr, with the initial pair interaction shown in the inset.}
    \label{fig:tpc_event}
\end{figure*}

The MeV gamma-ray sky remains largely unexplored due to the difficulty of detecting and reconstructing photons in this energy range. Figure~\ref{fig:tpc_event} illustrates the complexity of a Compton event, where a 1 MeV gamma-ray undergoes multiple interactions requiring precise kinematic reconstruction, alongside a 100 MeV pair production event, highlighting the scale of electron-positron tracks. We introduce GammaTPC, a novel gamma-ray instrument utilizing LArTPC technology, designed to accurately detect and reconstruct such events. Achieving good angular and energy resolution requires precise measurements of interaction positions, electron recoil directions, and deposited energy. These allow kinematic tests that can reconstruct the first two Compton scatters and define a potential source circle, which is further refined to an arc when the initial electron recoil direction is known, significantly improving pointing accuracy. Optimized detector design minimizes energy loss from gamma rays escaping or interacting with dead material, ensuring efficient reconstruction and high-resolution imaging. An efficient Compton Telescope has been a long-pursued goal in $\gamma$-ray instrumentation~\citep{Kierans2022}.
 
Pair production event reconstruction is more direct, as the source direction can be inferred from the initial $e^\pm$ tracks. Although multiple scattering affects these tracks, its impact diminishes with increasing energy. For gamma rays well above 10 MeV, the initial tracks evolve into extended electromagnetic showers, from which energy estimation is possible. Additionally, polarization information is extracted from the azimuthal scatter angle in Compton events~\citep{Bernard_2022_pol} and from the $e^\pm$ plane in pair events~\citep{Bernard_2013_pol}.

For fixed geometry, the angular resolution and reconstruction efficiency of a Compton instrument depend on energy resolution, $\sigma_E$, and spatial resolution, $\sigma_{xyz}$. In the pair regime, angular resolution depends primarily on $\sigma_{xyz}$, while $\sigma_E$ plays a lesser role. While Doppler broadening~\citep{Zoglauer2003} imposes a fundamental limit on $\sigma_E$, there is no foreseeable lower bound for improvements in $\sigma_{xyz}$ in either regime.

Gamma-ray instruments in the MeV-GeV energy range commonly feature a skyward tracker and a calorimeter beneath it. The tracker provides high spatial precision measurements of Compton scatters and pair production tracks, while the calorimeter, with its high-$Z$ material, ranges out and contains gamma-ray energy escaping the tracker. A distinctive feature of our design is a large-area, high-mass tracker, which boosts efficiency by maximizing the number of measured scatters and optimizing conversion efficiency in the pair regime. Since Earth's atmosphere is opaque to gamma rays in this range, space deployment is necessary. A low-inclination, low-Earth orbit (LEO) is then needed to minimize background and pile-up from charged particles, neutrons, and up-going albedo gamma rays.



The main advantage of a TPC is achieving $3D$ imaging using only $2D$ sensor arrays on the surfaces. For $N$ sensors per length, a fully segmented detector requires $N^3$ channels ($2N^2$ for $x$-$y$ strip sensors in $z$ layers), but a TPC reduces this to $N^2$ ($2N$), significantly cutting costs and power demands, especially when $N \sim 10^3-10^4$. Its uniform medium enables precise imaging and background reduction, benefiting dark matter~\citep{aalbers2024, xenonnt2023, pandax2023} and neutrinoless $\beta \beta$ decay~\citep{nexo_2018} searches. However, its slow charge drift ($\sim$1 $\mu$s/mm) and high particle rate in space necessitate heavy segmentation to prevent event pileup, at the cost of added dead material.

There is a history of TPC development for $\gamma$-ray astronomy (recently reviewed in~\citet{Bernard_2022}), with most effort focused on gas phase~\citep{Tanimori2015, Tanimori2017, Takada_2022, HUNTER2014, GROS2018}, though the earlier LXeGRIT effort used LXe~\citet{Aprile_2003}. Gaseous TPCs provide very high resolution imaging of tracks down to very low energy, and avoid the complication of cryogenics. But this comes at the substantial cost of very low stopping power compared to liquid phase, and hence much lower overall efficiency. Liquid phase TPCs have emerged as a leading technology for dark matter~\citep{Akerib_2020, xenonnt2023, pandax2023} neutrinoless $\beta \beta$ decay~\citep{nexo_2018}, and beam-based neutrino measurements, most notably DUNE~\citep[see, e.g.,][]{dunefd2_2023}. LAr is preferred over LXe for $\gamma$-ray detection for a number of reasons discussed in Appendix~\ref{sec:appendix_LAr}. These developments make it possible and useful to stage a $\gamma$-ray program with in-hand LAr TPC technology, as it is being pursued by the GRAMS~\citep{Aramaki_2020} balloon program, and the proposed MAST concept~\citep{Dzhatdoev_2019}. Our approach, by contrast, is to fully optimize LAr TPC technology for this application, including the development of GAMPix~\citep{Gampix}, a novel and powerful fine-grained pixelated charge readout architecture.

\begin{figure*}
    \centering
    \includegraphics[width=0.92\textwidth]{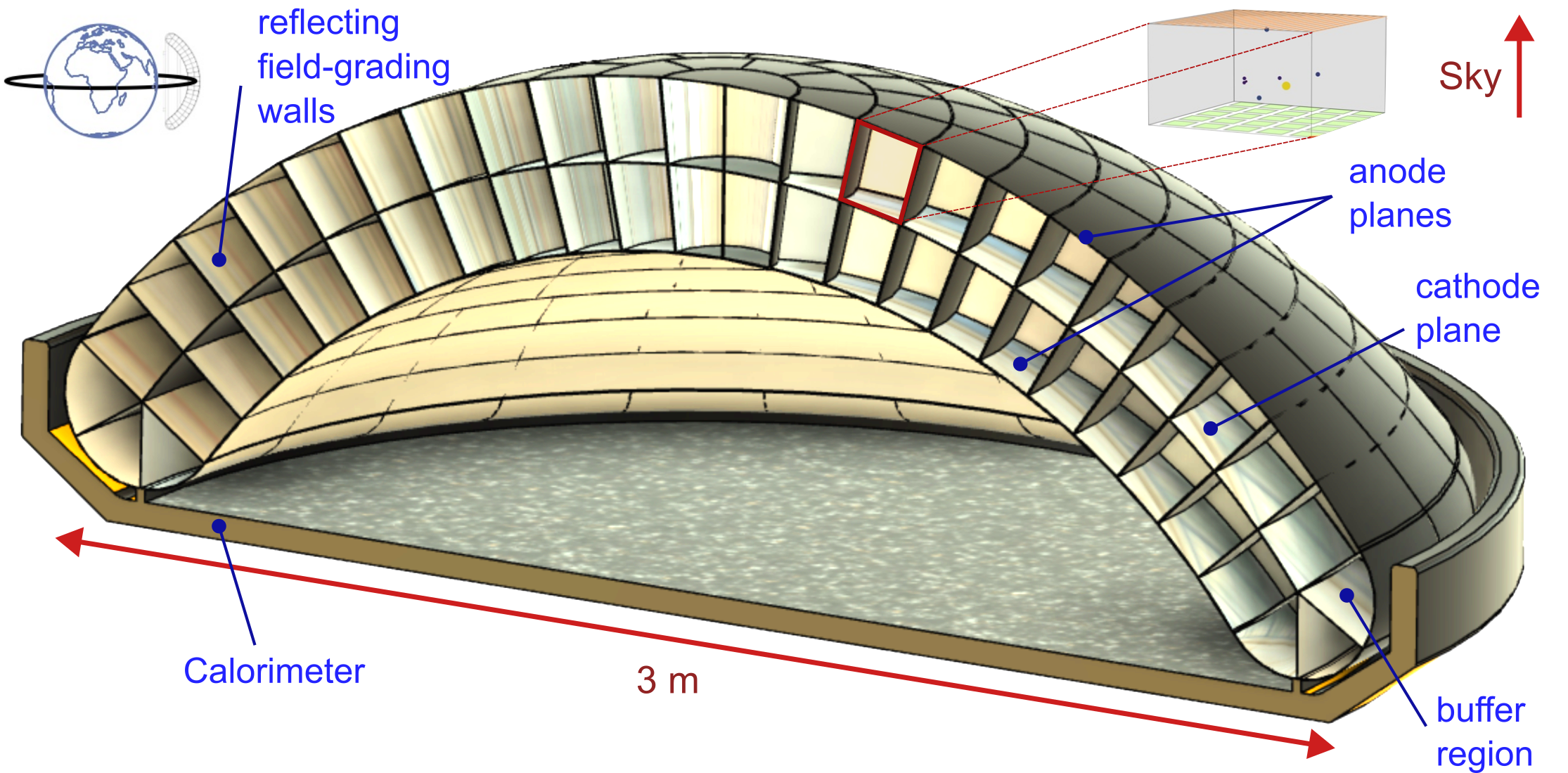}
    \caption{Schematic drawing of a large-scale implementation of the GammaTPC instrument concept, with the tracker a double-layered set of LAr TPC cells (right inset, and here shown as rectangular but in reality with hexagonal footprint), 10\,m$^2$ in area and 4 metric tons in mass contained in a thin-walled carbon fiber shell, backed by a CsI calorimeter. The tracker overall geometry is a section of a sphere, which provides 2$\pi$ $FOV$ (left inset), while the orbit, spacecraft orientation and profile of the tracker maximize the uniformity of the sky coverage.}
    \label{fig:gammatpc_schematic}
\end{figure*}

Figure~\ref{fig:gammatpc_schematic} presents the complete GammaTPC instrument concept, featuring a double layer of segmented LAr TPC cells, each 17.5 cm in height and width, along with a CsI calorimeter. The instrument’s large $10\,m^2$ tracker demonstrates the scalability of this technology. The tracker’s 35 cm thickness is optimized to efficiently contain Compton $\gamma$-rays and, at approximately two radiation lengths, ensures high conversion efficiency for $\gamma$-rays in the pair-production regime. The tracker adopts a hemispherical shape to contain LAr within a pressure vessel, where a curved geometry minimizes wall thickness and reduces unwanted $\gamma$-ray scattering off of inactive material that would add to the background. This design enables uniform $2\pi$ field-of-view (FOV) sky coverage when orbiting, as shown in Fig.~\ref{fig:tpc_event}, while the large area enhances etendue. Together, the shape and scale also mitigate the effects of incomplete event containment at the tracker’s edges. The TPC cells have hexagonal footprints, arranged in a Goldberg polyhedron pattern \citep[see Dual Geodesic Icosahedron Pattern 19 at][]{McCooey}.\footnote{There are also 12 smaller pentagons over a full sphere, but the section we employ likely contains only one central pentagon, which could be configured as either a simple scintillating cell or a complete TPC.} An anti-coincidence detector (ACD) is implemented as a 5-10 mm LAr layer between the TPC structure and the vessel, with SiPM-based scintillation readout for background rejection. The total internal dead mass due to segmentation is estimated to be around 5\%, as detailed in Sec.~\ref{sec:geometry_mechanics}.

GammaTPC's tracker is significantly larger and provides superior geometric information than competing technologies. The thickness in terms of stopping power at 1\,MeV (and at $T$\,=\,120\,K) is 5.8 times larger than the Si tracker of AMEGO-X~\citep{Fleischhack_2021, McEnery:2019tcm} or the similar eASTROGAM~\citep{De_Angelis_2017}, and 1.3 times the Ge tracker of COSI~\citep{cosi_2019}. A 10\,m$^2$ area is approximately 17 and 390 times that of the AMEGO-X and COSI respectively. The pitch of GAMPix's pixels is 500~$\mu$m, which is the same as eASTROGAM's strips~\citep{De_Angelis_2017} and AMEGO-X's pixels, and finer than the 2.0\,mm pitch of COSI's strips and GRAMS' wires. For equivalent spatial resolution, the better geometric information arises from large distances between scatters in the lower density of LAr compared to Si and especially Ge, as explored further in Appendix~\ref{sec:appendix_LAr}.

Our baseline assumption for the calorimeter is a CsI hodoscope similar to that of \emph{Fermi}~\citep{Atwood_2009}, though with a yet-to-be-determined thickness and level of segmentation. In addition to helping to contain high energy Compton events and measuring pair shower profiles, the calorimeter is a powerful active shield of the otherwise dominant up-going albedo gammas. CsI could be operated at the cryogenic temperature of the tracker for improved light yield~\citep[see, e.g.,][]{Mikhailik_2015, ding2020, Clark2018}. A sampling LAr calorimeter (such as that in ATLAS~\citep{Wilkens_2009}) is also possible and has the obvious advantage of using LAr, but has the disadvantage of energy loss in the interspersed W (or Pb) sheets.

In the next section of this paper, we describe in detail the technology needed to achieve the ambitious goals of this new instrument technology. We also look in some detail how the challenges of staging this in space will be overcome, and include a preliminary discussion of costs. The subsequent section presents a first study of the sensitivity of GammaTPC in the Compton regime, and also introduces new techniques for Compton Telescope analysis. A study of the sensitivity in the pair regime awaits a future publication.
\section{G\texorpdfstring{\MakeLowercase{amma}}{amma}TPC Tracker Technology}\label{sec:tracker_technology}

\subsection{Charge and Light Signals in LAr}\label{sec:charge_light}

Production of signals in liquid nobles is complex, with energy deposited by charged particle creating both scintillation light and pairs of free electrons and ions\footnote{Unless otherwise referenced, most LAr properties used in this paper are from the LAr properties database maintained by BNL at \url{https://lar.bnl.gov/properties/}}. The sum of scintillation and charge is proportional to energy deposited, as recombination results in the same excimer states that generate scintillation light, thus recombination effectively trades charge for light on an electron-per-photon basis. The partitioning between charge and light depends on the particle type, the applied electric field, and, especially for non-relativistic particles, the recoil energy, and often results in roughly comparable quanta of photons and electrons. Fluctuation in the strength of recombination greatly exceed what would be expected on the basis of binomial statistics, so measurement of either signal alone gives poor energy resolution~\citep{Conti_2003}. Instead, we estimate the initial number of electrons $q$ and photons $p$ from their measured values and from these estimate the energy $E = W(q + p)$ where $W$ is the combined $W$ value~\citep{Shutt2007}. This is unaffected by recombination fluctuations when both $q$ and $p$ are well measured. 

These effects have been studied in detail in LXe\citep[see e.g.][]{Doke_2002, Conti_2003, Dahl:2009nta, Szydagis_2011, Akerib_2019} in part because of its importance in dark matter and neutrinoless double beta decay experiments, and the energy resolution in a number of detectors~\citep{XENON2020} scales as expected from the above discussion with $1/\sqrt{E} \propto 1/\sqrt{N}$, with $N=q+p$. Signal production has also been measured in LAr~\citep{DOKE_1988, Aprile_1987, Doke_2002, Cao_2015, Washimi2018, Agnes2018, Kimura2019, Szydagis_2021}, extensively at high energies, but less so in the low energy range important for Compton $\gamma$-rays. In particular, energy resolution from combined light and charge and the strength of recombination fluctuations have been little studied. Nonetheless, the existing data show very similar behavior to LXe and provide a solid basis for estimating the LAr response.

A complicating factor for Compton events is that the measured light signal in a TPC cell is from the sum of the light from all scatters in the cell, while the charge signals from each scatter are separately measured. Thus, the sum of light and charge cannot be performed at the scatter level, which is unfortunate because this means recombination fluctuations are not fully canceled per scatter, which directly affects event reconstruction and pointing. However, we can partially recover as follows. For track $i$ we have $q_i$, and for the cell $q_{cell} = \sum{q_i}$, and the light signal $p_{cell}$ the sum of the unmeasurable $p_i$ with $E_{cell} = W(q_{cell} + p_{cell})$. Our best estimate of the energy of track $i$ is $E_i = \frac{q_i} {q_{cell}} E_{cell}$, where the ratio $E_{cell} / q_{cell}$ corrects $q_i$ for the fluctuation between the summed $q_{cell}$ and $p_{cell}$ in the cell. This procedure works well in the not uncommon case of the first scatter being dominant in a cell. These effects are included in simulations framework below (Sec.~\ref{sec:simulations}), with the strength of recombination fluctuations conservatively assumed to be equal to those in LXe\citep{Dobi2014}. The resulting predicted energy resolution is discussed in Sec.~\ref{sec:energy_resolution}.

\subsection{Charge Readout with GAMPix}\label{sec:gampix}

In addition to measuring the amount of charge, the charge readout must provide the best possible spatial resolution, $\sigma_{xyz}$. In the Compton regime this minimizes the geometric reconstruction performance metric $\sigma_{xyz}/\lambda_\gamma$ discussed in Appendix \ref{sec:appendix_LAr}.\footnote{Both energy resolution and geometric information jointly determine reconstruction performance. For a given energy resolution, there exists a limit below which improved geometric precision does not enhance reconstruction. However, the geometric precision is characterized by $\sigma_{xyz} / \left|\boldsymbol{r_{ij}}\right|$, where $\left|\boldsymbol{r_{ij}}\right|$ is the separation between scatters $i$ and $j$. Smaller $\sigma_{xyz}$ enables the utilization of smaller $\left|\boldsymbol{r_{ij}}\right|$, which are exponentially distributed, with smaller separations being more probable. Furthermore, reconstruction benefits from the full ensemble of scatters, amplifying the advantage of minimizing $\sigma_{xyz} / \left|\boldsymbol{r_{ij}}\right|$ for all $\left|\boldsymbol{r_{ij}}\right|$.} It also provides the best possible measurement of the initial directions of electron recoil tracks, which is a long-sought goal for Compton reconstruction (and further discussed in Sec.~\ref{sec:angular_res}), and is fundamental to event reconstruction and angular resolution in the pair regime.

\begin{figure*}
    \centering
    \includegraphics[width=0.47\textwidth]{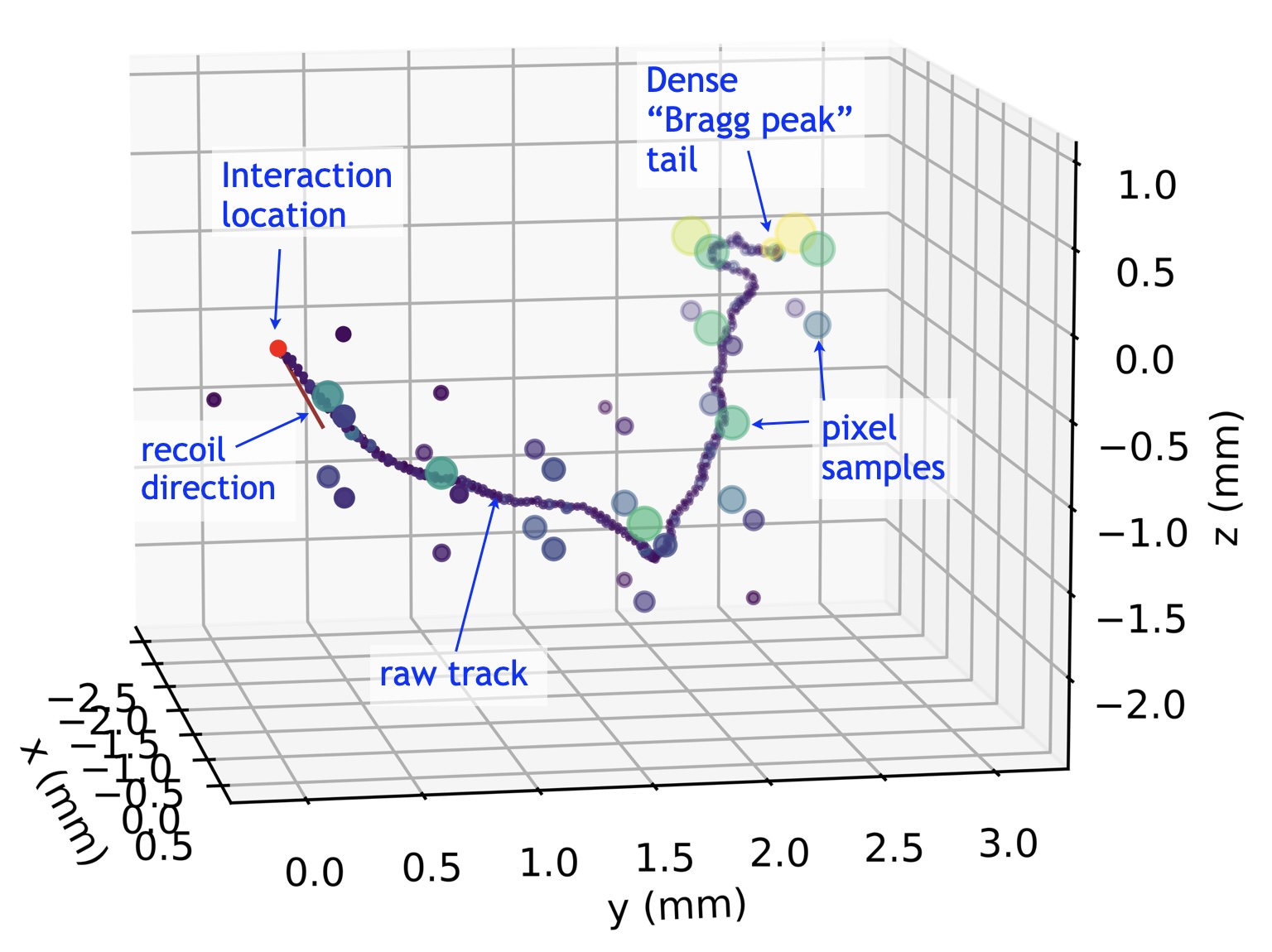}
    \includegraphics[width=0.47\textwidth]{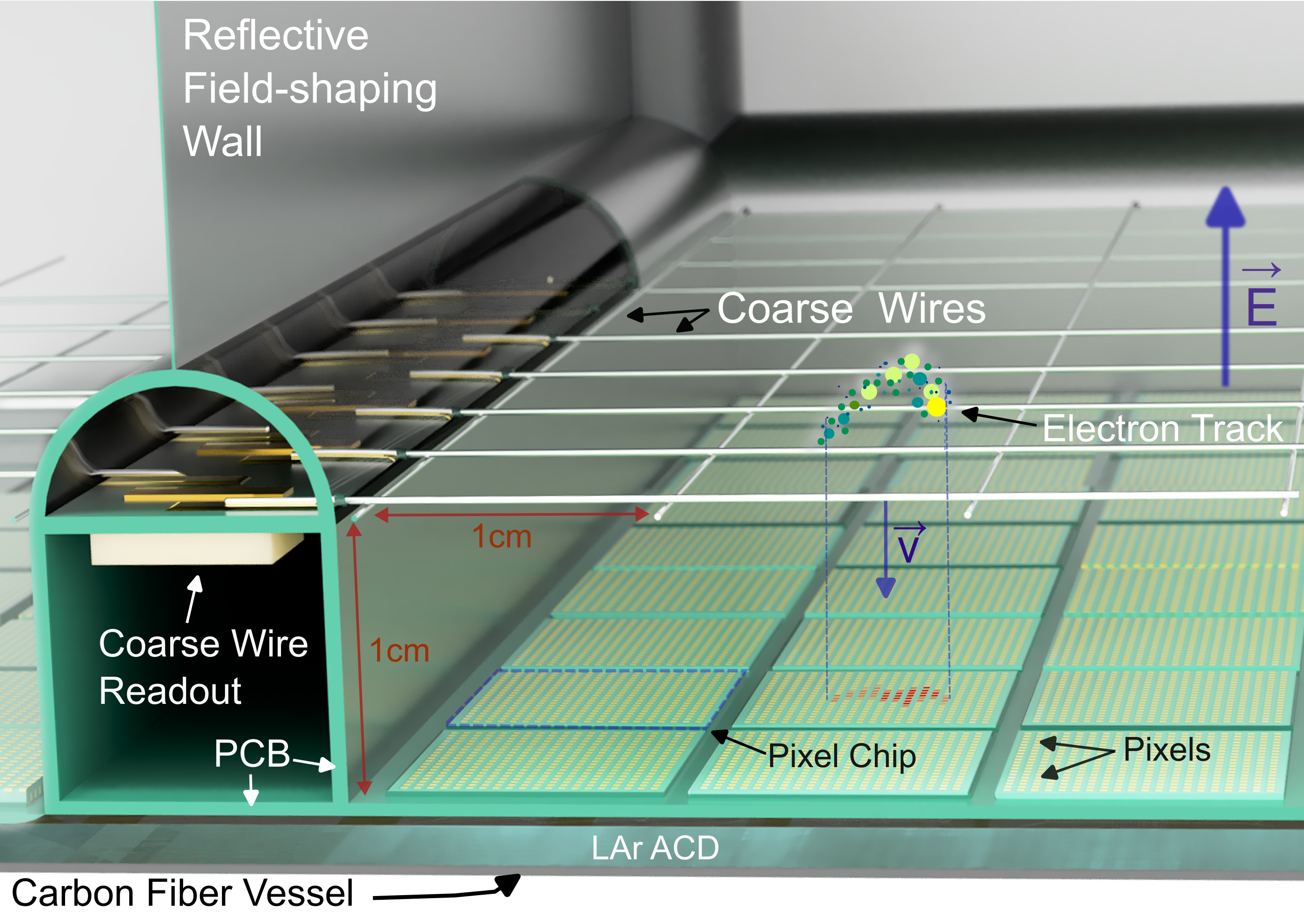}
    \caption{(left) A simulated 1.0 MeV electron recoil track with 21K e$^-$-ion pairs obtained from high fidelity PENELOPE \citep{penelope} simulation (heavy blue line), along with (circles) simulated samples from 500\,$\mu$m pitch pixels with 25\,e\,ENC and a 2 cm drift distance. (right) Diagram of the GAMPix architecture showing coarse grids and pixel chips, along with the drifting charge from an electron track and the resulting pixel signals indicated. Structural ribs house the coarse grids and anchor the reflective field cage walls, as described in Sec.~\ref{sec:geometry_mechanics}.}
    \label{fig:gampix}
\end{figure*}

Electron recoil tracks are effectively point-like at very low energy, but are extended structures at higher energies, as illustrated in Fig.~\ref{fig:gampix} (left). Good spatial resolution for extended tracks thus requires imaging the tracks sufficiently well to distinguish the head of the track (the interaction location) from the tail, which has higher energy deposition \( \text{dE/dx} \) due to the Bragg peak~\citep{pdg2022}. Also shown are simulated GAMPix pixel samples, for which by inspection we see the head of the track can distinguished, and a measure of the initial recoil direction provided.

The architecture of GAMPix, which stands for Grid Activated Multi-Scale Pixel readout, is illustrated in the right panel of Fig.~\ref{fig:gampix}. We give an overview of GAMPix in this section, with a full description in~\citet{Gampix}. Pixels provide true $3D$ imaging which is superior to that from commonly used crossed wires, or strips on a circuit board. Pixels also have the advantage of substantially lower sensor capacitance $C_s$ than for strips or wires and hence significantly lower noise, since the readout noise with an optimized charge sensitive amplifier (generally, $C_{amplifier}=C_s$) scaling as $\sqrt{C_s}$~\citep{Radeka, OConnor2000}. 

GAMPix solves two difficult problems inherent to a fine-grained charge readout in a TPC. The first is power due to the high pixel channel count, exacerbated in our application by the especially stringent $\sim$\,1\,W/m$^2$ power budget set by the spacecraft cryogenics (Sec.~\ref{sec:cryogenics_fluids}). Key is the power of front end amplifiers input MOSFET alone, for which there is a basic tradeoff between power and readout noise $\sigma_e$ ($\sigma_e$ is usually expressed in terms of electron equivalent noise, or ENC), which in CMOS scales as $\sigma_e\,\propto\,1/\sqrt{P}$~\citep{OConnor2000}. With a nominal 0.1\,W/ch amplifier power, the total system power of at least 4\,kW/m$^2$ is some 3 orders of magnitude too high. 

\begin{figure*}
    \centering
    \includegraphics[width=0.4\textwidth]{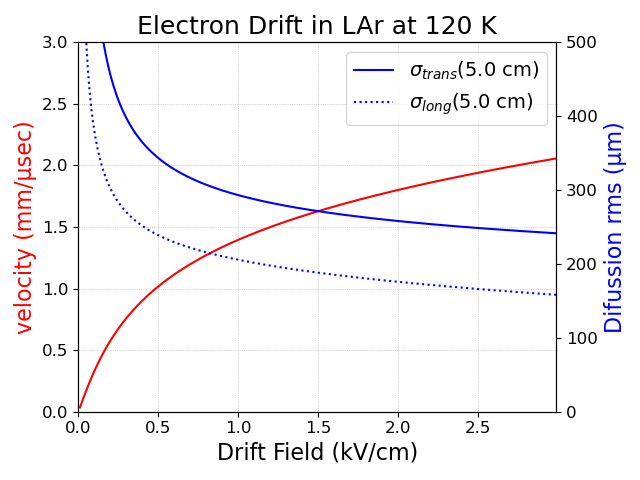}
    \includegraphics[width=0.4\textwidth]{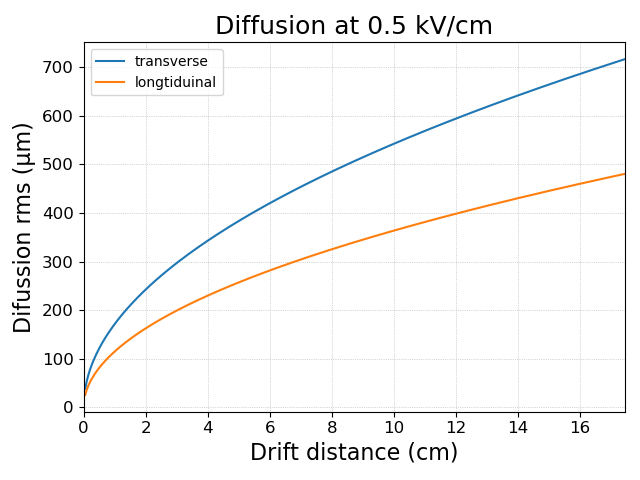}
    \caption{(left) Drift velocity and the rms spread due to diffusion at 5\,cm drift distance, both as a function of applied drift field. The knee at 0.5\,kV/cm leads to the widespread use of this field value. (right) Diffusion over the drift distance range in GammaTPC cells.}
    \label{fig:drift_diffusion}
\end{figure*}

The second problem is due to diffusion of the drifting charges. Diffusion washes out the detail of electron tracks and thus sets the ultimate limit to imaging in TPC. But, with rms spread at most comparable to the 500\,$\mu$m pixel pitch as shown in Fig.~\ref{fig:drift_diffusion}, this has a modest impact on track imaging. A more subtle but fundamental problem is that diffusion causes signal to be lost whenever the sensor pitch is comparable to the scale of diffusion. This happens at the periphery of the diffused charge distribution, where the amount of charge in many of the sensors readily falls below threshold for plausible values of $\sigma_e$. This effect is present for 1$D$ wire/strip readout, but is more severe for 2$D$ pixel readout, and in this application leads to an essentially fatal loss of signal, as detailed in~\citep{Gampix}.

GAMPix measures the drifting electron tracks twice: first as induction signals as the charges pass through a set of 1\,cm pitch coarse-grained and individually instrumented $x-y$ wires; and second when collected on the pixels. These are implemented as small ($\sim$\,200\,$\mu$m) bare pads on top of CMOS ASIC chips, which themselves are mounted on printed circuit boards (PCBs). Charges are directed to the pixel pads first by a modest focusing through the coarse electrodes, and then by strong pixel-pitch focusing from electrodes implemented on the CMOS cover layer.

The induction signals are depth independent because the coarse pitch is much greater than the diffusion scale, but they are highly spatially dependent. That spatial dependence can be corrected to better than 1\% using the track imaging from the pixels~\citep{Gampix}. This solves the diffusion problem. The power problem is solved by using coarse induction signals as a trigger to power up only the pixel chips which are expected to collect charge, with all pixel chips powered down by default. The coarse grid readout effectively voxelizes the event at the cm$^3$ scale, for which triggered voxels occupy only $10^{-3}-10^{-4}$ of the full volume. This suppresses the on-power by the same factor, and meets the strict power budget. As a bonus, the pixel data is also sparsified by the same factor.

GAMPix requires the development of a novel CMOS ASIC chip able to power up and stabilize within a few $\mu$s of triggers (corresponding to a few mm drift). We have developed a preliminary architecture for the complete system-on-chip ASIC, with switched capacitor analog memory, per pixel triggering, and multiplexed digitization. We have also completed an initial transistor-level design of the critical power-switched front end charge-sensitive amplifier (CSA) alone, finding $<\,0.5\,\mu$s settling time and ENC of $\leq$\,20\,e$^-$. This work builds directly on the design of the related the CRYO ASIC~\citep{CryoAsic1, CryoAsic2, CryoAsic3} at SLAC, a multi-channel cryogenic non power cycling system-on-chip charge readout being developed for the nEXO experiment~\citep{nexo_2018}, and takes place in the context of the development several related cryogenic ASIC readout systems developed for DUNE~\citep{CryoAsic1, Dwyer_2018, Adams_2020}.

\begin{figure}
    \centering
    \includegraphics[width=0.45\textwidth]{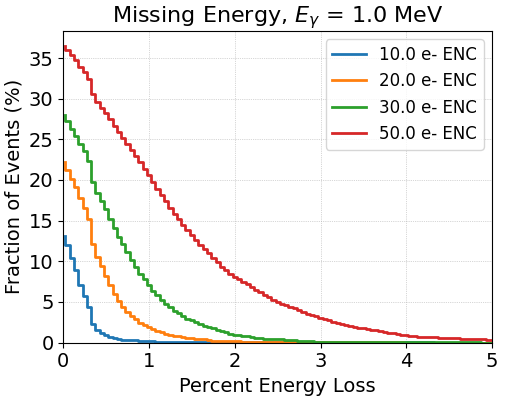}
    \caption{Fraction of events that have lost a given percentage of their energy to charge threshold, as a function of the readout noise in the coarse grids. For $\leq$\,30 e$^-$\,ENC this threshold effect has a small effect on the total measured energy, and typically causes 1 or 2 tracks near the end of the sequence to be missed which minimally affects reconstruction.}
    \label{fig:charge_thresholds}
\end{figure}

The coarse grid readout is simple by comparison but does have a demanding noise requirement of $\leq$\,30\,e$^-$ ENC per wire with a goal of 10\,$e^-$. This is driven primarily by the need for a low energy threshold for measuring electron recoil tracks, as shown in Fig.~\ref{fig:charge_thresholds}.\footnote{Note that the charge measurement will have a higher threshold than the scintillation measurement.} We estimate $\sim$\,2-3\,pF capacitance in the coarse wires, and note that a cryogenic CMOS ASIC for a similar capacitance has demonstrated $\sim\,20\,e^-$ ENC~\citep{Deng_2018}.

\begin{figure*}
    \centering
    \includegraphics[width=0.98\textwidth]{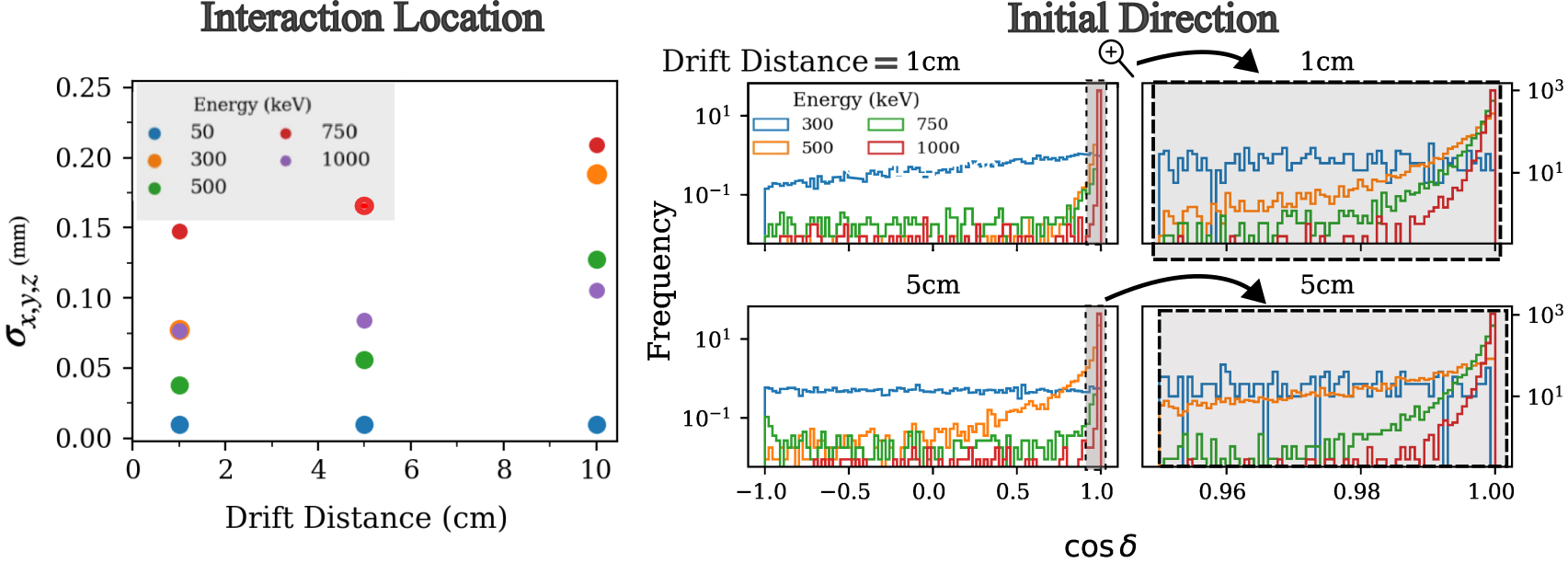}
    \caption{Electron track reconstruction results as a function of energy and drift distance, (modified) from \citep{Khek_2022}. (left) rms resolution of the interaction location, and (right) accuracy of the initial direction, measured as the cosine of the angle between the true and measured directions. Note that for energies $>$\,500\,keV the width of this distribution is a few degrees. }
    \label{fig:position_resolution}
\end{figure*}

We have studied the resolution $\sigma_{xyz}$ with which the interaction locations are measured, and the accuracy of initial track directions with machine learning (ML) techniques~\citep{Buuck_2023, Khek_2022}. This is summarized in Fig. \ref{fig:position_resolution}, where we find excellent $\sigma_{xyz}\,\le$\,250\,$\mu$m over all conditions and notably only modest dependence on drift distance. The accuracy of the initial track direction is also outstanding at high energy and still very useful down to fairly low energy, and has a strong drift distance dependence. While gaseous detectors achieve track imaging at much lower energies~\citep{Takada_2022}, we are unaware of comparable track direction measurement in any other solid or liquid detector technology. A final benefit of GAMPix is that the amount of diffusion can be accurately estimated by a comparison of the diffusion-independent coarse and diffusion-dependent pixel signals. As we show in~\citet{Gampix}, this provides a $\sim$\,5\,\% accurate measurement of the drift distance, a ``diffusion projection depth'', which reduces confusion from pile-up of events at high rates (see Sec.~\ref{sec:pileup_spacecharge}).

The 500\,$\mu$m pixel pitch was in part chosen for ease of implementation in the CMOS ASIC, and as a match to diffusion, and our studies for these $\leq$\,1\,MeV tracks finds little benefit to smaller pitch. However, this could be different at higher energies where tracks are much straighter, and so the pitch should be revisited. A more ambitious but highly speculative change to charge readout would be the addition of a ``cooling gas'' to the LAr to suppress diffusion, a technique commonly used in gaseous detectors~\citep[see, e.g.,][]{Blum2008}. Whether this is possible without either quenching or absorbing the extreme VUV scintillation light is unclear, but could be aided by the addition of Xe as an initial waveshifter.

In summary, GAMPix achieves both diffusion-limited $3D$ imaging and a diffusion-independent charge integral measurement, while simultaneously accomplishing an enormous reduction in pixel readout power and data volume. We believe the imaging capability this affords in a large, highly uniform and inexpensive detection medium will be a major advance in $\gamma$-ray instrumentation.

\subsection{Light Readout}\label{sec:light_readout}

The scintillation light comes from the decay of excimers, with a 128\,nm wavelength (9.7\,eV), with decay times of 6\,ns, and 1.6\,$\mu$s from singlet and triplet excimer states respectively. The primary requirement on the measurement of this light is to maximize light collection efficiency, or $LCE$, (the ratio of measured photo-electrons to emitted photons) to improve energy resolution as discussed in Sec.~\ref{sec:charge_light}. We also need timing information, but currently see little benefit from measuring position information.

Our design approach to achieve high $LCE$ is to make each cell a ``mirrored box'' with waveshifter-coated reflective materials on all possible surfaces, apart from the SiPM readout array, which is dense packed and also waveshifter-coated. The SiPM arrays are deployed on the cathode plane, as the low mass, field-grading cells walls are not compatible with any readout structures, and GAMPix fully occupies the anode plane. Waveshifting is needed because there are no known high reflectivity materials for the extreme UV 128\,nm light. This architecture is similar to that used in dark matter detectors, such as those of the DarkSide program~\citep{Aalseth:2017fik}, with achieved $LCE$s in the range of 10\%~\citep{AGNES2015456} to 20\%~\citep{Aoyama_2022}.

The baseline waveshifter choice for LAr is tetraphenyl butadiene (TPB) whose extensive use in LAr is thoroughly surveyed by~\citet{Kuzniak2021}. Peak emission is centered at 430\,nm and the re-emission lifetime is 2\,ns. To obtain high efficiency, the waveshifter must have high absorption probability, high photoluminescence quantum yield ($PLQY$), and high transparency to waveshifted light (i.e., a large Stoke's shift). While isolating and quantifying these parameters individually remains challenging and existing measurements exhibit significant variability (see data in~\citep{Kuzniak2021}), the measured high values of $LCE$ referenced above implies that these properties must be close to their theoretical maxima of unity. It is known that TPB dissolves at a low level into LAr, at least for some coating methods, and while to our knowledge this does not cause notable problems for detector performance (e.g., it does not cause scavenging of drifting electrons), it might require special handling in the purification system.

The reflector deployed (underneath waveshifter) on the field shaping walls and rib structures is a multilayer polymer "enhanced specular reflector" (ESR) film~\citep{Weber2000}\footnote{\url{https://multimedia.3m.com/mws/media/1389248O/application-guide-for-esr.pdf}}, which has remarkably high ($\sim$\,99\%) and largely angle-independent reflectivity over the optical range\footnote{PTFE is another highly reflective material often used in liquid noble TPCs, but is extremely electronegative and could thus accumulate surface charge which distorts the drift field. It also has mechanical properties less suited for this application.} The CMOS cover layer of GAMPix's pixel ASIC and the coarse grids can be coated with reflective Al or Ag cover layers (again underneath waveshifter), with reflectivities near 90\%. 

SiPMs have a number of advantages of over PMTs, but the driving feature here is their much smaller mass and volume that enables a low-dead mass central cathode plane. SiPMs have $QE$s (photo-electrons per photons absorbed) at 430\,nm that are now exceeding PMT $QE$s, and a total photon detection efficiency (including both SiPM $QE$ and the active area fill fraction) of $>\,40$\% has been demonstrated~\citep{Gallina2024}. The use of SiPMs in liquid noble detectors has been expanding rapidly, with large, dense SiPM arrays under development of $>\,4.5$\,m$^2$ in LXe for nEXO~\citep{Gallina_2022}, and $>\,20$\,m$^2$ and in LAr for DarkSide-20k~\citep{Aalseth:2017fik}. A large array is currently operating in LXe as part of the MEG II experiment~\citep{Baldini_2018}. The dark rate in SiPMs is much higher than PMTs, but at temperatures less than $\sim$\,150\,K is below 0.1\,Hz/mm$^2$, which should allow a 2 photo-electron threshold and thus an energy threshold (assuming an $LCE$ above 10\%) well below that of the charge measurement. Cross-talk between channels is significant in SiPMs~\citep{Gibbons_2024}, but seems unlikely to present a serious problem in this non-imaging application.

While for these reasons, SiPMs are well-matched to this application, there are issues to address. The power dissipation for both the nEXO and DarkSide-20k arrays is expected to be $\sim\,20$\,W/m$^2$, which must be decreased by an order of magnitude (see Sec.~\ref{sec:cryogenics_fluids}). A likely path forward is some version of ``digital SiPMs''~\citep{Frach2009, Fischer2022, diehl2024digital}, which are a new class of CMOS ASIC readout that have separate channels for each of the SiPM diode arrays, and promise several benefits including the needed power reduction~\citep{Retiere2017}. SiPMs are beginning to be used in space and are affected by radiation damage~\citep{Garutti2019, Zheng2022, Acerbi2023, Altamura2023, POLAR-2:2022tiw}, primarily in the form of an increase in dark current increase, an effect which is minimized by cryogenic operation. Finally, our baseline design is to have the waveshifter-coated SiPM array itself serve as the cathode plane upon which ions are collected. If this creates problems due to positive charge build-up, the ions could instead be collected on a high transparency mesh electrode placed directly in front of the SiPMs. This comes at the cost of a thin layer with close to zero charge collection, but whose volume is a modest $\sim$\,0.6\% of the cell volume per mm of thickness.

Timing information from the scintillation signal has several uses.\footnote{In principle, it may be possible to obtain some measure of timing and hence sequence of scatters from a fast singlet signal. However, there are significant obstacles to this approach: the use of waveshifter which further delays the signal, the fact that only a small fraction of the light arrives promptly at the SiPMs, the high average hit count in a cell which determines the number of summed light signals to disentangle, and the need for high speed digitization which would require significant power.} If a science case drives it, resolution on the scale of a ns or less could be obtained based on the excimer singlet decay time and waveshifter lifetime. Otherwise, the TPC depth reconstruction must at least be as good as the $\sim$\~500\,ns pixel sampling time, and thus likely $\sim$\,100\,ns or better. This time scale also allows robust measurement of the relative populations of single and triplet decays to tag background neutron recoils (see Sec.~\ref{sec:backgrounds}), and allows tagging of event pile up from the presence of multiple scintillation signals (see Sec.~\ref{sec:pileup_spacecharge}). The DAQ system will likely provide only the integral signal integrals and pulse times, as waveform digitization will be power hungry, especially for $\leq$\,100\,ns timing.

We have studied the $LCE$ using {\tt LightGuide}, a custom ray-tracing simulation package which includes waveshifting and detailed treatment of interactions with grid wires. Based on those studies (and consistent with~\citep{AGNES2015456} and~\citep{Aoyama_2022}) we adopt pessimistic, nominal and optimistic $LCE$ values of 5, 10 and 30\% for our instrument performance studies (Sec.~\ref{sec:performance}). The optimistic value will require high SiPM $QE$ and fill factor, and achieving very high reflectivity in the GAMPix readout and other structures. Due to the mirrored box design, the spatial uniformity of the response is better than $\sim$\,10\% despite the asymmetry imposed by the use of a single light readout plane. Non-uniform light response is readily corrected in analysis, but recombination fluctuations on each scatter impose a dispersion term that cannot be corrected when combining light and charge signals to find energy (Sec.~\ref{sec:charge_light}). We assess this effect to be sub-dominant with $\sim$\,10\% spatial uniformity.

\subsection{Cryogenics and Fluids}\label{sec:cryogenics_fluids}

The cryogenic portions of the satellite are the LAr tracker and associated LAr circulation and purification equipment, and the calorimeter if it is cold. These will be insulated by multi-layer-superinsulation (MLI), and covered with and a micro-meteor and orbital debris (MMOD) shield and a solar reflecting surface. The Earth side of the cooled assembly will have a constant infrared radiation load from the Earth (or other $\sim$\,300\,K parts of the spacecraft), while the space side will alternately face the Sun and deep space during orbit. Cooling will be provided by a cryocooler\footnote{We also considered passive cooling, but this would require very large Earth and Sun shields which appear prohibitively large.} applied to a circulation loop through which LAr is circulated. The loop also allows online purification of LAr to remove electronegative impurities from outgassing of the TPC and vessel.

\begin{figure}
    \centering
    \includegraphics[width=0.45\textwidth]{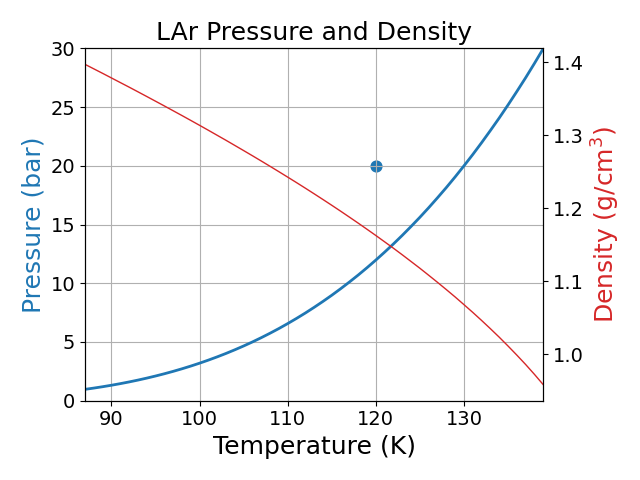}
    \caption{The equilibrium pressure and density of LAr as a function of temperature. The circle indicates a possible operating point with 10\,K of sub-cooling.}
    \label{fig:density_pressure}
\end{figure}

The cooling requirements and the power budget for electronics and other detector components are coupled, and we outline the contours of a final design here. Because the available cooling power increases and input power decreases with increasing cold temperature, the detector will likely be operated above the 87\,K atmospheric pressure boiling point. This elevates the pressure as shown in Fig.~\ref{fig:density_pressure}, requiring an increase in vessel wall thickness (see Sec.~\ref{sec:geometry_mechanics}), and suggests operating in the 120--130\,K and 12--20\,bar range. Existing space-ready cryocoolers of pulse tube (Northrup-Grumman HEC\footnote{\url{https://www.northropgrumman.com/space/cryocoolers/}}) and Stirling (Sunpower CryoTel\textsuperscript{\textregistered} CT-S\footnote{\url{https://www.sunpowerinc.com/products/stirling-cryocoolers/cryotel-cryocoolers/spacecryocoolers}}) designs provide roughly 20\,W cooling power at 120\,K while requiring less than 200\,W input power. Much higher powered but not yet-space-qualified versions of these technologies are on the market (e.g., the Sunpower CryoTel\textsuperscript{\textregistered} DS\,30 which has 60\,W cooling power at 120\,K with 480\,W input power). We may benefit from developments for "Zero-boil off" liquid oxygen storage, including a high energy-efficiency reverse-Brayton cryocooler with 150\,W cooling at 90\,K~\citep{Plachata2018}.

The maximum radiant heat load through MLI can be below 0.5\,W/m$^2$\footnote{See, e.g., Quest Thermal Products IMLI,~\url{https://questthermal.com/products/imli/}}, and will vary as the sun exposure changes during orbit. The mechanical connection between the cryogenic elements and the warm part of the spacecraft must be designed for low thermal conductance, with less than $\sim$\,10\,W power appearing feasible. Based on these effectively irreducible radiation and mechanical heat loads, and an expected cryocooler cooling power in the range of 30-100\,W, we have adopted provisional power requirements of $\lesssim$\,1\,W/m$^2$ for each layer of charge and light readout, and any other power dissipating elements of the detector, including the power dissipated in the divider chains of the high voltage (HV) field cage(s).

The LAr will need to be circulated both for cooling and purification. Assuming a conservative total power of 10\,W per m$^2$ of tracker area, and that the cryocooler cools the liquid by 5\,K, this can be provided by a modest LAr flow of $\sim$\,0.25\,lpm per m$^2$ of tracker. However, because there will be no convection in zero gravity, the flow of the LAr needs to be engineered to effectively bring the cooled liquid in contact with the power dissipating elements. The main power sources in each TPC cell are GAMPix on the anode plane, SiPMs on the cathode plane, and the resistors of the field cage. The anode plane is readily cooled by flowing cold through the thin LAr layer (used as the ACD) between anode and vessel wall. Cooling can be provided to the perimeter of the cathode plane in each cell by flowing cooled LAr through the inside of the ribs which form the structural backbone of the TPC cells structure. Lateral transport in the plane to the perimeter will require likely require a metallic layer in the circuit boards, with, for example, a $\sim$\,200\,$\mu$m thick Be sheet reducing any gradient to below a Kelvin. The cooling flow through the ribs directly cools the resistor divider chains for the field cage walls, as the interior of the rib is the natural location for the divider chains. Finally, while the pixel ASIC on-power is a very high $\sim$\,W/cm$^2$, the short ($\leq$\,100\,$\mu$s) on time and chip heat capacity combine to give only a mK-scale temperature rises per cycle.

Purification is needed to remove dissolved gasses which can degrade both charge and light collection, and which arise from outgassing of plastics and other materials. Techniques for removing dissolved gasses have been extensively developed for both LAr~\citep[see, e.g.,][]{Adamowski_2014} and LXe. The primary concern is O$_2$ and H$_2$0, which are now routinely suppressed to well below 0.1\,ppb, resulting in no appreciable light loss and little charge loss over multi-meter distances. We plan continuous liquid circulation through an appropriate getter or similar filter material, which is the baseline design in most LAr systems. The system requirements are modest given our small $\sim$\,20\,cm TPC cells, and the LAr circulation rate through the detector is probably driven by the cooling requirements instead of purification. 

There is a broad range of commercially available spacecraft fluid-handling components including pumps, valves and instrumentation, much of which is used to handle propellants such as LO2. We have more stringent leak tightness requirements than propellant handling, but much easier chemical compatibility requirements. There are two commonly used purifier materials, the first a commercial zirconium alloy non-regenerable getter (SAES ST707) usually used in gas phase at elevated temperature, but recently demonstrated in LXe~\citet{Plante_2022}. The other, used routinely for LAr, is Cu impregnated alumina spheres (BASF Cu-0226 S) that can be regenerated at high temperature under a H2/Ar gas stream. The capacity of the getter, or inventory of regeneration gas, will presumably be minimized by pre-outgassing of the detector.

There is one significant way in which the combined cryogenic and fluid handling system in space must differ from a ground-based system, namely the absence of gravity. The key issue here is that if the only control were the liquid temperature fixed by the cryogenic system, the Ar would naturally sit at its equilibrium saturation pressure (see Fig.~\ref{fig:density_pressure}), with the total inventory of Ar present determining the amount of Ar in the liquid and gas phases. A liquid-only state would require an unattainable exact matching of the actual Ar inventory to that needed to exactly fill the system with liquid. On the ground, gravity separates any gas from the liquid, but in space the location of the gas will not be controlled, which is unacceptable for the TPC operation. 

We propose to solve this by use of an expandable membrane, like that used in propellant tanks~\citep{Hartwig2016}. One side is in contact with the LAr while the other is pressurized with a gas (likely with N$_2$) to enforce a single phase of liquid argon at fixed pressure. With the pressure set at a higher value than the equilibrium pressure corresponding to the fluid temperature established by flow through the cryogenic system, the LAr is in a sub-cooled state. This suppresses boiling at the electronics and other power dissipating elements. As shown in Fig.~\ref{fig:density_pressure} sub-cooling well above the sub-kelvin gradient expected in the SiPM array (the likely hottest part of the detector) can readily be achieved, and in fact is larger and more uniform than what obtains in ground based detectors due to the $\Delta P = \rho g h$ pressure increase with depth. A final fluid issue is how the liquid is handled during launch. If the LAr is in the detector, any sloshing would readily damage or destroy the lightweight TPC structure. A more attractive option is to store the liquid in a separate tank (possibly a large toroid below the main instrument) during launch and to transfer it to the detector once in orbit, possibly using the same membrane system used to establish the fluid state.

This cryogenic system will be integrated into the spacecraft's full thermal control system which handles all $\sim$\,300\,K power loads from electronics and other spacecraft systems, as well as the cryocooler. That conventional system is beyond the scope of this paper.

\subsection{Tracker Geometry and Mechanics}\label{sec:geometry_mechanics}

The radius of the hemispherical section (of fixed thickness) can be adjusted to optimize the sky coverage uniformity and etendue. Competing factors drive the tracker thickness, with increased thickness given higher Compton event containment, but shorter cell height reducing both diffusion (Sec.~\ref{sec:gampix}) and event pile-up and space charge (Sec.~\ref{sec:pileup_spacecharge}). This leads to the double-layer geometry, which also naturally isolates the large negative HV cathode layer from the grounded vessel wall\footnote{The vessel wall should be a Faraday cage if for no other reason than coping with the large particles flux in orbit.}, though at the cost of doubling the readout systems and adding dead material. The buffer volume at the perimeter of the hemisphere provides a natural location to stage the HV feedthrough, fluid handling elements, and other services, and is readily instrumented with SiPMs as an active shield.

The TPC cell structure should have minimum mass, but must be robust enough to survive launch, accommodate a wide range of temperatures, and maintain dimensional accuracy for imaging. PCB ribs line all edges of the TPC cells, and the PCB planes are rigidly fixed to this (see Fig,\ref{fig:gampix}), forming a stiff lightweight hemispherical honeycomb structure. The thermal expansion of that structure will likely not match that of the field cage walls and vessel. The thin cell walls, likely polyimide sheet coated with ESR reflectors, could be held taut using leaf springs in the ribs, and the overall TPC structure could likewise be suspended in the vessel with a stiff spring arrangement. The interior of the ribs will house the GAMPix's coarse grid mechanics and readout (Fig.~\ref{fig:gampix}), the SiPM array cabling, resistors for the field cage, and mechanical items, and will route fluid flow for cooling and purification (Sec.~\ref{sec:cryogenics_fluids}). 

The readout plane PCBs are lined with CMOS or SiPMs on one side, and on the other mostly Si capacitors. Signal and power traces are routed in buried layers to interconnects on the board edges housed in the ribs. Standard PCB manufacturing a thickness between $\sim$\,0.7 and 1.8\,mm (6 - 14 board layers). The many capacitors needed for buffering power could be low mass and temperature-stable Si capacitors\footnote{From, e.g., Murata Capacitors, \url{https://www.murata.com/en-us/products/capacitor/siliconcapacitors}} with thicknesses between 60 and 400\,$\mu$m.

The vessel wall thickness and the areal density of the micro-meteor and orbital debris (MMOD) shield will be minimized to reduce dead mass. The vessel wall thickness is determined by the strength of the vessel material, the radius of curvature, and the internal pressure. Conservatively using the ultimate tensile strength of generic carbon fiber composite (higher performance composite materials are available), and the NASA guideline pressure safety factor of 1.5~\citep{Holladay_2004}, we have 2.2 and 4.3\,mm wall thicknesses at 10 and 20\,bar, respectively. The MMOD design and requirements will depend on factors such as the orbit and orientation of the tracker, with a likely mass in the range of 0.2\,-\,1.0\,g/cm$^2$~\citep[see, e.g.,][]{Arnold_2009}. At the upper end of this range, it would be the dominant dead mass, and so will need to be carefully optimized.

In Table~\ref{tab:dead_mass} we show preliminary estimates of the dead mass in several subsystems. Events with lost energy in internal ribs, planes and walls have compromises reconstruction, whereas events with energy lost in the vessel and MMOD have unaffected reconstruction, but form a background of incorrectly pointed $\gamma$-rays. In the simulations below we use less extreme values (Sec.~\ref{sec:simulations}) than the high and low totals here, as the probability of all component values being at their extrema is small.

\begin{table}[h]
    \centering
    \caption{Fractional Dead Mass Across Components and Scenarios} 
    \label{tab:dead_mass}
    \begin{tabular}{lccc}
        \toprule
        & \multicolumn{3}{c}{Dead Mass Percent} \\ 
        Component & Low & Medium & High \\
        \hline
        Ribs              & 0.15 & 0.59  & 1.33  \\
        Planes + Walls    & 0.79 & 1.57  & 3.93  \\
        Vessel Wall + MMOD & 1.20 & 2.60  & 6.00  \\
        Total             & 4.10 & 8.80  & 16.60 \\
        \hline
    \end{tabular}
\end{table}

\subsection{Fields, High Voltages, and Particle Flux}\label{high_voltage_particles}

High fidelity imaging requires careful control of the electric field throughout the TPC, achieved by field grading structures that cover the cell walls and ribs. These focus electrons away from the ribs and distribute them uniformly over the GAMPix plane. They also shield the drift region from the interior of the ribs, where conductors in the readout and mechanical elements will result in highly non-uniform fields. The field grading will likely be implemented as metalized traces on a polyimide film layer, with resistor divider chains housed in the ribs. With 1\,mm  trace pitch, the resulting volume with non-uniform fields which can lead to charge loss is only $\sim$\,0.5\% of the total\footnote{An alternative scheme with resistive films and no ladder has been studied for DUNE~\citep{Berner:2019uvt} and could be used here. But the very high resistivity needed to manage heat loads (likely $\geq$\,10$^{10}$\,G$\omega$/sq) is challenging, and cooling appears difficult.} 

Our 0.5 kV/cm drift field target is now regularly achieved in various large DUNE prototypes, but here only requires handling $\lesssim$\,10\,kV bias voltage compared for instance to 180\,kV in ProtoDUNE~\citep{protodune_2021}, and as noted above the geometry is favorable for implement this. Electronics for pixels, wires and SiPMs are staged at different -HV levels with fiber and cable interconnects. Fields are highest where focused onto the pixels, which could readily be damaged by sparking, but we note successful operation of large arrays of LArPix readout~\cite{Dwyer_2018} which has a similar field configuration and CMOS readout.

The high particle flux (Sec.~\ref{sec:pileup_spacecharge}) places special demands on the design, including careful consideration of component aging. In separate sections, we discuss activation (Sec.~\ref{sec:backgrounds}), SiPM degradation (Sec.~\ref{sec:light_readout}), pile-up and space charge (Sec.~\ref{sec:pileup_spacecharge}) and backgrounds (Sec.~\ref{sec:backgrounds}). With on-board purification (Sec.~\ref{sec:cryogenics_fluids}) LAr is effectively immune to radiation damage, and CMOS circuitry is extensively used at much higher doses than LEO in collider experiments. Aging has not been an important factor in LAr scintillation readout over many experimental configurations to date~\citep{Kuzniak2021}, but this will have to be verified for the flux expected in orbit.

 De  spite the large number of pixels in the instrument, the amount of data produced, processed and communicated to with the ground does not appear prohibitive. We estimate a triggered pixel sample rate for $\>$\,100\,keV cosmic $\gamma$-rays at $\sim$\,300\,K-samples/m$^2$/s, and a triggered coarse wire sample rate a factor of several smaller. We expect the data samples rates from charged particles and albedo gammas to be lower than this because of online event vetoing from the ACD and calorimeter, respectively. On board processing, such as reducing the pixel samples to interaction locations and electron track directions, could substantially reduce the data volume.

\subsection{Costs and Development}\label{sec:costs_development}

We expect the development of the GammaTPC technology to follow the track record of liquid noble TPCs, quickly and economically scaling to large size after the core technology is developed and proven at a small scale. For example, core aspects of LXe TPC dark matter detectors were developed at the 100\,gm scale,~\citep{Aprile_2006, Dahl:2009nta} while the first world-leading experiments were contemporaneously staged at the $\sim$\,10\,kg scale~\citep{Angle_2008, Alner_2007, Lebedenko_2009}. This mass increased by a factor of 1000 to 10\,tons in just over 15 years~\citep{LZ2023, xenonnt2023}, with closer to an order of magnitude increase in detector cost. Here, much of the instrument cost will be in the development of the CMOS ASICs for GAMPix and the probable CMOS ASICs for SiPM readout, and in creating space-ready versions of otherwise conventional tracker readout mechanical structures and LAr fluid handling and cryogenics. Demonstration of zero-g operation of the tracker can be done with a single TPC cell in a small launch vehicle.

The production cost of the 130\,nm CMOS process used for GAMPix is currently $\sim$\,0.3\,M\$/m$^2$, and we expect similar cost/area for the SiPM arrays and readout, such that the hardware cost of the tracker is likely $\sim$1-2\,M\$/m$^2$. Given semiconductor trends, these cost are if anything likely to fall over time. The remaining hardware costs are in the less complex CsI calorimeter, cryogenics and fluid handling, along with conventional electronics and the spacecraft. The calorimeter cost will scale with mass, but we expect the cryogenics and fluid handling costs to scale only weakly with detector mass. Note that while the cryocooler may need to be adapted from a conventional model (see Sec.~\ref{sec:cryogenics_fluids}), the NASA Instrument Costing Model~\citep{NICMCryocooler_2017} estimate for this with a 120\,K base temperature is only $\sim$\,2\,M\$.

While the scale and cost of an eventual mission is not yet clear, a key performance metric efficiency of event containment in the tracker, set by its thickness and area. The cost scales with area, but is independent of thickness. Events within $\delta r \sim$\,20\,cm of the lateral edges will typically not be contained. The fractional area with compromised containment is $2 \delta r /r$ (approximating the geometry as planar), which for instrument diameters of 1, 2 and 3\,m is 40, 20 and 14\% respectively. This favors a diameter well above 1\,m, for which the mass is on the scale of tons. 

Historically, space mission costing assumes a strong dependence with mass, with multi-ton missions well outside the MIDEX scale. However most of our tracker mass is in the inert and essentially zero cost LAr, with the cost-driving internal structures only $\sim$\,5\% of the of LAr mass (see Sec.~\ref{sec:geometry_mechanics}). Moreover, the cost of launching both large and small masses to space has been dramatically declining with the advent of commercial launch capabilities. Low cost LAr TPC technology is uniquely suited to take advantage of this, both during development and in the final instrument, whose size can be readily tailored to the scientific opportunities and funding landscape.

\section{Instrument Performance}\label{sec:performance}

\subsection{Simulation Framework}\label{sec:simulations}
To assess the performance of the GammaTPC instrument for Compton events up to 10 MeV, we conducted detailed simulations combining multiple tools and methodologies. Event generation was performed using {\tt Cosima}, a Geant4-based Monte Carlo (MC) module within the {\tt MEGAlib} software suite \citep{MEGALIB}. Background fluxes of photons and particles in Low Earth Orbit (LEO) were based on the average values from {\tt LEOBackground} \citep{Cumani_2019}, while activation-induced radioactivity was simulated using {\tt Cosima}. We have not yet incorporated time dependence of the particle flux.

\begin{figure}[tb] 
\centering 
\includegraphics[width=0.475\textwidth]{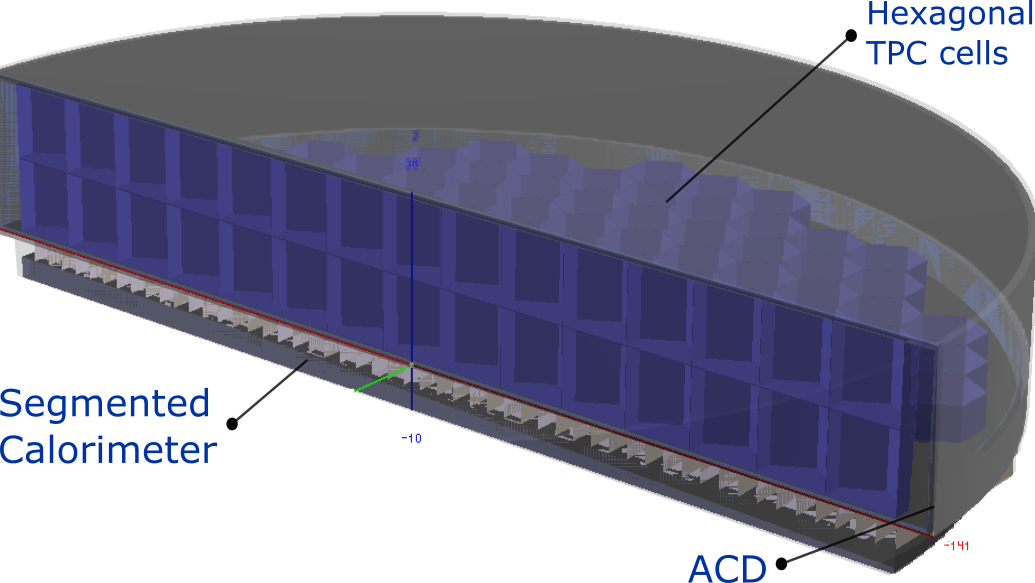} 
\caption{Cross-sectional view of the $3D$ model of the instrument used in simulations, with dimensions given in Table~\ref{tab:simulations_parameters}.} 
\label{fig:sims_geometry} 
\end{figure}

The detector response was modeled with {\tt GAMPy}\footnote{\url{https://github.com/tashutt/Gampy/}}, a custom Python-based framework incorporating detailed treatments of charge and light signals as described in Sec.~\ref{sec:charge_light}. The tracker’s spatial resolution and the directionality of electron tracks were parameterized based on the simulation results shown in Fig.~\ref{fig:position_resolution}. A uniform light collection response was assumed for the tracker and anti-coincidence detectors (ACDs). The simulated geometry, illustrated in Fig.~\ref{fig:sims_geometry}, includes a double layer of hexagonal TPC tracker cells and a CsI calorimeter. The tracker is modeled with a simplified planar geometry, which gives only an approximate treatment of the sky coverage and $FOV$. Also, for simplicity, the inert materials within the tracker were modeled as inert LAr, with dimensions adjusted to preserve the correct dead mass. The spatial information in the calorimeter signals was not yet used in this study.

The major parameters of the simulated detector are shown in Table~\ref{tab:simulations_parameters}. Individually varying all parameters is impractical, so we adopt a set of pessimistic, nominal and optimistic values. While with effort we may approach or achieve some optimistic values, the probability of the full set of parameters being at the optimistic or pessimistic values is presumably low.

\begin{table*}[htbp]
    \centering
    \caption{Varied Instrument Geometry and Response Parameters for Different Scenarios.}
    \begin{tabular}{lccc}
        \toprule
        \textbf{Simulation Parameters} & \textbf{Optimistic} & \textbf{Nominal} & \textbf{Pessimistic} \\
        \hline
        \textbf{Geometry} & & & \\
        Instrument diameter (m) & ... & 3 & ... \\
        Tracker cell height and flat-to-flat width (cm) & ... & 17.5 & ... \\
        Number of tracker cells & ... & 374 & ... \\
        LAr mean ACD thickness (mm) & ... & 9.5 & ... \\
        CsI calorimeter thickness (cm) & ... & 10.0 & ... \\
        Anode and cathode plane thicknesses (mm) & 0.7 & 1.2 & 2.0 \\
        Ribs + planes + walls mass fraction (\%) & 1.4 & 2.2 & 4.0 \\
        Total dead mass fraction (\%) & 2.5 & 5.6 & 9.7 \\
        \hline
        \textbf{Readout} & & & \\
        Spatial resolution multiplier & 0.75 & 1.0 & 1.5 \\
        e$^-$ direction error multiplier & 0.8 & 1.0 & 1.2 \\
        Light collection efficiency & 0.3 & 0.1 & 0.05 \\
        Recombination fluctuations strength parameter & 0.04 & 0.05 & 0.06 \\
        Coarse grids noise ENC (e$^-$) & 10 & 20 & 40 \\
        \hline
    \end{tabular}
    \label{tab:simulations_parameters}
\end{table*}

\begin{figure}
\begin{center}
\centering
\begin{tikzpicture}[
    every node/.style={font=\footnotesize, align=center},
    block/.style = {rectangle, draw, text centered, rounded corners, minimum height=2em, minimum width=10em},
    process/.style = {rectangle, draw, text centered, rounded corners, fill=orange!20, minimum height=3em, minimum width=11em},
    parser/.style = {rectangle, draw, fill=yellow!20, text centered, rounded corners, minimum height=2em, minimum width=12em},
    lightread/.style = {rectangle, draw, fill=orange!40, text centered, rounded corners, minimum width=9em},
    output/.style = {rectangle, draw, fill=green!20, text centered, rounded corners, minimum height=2em, minimum width=10em},
    arrow/.style = {thick,->,>=stealth},
    dashedarrow/.style = {thick,dashed,->,>=stealth},
    gampy/.style = {draw, dashed, inner sep=0.5em, rounded corners, line width=0.8pt}
]

\node[block, fill=blue!20 ] (GammaTPCGeom) {\textbf{Cosima}\\GammaTPC (Fig. \ref{fig:sims_geometry})\\ Geometry \& Materials};

\node[process, above=of GammaTPCGeom, xshift=-2.2cm , yshift=-0.6cm] (lineSources) {\textbf{MEGAlib}\\ Far Field Monoenergetic \\ Point Sources};

\node[process, above=of GammaTPCGeom, xshift=2.2cm , yshift=-0.6cm] (backgrounds) {\textbf{MEGAlib  \& LEOBackground}\\Activation $\&$ Cosmic Backgrounds};

\node[parser, below=0.6cm of GammaTPCGeom ] (eventParser) {\textbf{Event parser}\\ - (x,y,z) location\\ - Energy, Time\\ - Electron Recoil Direction};

\node[process, below=0.2cm of eventParser, minimum width=4em, fill=red!50 ] (antiCoincidence) {Anti-\\Coincidence\\ Detector};

\node[lightread, below left=0.3cm and -0.8cm of antiCoincidence, minimum height=1.4cm ] (lightReadout)
{
  \textbf{Light Readout}\\
  (Simple treatment)\\
  - Energy in light\\
  - Time
};

\node[lightread, fill=purple!20, below=0.2cm of lightReadout, minimum height=1.2cm ] (calorimeter)
{
  \textbf{Calorimeter }\\
  - Energy\\
  - Position*
};

\node[lightread, below right=0.3cm and -0.55cm of antiCoincidence, minimum height=2.8cm ] (chargeReadout)
{
  \textbf{GAMPix}\\
  \textbf{Charge Readout}\\
  - $3D$ $e^-$ Track Shape\\
  - $e^-$ Recoil Direction\\
  - Total Charge\\
  - Drift length
};

\node[process, below=3.6cm of antiCoincidence, minimum width=18em, fill=orange!50 ] (energyCombination)
{\textbf{Scatter Energy Calculation}\\ - Combine Light, Calorimeter, and Charge\\ - Estimate Total Energy Per Scatter};

\node[process, below=0.5cm of energyCombination, minimum width=18em, fill=orange!60 ] (reconstruction)
{\textbf{GammaTPC Reconstruction}\\ - CKD Test and Ordering\\ - Energy and Incoming Direction Calculation\\ - Classification};

\node[output, below left=0.4cm and -3cm of reconstruction, minimum height=1.2cm ] (energyResolution)
{- Energy Resolution\\ - Angular Resolution\\ - Effective Area};
\node[output, below right=0.4cm and -3cm of reconstruction, minimum height=1.2cm ] (backgroundFlux)
{- Background Flux\\ - Pileup Rate};

\node[output, below=2.6cm of reconstruction , fill=yellow!40, minimum width=15em, , yshift=0.4cm] (performance)
{\textbf{Performance}\\ - Line Sensitivity\\ - Continuum Sensitivity};

\begin{scope}[on background layer]
    \node[gampy, fit=(eventParser) (antiCoincidence) (lightReadout) (calorimeter) (chargeReadout) 
                 (energyCombination) (reconstruction) (energyResolution) (backgroundFlux)] (gampyBox) {};
\end{scope}

\node at (gampyBox.north west) [left, font=\ttfamily, anchor=north west, xshift=2mm, yshift=-1mm] {GAMPy};

\draw[arrow] (lineSources) -- (GammaTPCGeom);
\draw[arrow] (backgrounds) -- (GammaTPCGeom);

\draw[arrow] (GammaTPCGeom) -- (eventParser);
\draw[arrow] (eventParser) -- (antiCoincidence);

\draw[arrow] (antiCoincidence) -- (lightReadout);
\draw[arrow] (antiCoincidence) -- (chargeReadout);

\draw[arrow] (lightReadout.east) -- ++(0.125,0) -- (energyCombination);

\draw[arrow] (antiCoincidence.west) -- ++(-2.3,0) -- ++(0,-2.9) -- ++(0.0,0) -- (calorimeter.west);

\draw[arrow] (calorimeter) -- (energyCombination);

\draw[arrow] (chargeReadout) -- (energyCombination);

\draw[arrow] (energyCombination) -- (reconstruction);
\draw[arrow] (reconstruction) -- (energyResolution);
\draw[arrow] (reconstruction) -- (backgroundFlux);

\draw[arrow] (energyResolution) -- (performance);
\draw[arrow] (backgroundFlux) -- (performance);

\end{tikzpicture}

\caption{Key stages of the performance simulation pipeline. Simulated monoenergetic sources and cosmic/activation backgrounds from MEGAlib are parsed for spatial and temporal information, filtered for ACD anti-coincidence, followed by simulation of signal production and readout processes including fluctuations, efficiencies and readout noise. Tracker, calorimeter and ACD signals are feed into the reconstruction phase, which includes the CKD test, energy-direction calculations, and classification. The outputs include energy and angular resolutions, background flux, pileup rate, and effective area, leading to the evaluation of line and continuum sensitivities.}
\label{fig:sim_flow}
\end{center}
\end{figure}
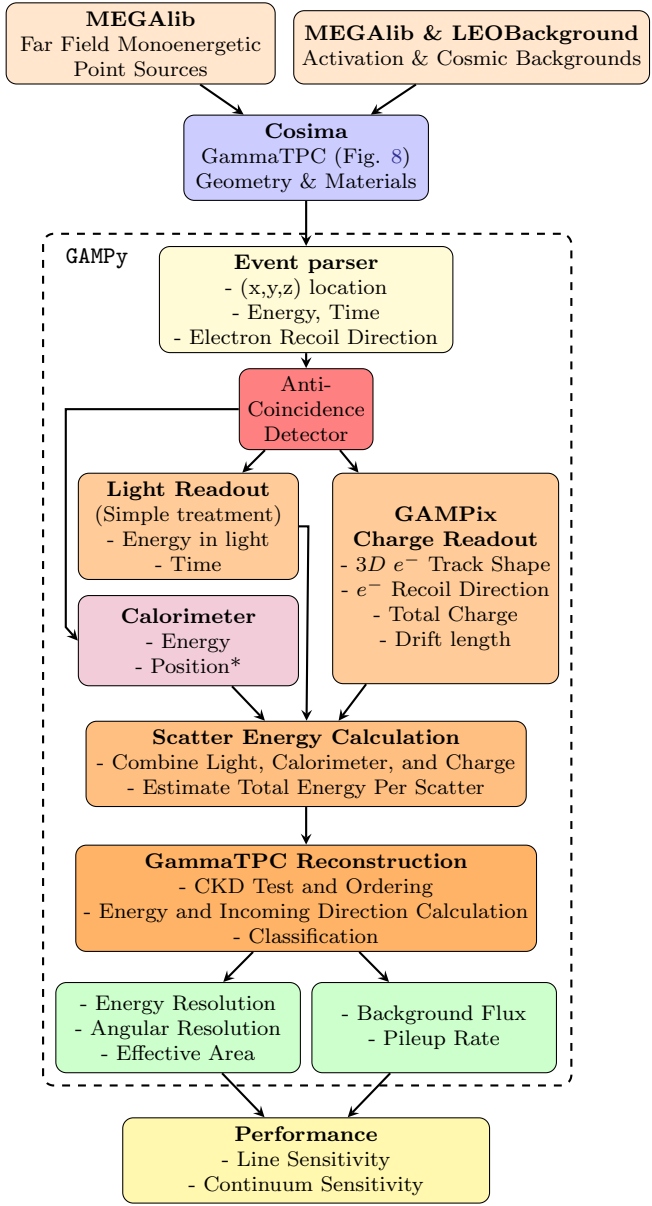

The simulation workflow is illustrated in Fig.~\ref{fig:sim_flow}. Far-field monoenergetic point sources at various energies and incident angles were simulated, along with contributions from cosmic, albedo, and activation-generated photons. {\tt GAMPy} was used to apply the readout response, providing the position and energy of each scatter, as well as the recoil electron directions. Signals in the ACD were used to suppress background particle events.

\vspace{3em}

\subsection{Reconstruction}\label{sec:reconstruction}

The primary objective of the reconstruction process is to identify which of the multiple Compton scatters in a single event occurred first and second, as these initial interactions determine the incident $\gamma$-ray direction. This process is commonly achieved through Compton Kinematic Discrimination (CKD), which systematically evaluates every possible sequence of Compton scatters in search of the most probable one. Early implementations, such as those by \citet{Aprile1993}, relied on analytic constraints derived from Compton scattering kinematics, while \citet{Boggs_2000} introduced more sophisticated statistical approaches to account for uncertainties in detector response. More recently, \citet{NN_CKD} demonstrated neural network solutions that learn complex scattering patterns from simulated data, leading to further improvements in reconstruction accuracy.

At the reconstruction stage, we chose not to adopt machine learning methods in order to maintain a transparent analytic approach that can be thoroughly validated. Early testing revealed that purely angular-based error minimization is not robust near scattering angles of 0 or $\pi$. There, our instrument’s positional (i.e., geometric) readouts are often more precise than its energy measurements, making an energy-centric error formulation more reliable. Consequently, we developed a modified CKD algorithm that minimizes discrepancies between measured and predicted energies at each scatter, while penalizing sequences deemed statistically unlikely or incompatible with the electron track readout. Details of this energy-based minimization strategy and its regularization are presented in Appendix~\ref{sec:appendix_minCKD}.

To further enhance the reconstruction, we exploit a key advantage of our instrument: many events are fully contained within the detector. That allows comparison of the total measured energy from all scatters to the energy calculated from the first three interactions. Following \citet{Aramaki_2020}, we define
\begin{equation} 
E_{\text{calc}} = E_1 + \frac{E_2}{2} + \sqrt{\frac{E_2^2}{4} + \frac{m_e c^2 E_2}{1 - \cos \theta_2}},
\label{calc_ene} 
\end{equation}
where \( E_1 \) and \( E_2 \) are the energies deposited at the first and second Compton scatter sites, and \( \theta_2 \) is the Compton scattering angle at the second interaction, defined by the vectors connecting the first-to-second and second-to-third scatter locations.

\subsection{Classification}\label{sec:classification}
Comparing \( E_{\text{calc}} \) with the total measured event energy helps identify invalid sequences that fail basic energy consistency checks. However, this alone does not resolve all ambiguities, as some invalid sequences can closely mimic valid ones. Additionally, we aim to categorize events by origin—whether they are cosmic \(\gamma\)-rays or backgrounds from the atmosphere or the detector itself. Consequently, we introduce a dedicated classification stage that integrates multiple quality metrics in addition to the energy comparison.

\paragraph{Quality Metrics}
Once the most probable scatter sequence is identified for each \(\gamma\)-ray event, we compute a set of quality metrics, which serve both as quality cuts and as inputs to the classification algorithm. These metrics capture event characteristics beyond simple energy matching, and help to distinguish legitimate, correctly reconstructed \(\gamma\)-ray interactions from non-cosmic background events.

\begin{enumerate}
    \item \textbf{Cumulative Energy Error}: The square root of the sum from Equation \ref{eq:sum_error} (Appendix~\ref{sec:appendix_minCKD}), representing the total discrepancy between measured and calculated energy of the incoming $\gamma$-ray.

    \item \textbf{Calculated Energy}: The energy derived from the energies of the first two scatters and the geometric scattering angle at the second scatter \( \theta_2 \). This is computed using Equation~\ref{calc_ene}.

    \item \textbf{Klein-Nishina Probability \( P_{\text{K-N}} \)}: The product of the Klein-Nishina probabilities for each scatter in the event, assessing the physical plausibility of the observed sequence:
    \begin{equation}
    P_{\text{K-N}} = \prod_{i=0}^{N} P_{\text{K-N}}^{(i)}.
    \end{equation}

    \item \textbf{Electron Recoil Angle Mismatch}: The squared sum of angular errors between the plane defined by three consecutive scatters and the direction of the electron recoil track at the middle scatter (Appendix~\ref{sec:appendix_minCKD}, Equation \ref{eg:electron_track_penalty}).

    \item \textbf{Minimum Scatter Distance}: The shortest distance between any two consecutive scatters in the sequence. 

    \item \textbf{Initial Scatter Distance}: The distance between the first and second scatters. This metric is crucial for angular resolution, as errors in this distance propagate to the inferred Compton scatter angle.

    \item \textbf{Calculated Incoming Compton Angle}: The Compton scattering angle at the first scatter, \( \theta_1 \), determined using the Compton formula:
    
    \begin{equation}
    \theta_1 = \arccos \left[ 1 - \frac{m_e c^2}{E_{\text{T}}} \left( \frac{1}{E_{\text{T}} - E_1} - \frac{1}{E_{\text{T}}} \right) \right]
    \end{equation} 
    where $E_{\text{T}}$ is the total energy obtained by summing the energies of all the hits in the event and $E_1$ is the energy of the first scatter. 

\end{enumerate}

For classification, we use a Random Forest Classifier trained on a subset of data annotated by source type (including line sources, backgrounds, and activation) and evaluated on a separate test set. The detector records events from both cosmic \(\gamma\)-rays and various background sources. Our goal is to accept only those events that (i) originate from cosmic \(\gamma\)-rays and (ii) are reconstructed correctly.

A \emph{valid} event thus meets all of the following criteria:
(1) The event is confirmed to be a cosmic \(\gamma\)-ray rather than a non-cosmic background source; 
(2) The first three scatters are correctly identified;
(3) The total measured and calculated energy are consistent within expected uncertainties;
(4) The sequence adheres to physically plausible Compton kinematics and electron track topology.

Any event failing any of these conditions is considered \emph{invalid}, which includes correctly reconstructed but non-cosmic events, as well as misreconstructed cosmic events.

Under this definition, the classifier achieves an overall accuracy of approximately 93\%, where accuracy is the fraction of all events correctly identified as valid or invalid. Furthermore, it attains a \emph{contamination rate} of 5\% (i.e., 5\% of events labeled as valid are actually invalid) and a \emph{misclassification rate} of 3\% (i.e., 3\% of genuinely valid events are mistakenly flagged as invalid). Table~\ref{tab:features} shows the relative importance of each classification feature used in the classifier.

\begin{table}[h]
\centering
    \caption{Relative Importance of Classification Features}
    \begin{tabular}{lc}
        \toprule
        \textbf{Feature} & \textbf{Importance} \\
        \hline
        Energy From Sum \(E_T\) & 0.23 \\
        Cumulative Energy Error & 0.15 \\
        Minimum Scatter Distance & 0.15 \\
        Calculated Energy Using Eq.~\ref{calc_ene} & 0.14 \\
        Klein-Nishina Probability & 0.14 \\
        Incoming Compton Angle & 0.13 \\
        Number of Scatters & 0.06 \\
        \hline
    \end{tabular}
    \label{tab:features}
\end{table}

\subsection{Pile-up and Space Charge}\label{sec:pileup_spacecharge}

\begin{figure}
    \centering
    \includegraphics[width=0.475\textwidth]{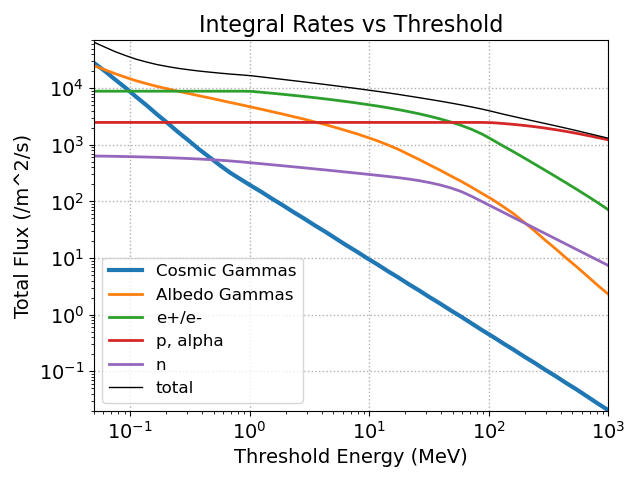}
    \caption{The average flux of particles and $\gamma$-rays in equatorial 550 km LEO from~\citep{Cumani_2019}, shown as the integral flux above a given threshold energy.}
    \label{fig:particle_flux}
\end{figure}

We now use this framework to study the effects of the charged particle flux in equatorial LEO, which is shown in Fig.~\ref{fig:particle_flux}. One of the core challenges to cosmic $\gamma$-ray detection is readily apparent: the cosmic $\gamma$-ray flux is subdominant to all other fluxes except at the lowest energies, creating a large set of backgrounds, and potentially causing event pile-up.

Pile-up manifests itself in two primary ways. First, Compton imaging becomes unreliable when two or more events produce hits in the same cell within the 170\,$\mu$s maximum charge drift time, because assigning the correct scintillation signal and hence start time and depth to individual scatters becomes ambiguous. The probability of this happening depends on event rate, and on the number of cells with hits from an individual event. The latter is a rapidly increasing function of energy, with high energy particle showers occupying a significant fraction of the detector. The depth estimate from GAMPix's diffusion projection (see Sec.~\ref{sec:gampix}) resolves much of those confusions when the level of pile-up is modest. Combining these effects, we find pileup over all energies of $\sim$\,5\%, which is reduced to $\sim$\,1\% using the diffusion-projection depth discussed in Sec.~\ref{sec:gampix} for the average fluxes in Fig.~\ref{fig:particle_flux}.

Secondly, event pile-up causes Compton reconstruction and hence pointing to fail. In this case, the energies and times of the piled-up events is still provided by the scintillation signals alone, though with a modest loss of energy resolution (see Sec.~\ref{sec:charge_light}). This means that the temporal flux and energy distribution will be measured up to rates on the scale of 10$^7$\,Hz for bright $\gamma$-ray bursts (assuming $\sim$\,100\,ns time resolution, see Sec.~\ref{sec:light_readout}). Pointing we be measured when the flux falls below $\sim$\,100\,times the average total particle flux. Thus, despite the TPC's relatively slow readout, very bright transients will be well measured.

Another potential concern arising from the particle flux is the buildup of space charge, given the very slow drift velocity of ions at 8\,mm/$\mu$s. This effect has been observed in the large MicroBooNE LAr TPC operated on the Earth's surface~\citep{mooney_2015}, with a combined cosmic ray and neutrino beam particle flux that is lower than in low Earth orbit (LEO). However, the effect of space charge is strongly dependent on the TPC cell size, which is much smaller in GammaTPC. Our simulations find that the field generated by space charge is less than 10$^{-3}$ of the applied 500\,V/cm drift field. This ratio is comparable to the maximum possible geometric information from $\sigma_{xyz}\sim$\,100\,$\mu$m measured over a maximum $\sim$\,20\,cm cell dimension, and we expect any errors to be further reducible by in-situ calibrations~\citep[see, e.g.,][]{mooney_2015}. Thus, we anticipate the impact on event reconstruction to be minimal.

\subsection{Backgrounds}\label{sec:backgrounds}

Backgrounds are the other challenge posed by the various fluxes shown in Fig.~\ref{fig:particle_flux}, all of which exceed the cosmic $\gamma$-ray flux above 100\,keV. Charged particles are not an important background in the Compton regime because they will be tagged with high efficiency by the ACD. Also, they are mostly outside that energy window, and have topologies that are distinct from $\gamma$-rays. A much more significant background comes from radioactive activation of the detector materials by the hadronic particles (p, n, $\alpha$), which subsequently decay and in general emit $\gamma$-rays in the 0-few\,MeV energy range. Neutrons with energies below $\sim$\,10\,MeV will mostly elastically scatter on nuclei with interaction lengths comparable to those of Compton $\gamma$-rays. These events have similar geometries to $\gamma$-rays, though with lower deposited energy due to the mass mismatch between neutrons and Ar nuclei, and should be powerfully rejected by the very different light pulse shapes of nuclear recoils from neutrons in LAr~\citep{Ajaj:2019imk} (a fact that is central to the use of LAr for dark matter searches). Inelastic scatters of higher energy neutrons can also induce prompt $\gamma$-rays. Upward going albedo $\gamma$-rays from the atmosphere exceed the cosmic $\gamma$-ray rate by 1-2 orders of magnitude. These are actively vetoed by the calorimeter, and any untagged energy that manages to punch through the calorimeter is mostly reconstructed as coming from below. 

Any cosmic $\gamma$-ray that is incorrectly reconstructed becomes a background, as do cosmic $\gamma$-rays which scatter in the inert vessel or MMOD material, altering their trajectory. Finally, any otherwise good $\gamma$-ray serves as a background for other $\gamma$-rays due to the relatively poor pointing of the Compton telescope technique, especially at low energy where the electron recoil direction measurement is absent or poor. That background, which is ultimately dominant, is intrinsic to a Compton telescope and can only be reduced by improving the instrument's angular resolution.

\begin{figure}
    \centering
    \includegraphics[width=0.475\textwidth]{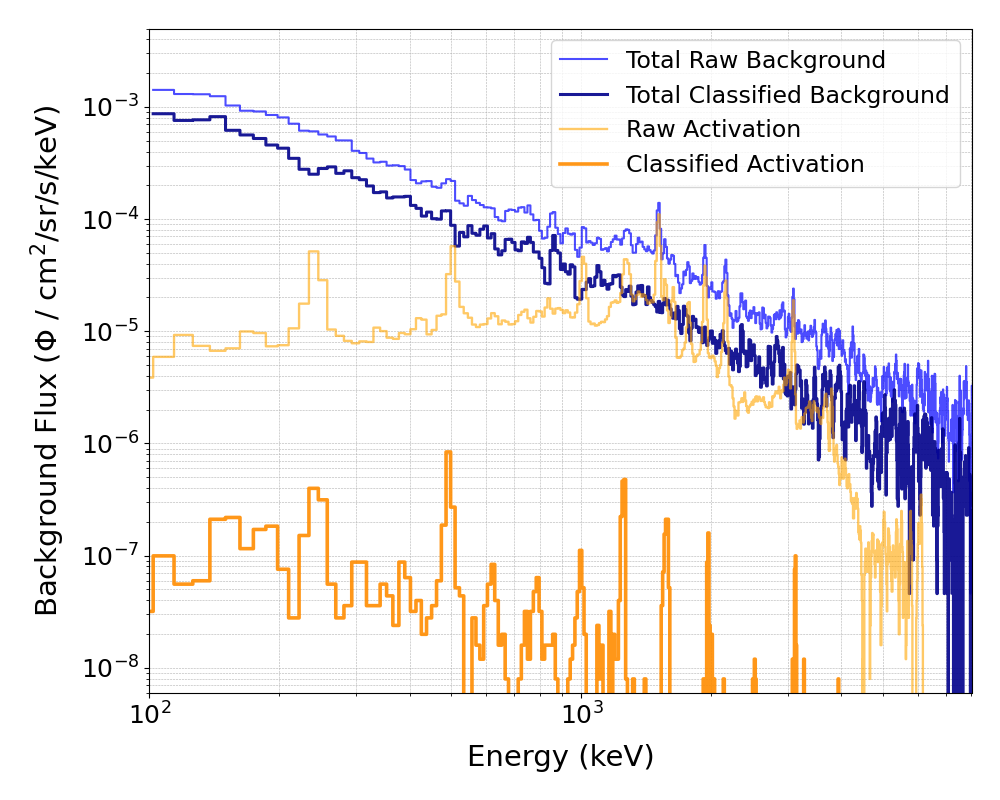}
    \caption{Background flux as a function of energy for cosmic rays, albedo photons, and activation decays, both before (raw) and after (classified) application of event classification. The final total background (dark blue) is significantly below the raw total background (light blue). Backgrounds from activation alone are separately shown (orange), where the suppression power of classification is evident. The final backgrounds is mostly featureless.}
    \label{fig:backgrounds}
\end{figure}

We have completed a Monte Carlo of the resulting backgrounds for a point source for up to 10\,MeV, with the results shown in Fig.~\ref{fig:backgrounds}. This background is the set of events reconstructed (Sec.~\ref{sec:reconstruction}) as $\gamma$-rays, both before and after the application of event classification (Sec.~\ref{sec:classification}). Note that while activation $\gamma$-rays are a significant (though not fully dominant) in the raw spectra, they are rendered fully sub-dominant after the event classifier described below is applied. This leaves a mostly featureless background, which means that this instrument will be powerful for measuring nuclear lines and 511 $\gamma$-rays.

Activation backgrounds merit further discussion since they can be the dominant background in a Compton telescope. Activation creates radioactive daughters that subsequently beta decay (primarily $\beta^-$, but some $\beta^+$), usually accompanied by one or more $\gamma$-rays. The total energy of the decay, or Q value, depends on the nucleus but is typically in the few to $\sim\,5$ MeV or higher range, and is shared between the invisible neutrino, the $\beta$, and the $\gamma$-rays. The amount of activation varies by material, but in general higher $Z$ materials have higher activation. Here, activation in Ar and CsI dominates, since these are the largest masses in the instrument. However, activation events from CsI are largely self vetoed, and for those that are not, their non-Cosmic origin is recognized by reconstruction. While our simplified geometry does not yet accurately include all the materials in TPC structures, most notably fiberglass, we don't expect a more accurate treatment to substantially alter these results given the small masses and generally low $Z$ of those components.

For the LAr target, there is the additional question of where the radioactive isotopes are located, as they are mostly ionized with positive charge following the initial decay, and live long enough to be neutralized or even negatively charged via electron capture. If charged, they will drift to one of the readout planes, and the circulation of LAr can also transport these daughters. Here we have assumed the daughters remain at their creation location. In this case, the $\sim$\,MeV energy deposited by the beta particle will be measured for decays in LAr or CsI. This extra energy causes the events to be rejected as external-origin $\gamma$-rays by the classifier. The many isotopes with multiple $\gamma$-rays per decay are even more powerfully kinematically distinguished from a single $\gamma$-ray. Together, these effects explain the powerful reduction of activation backgrounds in Fig.~\ref{fig:backgrounds}. If the daughters are instead transported to either the anode or cathode, the location of that decay provides an additional powerful tag, offset by the loss of beta energy deposited in the readout plane. A rough estimate of these competing effects suggests rejection on the scale of that in Fig.~\ref{fig:backgrounds}. 

The full set of possible spallation interactions that generate radioactive daughters are often not measured, and the treatment in code relies on semi-empirical predictions. Future work will include cross-checking these results against ACTIVIA~\citep{Back_2008}, a purpose-built activation code used widely by low background experiments. Careful attention will also be given to the related process of prompt $\gamma$-rays being created by inelastic scattering of the substantial $>$\,MeV neutron flux. Fortunately, the rates for this process on Ar for 1-30 MeV neutrons have recently been measured~\citep{MacMullin2014}. 

In the higher energy pair regime, the cosmic $\gamma$-ray rate is increasingly sub-dominant, but the ACD and the different topologies of particle-generated events and gammas will powerfully tag these backgrounds. Rare processes, such as those involving imperfections in the ACDs, or bremsstrahlung $\gamma$-rays from glancing particle interactions in the vessel, will be comparatively more important, and the study of these effects awaits a future publication.

\subsection{Results}\label{sec:results}

\subsubsection{Energy Resolution}\label{sec:energy_resolution}

\begin{figure}[h!t]
    \centering
    \includegraphics[width=0.475\textwidth]{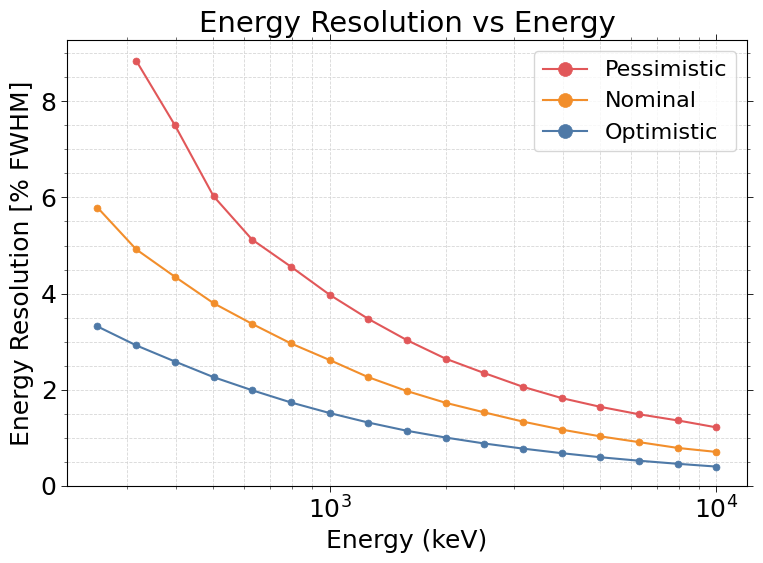}
    \caption{Energy resolution as a function of $\gamma$-ray energy for monoenergetic point sources with normal incidence. Results for off-axis incidence angles are consistent with those shown.}
    \label{fig:energy_res}
\end{figure}

The detector's energy resolution depends on the combined performance of the light and charge readout, as described in Sec.~\ref{sec:charge_light}, \ref{sec:gampix} and~\ref{sec:light_readout}. In Fig~\ref{fig:energy_res} we show the simulated resolution for the three sets of readout performance parameters listed in Table \ref{tab:simulations_parameters}.
 
Energy resolution plays a critical role in the study of nuclear $\gamma$-ray lines, as it determines the precision with which individual lines can be identified and their properties analyzed. For astrophysical applications, such as supernova (SN) remnants, the ability to resolve lines with high precision enables the investigation of isotope production and decay processes. Percent level resolution is sufficient to identify prominent nuclear lines, and as seen if Fig.~\ref{fig:backgrounds} we expect few backgrounds lines from activation. This technology does not, however, approach the 0.1\%-level resolution possible with high purity Ge of COSI (or CZT) detectors and thus have access to the science enabled by that~\citep{tomsick2019}.

\subsubsection{Angular Resolution}\label{sec:angular_res}

The angular resolution of a Compton telescope is commonly described by the Angular Resolution Measure (ARM), traditionally defined as the Full Width at Half Maximum (FWHM) of the angular deviation distribution between the reconstructed and true photon directions. However, as demonstrated by \citet{Tanimori2015}, incorporating the electron track direction measured provides an additional kinematic constraint and can significantly improve the angular resolution - even when measurements of the electron recoil (quantified by the Scatter Plane Deviation, SPD) are suboptimal.

In this work, we introduce a modified ARM definition that explicitly accounts for the electron recoil SPD. A detailed derivation of this modified ARM, along with its consistency with image-based reconstruction methods, is presented in Appendix~\ref{sec:appendix_ARM}. While the standard ARM formulation remains valid at lower $\gamma$-ray energies, the impact of incorporating the electron recoil becomes increasingly significant above $\sim$2~MeV.

\begin{figure}[h!t]
    \centering
    \includegraphics[width=0.475\textwidth]{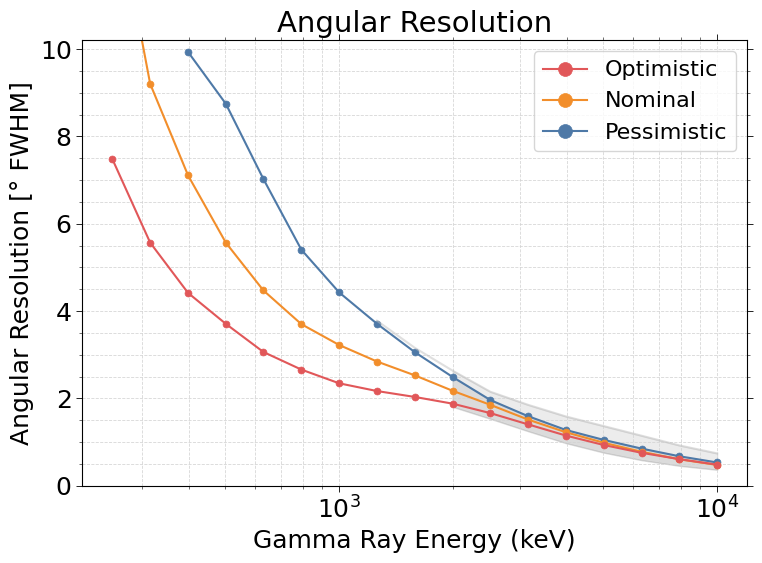}
    \caption{Angular resolution as a function of incoming $\gamma$-ray energy. The shaded region indicates the range of potential systematic error in treatment of the electron recoil direction measurement as discussed in the text.}
    \label{fig:angular_res}
\end{figure}

As the energy of the incoming $\gamma$-rays increases, the uncertainties in the Compton scattering angle, $\theta_C$, decrease, resulting in a more precise determination of the source direction and consequently an enhanced angular resolution. This improvement stems from two key factors: first, the relative energy resolution, $\Delta E / E$, improves approximately as $1 / \sqrt{E}$, as discussed in Sec.~\ref{sec:charge_light}; second, higher-energy $\gamma$-rays are more forward-scattered, reducing the angular spread. When the energy of the initial electron scatter exceeds $\sim$300~keV, the inclusion of electron recoil direction readout leads to a significant further improvement in the ARM. Note that the energy shown on the x-axis of Figure \ref{fig:angular_res} represents the total energy of the incoming $\gamma$-ray, which is typically a factor of 2 or more greater than the energy of the first scatter.

The parameterization of the electron recoil track direction as a function of energy and drift distance used in this study is based on the machine learning (ML) analysis described in Sec.~\ref{sec:gampix}. These parameterizations rely on simulations limited to track energies of up to 1\,MeV. As computationally intensive simulations for higher-energy tracks have not yet been performed, we extrapolate the results to higher energies as follows. In Appendix~\ref{sec:appendix_LAr}, we find a scaling of $\Delta \theta \propto (\beta p)^{-2/3}$ with a simple analytical argument, while \citet{Bernard2013_tpc} argues for a similar scaling of $\Delta \theta \propto p^{-3/4}$. Here $p$ is the electron momentum, $\beta$ is the relativistic velocity, $p \propto E$ in the relativistic regime, and $\beta p \propto E$ in the non-relativistic regime. $X_0$ is the radiation length of the material. These are close to $\emph{Fermi}$'s result of $E^{-0.78}$ at higher energy.

We thus adopt $E^{-3/4}$ scaling as a baseline, and show this with the solid lines in the graph, both here and in the sensitivity results below. The shaded region represents possible uncertainty in this extrapolation, ranging from $E^{-1}$ to $E^{-1/2}$. Moreover, as shown in Appendix~\ref{sec:appendix_ARM} and Fig.~\ref{fig:arm_comparison}, even if the electron track readout accuracy does not improve above 1\,MeV, the orthogonal geometric constraints provided by electron tracking still lead to significant improvements in angular resolution and sensitivity. 

\subsubsection{Effective Area}\label{sec:effective_area}

The effective area of the instrument is the product of the geometric area and the reconstruction efficiency, defined as the ratio of correctly reconstructed to total simulated $\gamma$-rays. For the $7~\text{m}^2$ design (illustrated in Fig.~\ref{fig:sims_geometry}) and a reconstruction efficiency of approximately $25\%$ below a few MeV, the resulting effective area for continuum sensitivity is estimated to be $18{,}000, \text{cm}^2$.

The current reconstruction framework is optimized for $\gamma$-rays near $1 \, \text{MeV}$. Efficiency declines at higher energies due to increasing event complexity. The number of Compton scatters increases with energy, and especially above $\sim$\,$2 \, \text{MeV}$ leads to a significant number of events with more than 10 scatters, for which reconstruction is difficult. Additionally, higher-energy $\gamma$-rays are more likely to penetrate deeper into the detector, with the initial interactions occurring in the calorimeter rather than the tracker. These events are currently excluded from reconstruction due to the current framework limitations. Finally, bremsstrahlung photons and pair production become increasingly significant at high energies, and while properly simulated, neither is yet properly handled by our reconstruction algorithms. 

\begin{figure}[h!t] 
\centering 
\includegraphics[width=0.485\textwidth]{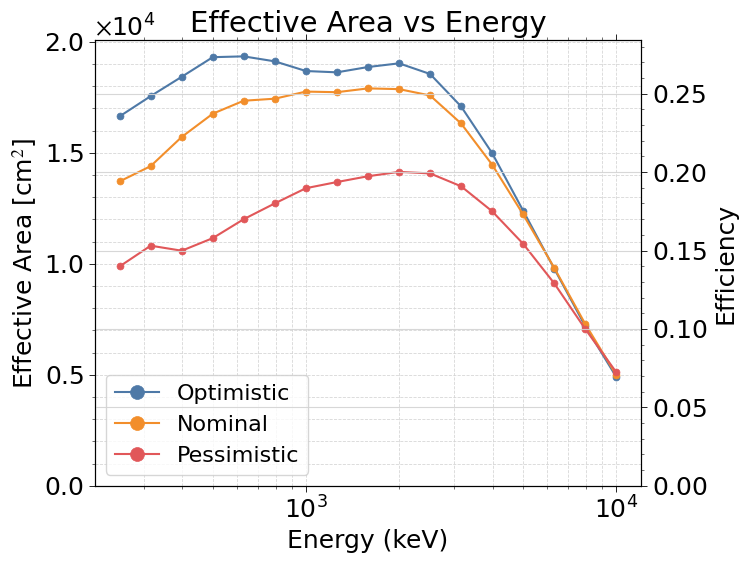} 
\caption{Effective area as a function of $\gamma$-ray energy. The result for the nominal scenario around 1\,MeV is roughly $25\%$ of the tracker's area. At higher energies, a decline in efficiency is observed due to both physical effects and the current reconstruction framework's limitations. } 
\label{fig:efficiency_acceptance} 
\end{figure}

\subsubsection{Sensitivity}\label{sec:sensitivity}

The sensitivity of an instrument refers to its ability to detect faint sources of $\gamma$-rays. This property is typically described using two main metrics: \textit{continuum} and \textit{line sensitivity}. 

\textit{Continuum sensitivity} quantifies the instrument's ability to detect a broad, continuous spectrum of $\gamma$-rays over a range of energies. It assumes a background-limited observation and is expressed as:

\begin{equation}\label{eq:continuum_sensitivity}
S_{\text{cont}}(E) = 3 \, \sqrt{\frac{\Phi_B \Delta \Omega}{A_{\text{eff}} T_{\text{eff}} \Delta E}},
\end{equation}
where $\Phi_B$ is the background flux (illustrated in Fig.~\ref{fig:backgrounds}), $\Delta \Omega$ is the solid angle corresponding to the angular resolution (depicted in Fig.~\ref{fig:angular_res}), $A_{\text{eff}}$ is the effective area (described in Fig.~\ref{fig:efficiency_acceptance}), $\Delta E$ is the energy range under consideration (often taken as $0.5E$, where $E$ is the central energy), and $T_{\text{eff}}$ is the effective observation time ~\citep{Aramaki_2020}.

 \textit{Line sensitivity} characterizes the instrument's capability to detect narrow spectral lines and is given by:

\begin{equation}\label{eq:line_sensitivity}
S_{\text{line}}(E) = 3 \, \sqrt{\frac{\Phi_B \Delta \Omega \Delta E^{\prime}}{A^{\prime}_{\text{eff}} T_{\text{eff}}}},
\end{equation}
where $A^{\prime}_{\text{eff}}$ is a modified effective area that accounts for the additional requirement of correctly reconstructing the $\gamma$-rays' energy within the full width at half maximum (FWHM) range. That energy resolution $\Delta E^{\prime}$ is shown in Fig.~\ref{fig:energy_res} ~\citep{Aramaki_2020}.

The effective observation time, $T_{\text{eff}}$, considers the field of view, estimated to cover $40\%$ of the sky, and an observation duration of 3 years. While the planar geometry used in these simulations does not fully account for uniform response across all source locations, the impact is limited. Apart from regions near the detector's outer edge, where the approximate planar geometry performs worse than the true hemispherical geometry, the performance metrics show only a weak dependence on the incidence angles of $\gamma$-rays. We attribute this to the TPC's uniform 3D imaging. The shaded regions again represent the uncertainty in extrapolating electron recoil direction accuracy described in Sec.~\ref{sec:angular_res}.

\begin{figure*}[htp!]
  \centering
  \begin{minipage}[b]{0.5\textwidth}
    \centering
    \includegraphics[width=0.95\textwidth]{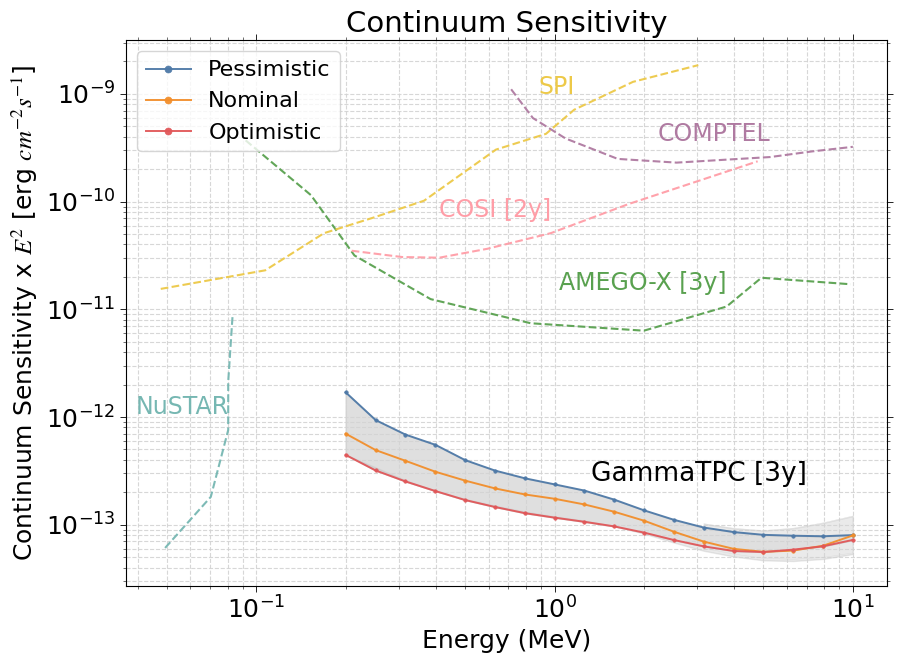}
    \label{fig:continuum_sens}
  \end{minipage}%
  \begin{minipage}[b]{0.49\textwidth}
    \centering
    \includegraphics[width=0.95\textwidth]{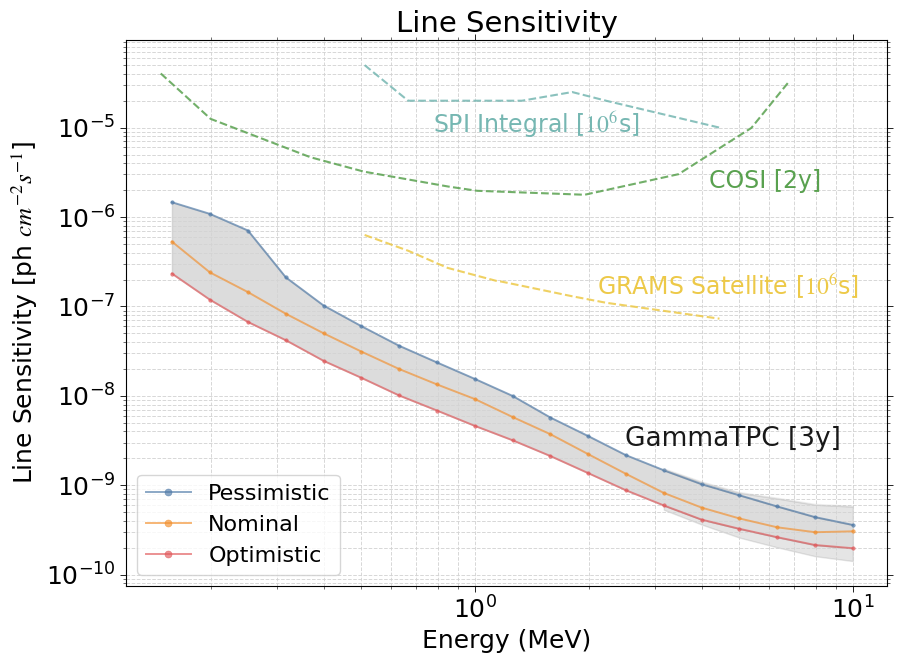}
    \label{fig:sensitivity}
  \end{minipage}
  \caption{Continuum (left) and line sensitivity (right) in the Compton regime, along with sensitivities of other instruments, demonstrating the transformative potential of the GammaTPC instrument concept.}

\end{figure*}\label{fig:cont_and_lin_sens}

These remarkable results in the largely unexplored Compton regime highlight the transformative potential of this technology. For comparison, at 1\,MeV, our continuum sensitivity is approximately 40 times better than AMEGO-X, implying a net improvement of up to 1600-fold when factoring in $A_{\text{eff}}$, $T_{\text{eff}}$, $\Phi_B$, and $\Delta \Omega$ (as defined above). Strong suppression of activation at 1\,MeV—reducing the total background by half—and the implementation of electron tracking (which limits background contributions from a full ring to an arc) together yield an approximately fivefold improvement in background rejection. Additionally, enhanced angular resolution further reduces the background flux, while the explicit angular term in the sensitivity equations provides an additional factor of $\sim$2. 

Our hemispherical design, covering nearly half the sky compared to AMEGO-X's $\sim$20\% sky coverage, effectively doubles the observation time. Furthermore, the instrument’s large active area — eight times larger than AMEGO-X and 400 times larger than COSI in this study — combined with a thick tracker volume, ensures a high fraction of fully contained events that are efficiently reconstructed. The uniform tracker geometry further enhances reconstruction efficiency, while the ability to measure the direction of electron tracks significantly sharpens the point-spread function, especially above $\sim$2\,MeV (Appendix~\ref{sec:appendix_ARM}). Background contamination from mis-reconstructed $\gamma$-rays, albedo photons, and activation is also minimized through a combination of efficient reconstruction techniques, calorimeter measurements, and advanced classification and tagging.

Since these sensitivities are background-limited, they scale with the square root of the factors driving detector performance (Eqs.~\ref{eq:continuum_sensitivity} and \ref{eq:line_sensitivity}). However, short-duration transients will not be background-limited, meaning that the $\sim10^2$-fold improvement in long-exposure sensitivity over AMEGO-X or COSI should translate into an impressive $\sim10^4$-fold gain in sensitivity for transient events.

\subsubsection{Future Studies}\label{sec:future_studies}

Work remains to fully characterize the instrument’s capabilities, with both improvements to the simulations and analysis framework, and further sensitivity studies. The main improvements to the simulations are implementing segmented readout of the $\textsc{CsI}$ calorimeter, and working with of the full hemispherical $3D$ geometry needed to correctly estimate sky coverage. The current reconstruction framework is an initial prototype which provides a conservative performance baseline, but has ample room for improvement.

Here one focus is implementing systematic optimization methods, similar to the {\tt SensitivityOptimizer} tool in {\tt MEGAlib}~\citep{MEGALIB}, which iterates over analysis parameters to identify optimal configurations. The current approach is designed to maximize the effective area, however, that is not necessarily the optimal approach for maximizing sensitivity. For instance, limiting acceptance of photons with large scatter distances between their first two interactions can boost angular resolution and sensitivity, but decrease the effective area. It's worth noting that such events are more likely in a curved, hemispherical geometry (see Fig.~\ref{fig:gammatpc_schematic}), where photons may traverse vacuum between interactions. Hence, we expect that implementing the full 3D geometry in the simulations will further improve the performance metrics. Similarly, selecting events with significant energy deposition near the pixel readout plane can leverage superior electron-tracking capabilities. 

At higher energies, the challenge of using CKD-based methods becomes increasingly complex to compute ($O(N!)$) due to the higher multiplicity of Compton scatters, as well as the additional presence of bremsstrahlung and pair-production interactions. Initial studies leveraging electron recoil directions to deduce the sequence of scatters, with a computational complexity of $O(N^2)$, have shown promising results but require further refinement. Additionally, with sufficiently precise measurement of the angle between the electron recoil direction and the incident $\gamma$-ray trajectory, both energy resolution and source localization can be improved.

The most important future work is to study the sensitivity of the instrument in the pair regime. The high spatial resolution of the tracker should provide powerful imaging of e$^\pm$ tracks, and we can expect a large effective area. We note that \citet{Bernard2013_tpc}, in a comprehensive study of liquid and gas TPCs using a simplified readout model, found the angular resolutions of LAr detectors to be 3–5 times better than \emph{Fermi}. Finally, we need to study the sensitivity to polarization.

\section{Conclusion}\label{sec:conclusion}

We have introduced a powerful new 0.1-1000 MeV $\gamma$-ray all sky imager based on LAr TPC technology. The sensitivity in the Compton regime shown in Fig.~\ref{fig:sensitivity} for a 10\,m$^2$-class instrument is transformative, and will if anything be comparatively more impressive for short duration transients. While not yet studied, we can expect good polarization sensitivity in the Compton regime, good sensitivity in the pair regime. Such an instrument will have enormous discovery potential, and will enable a broad and powerful science program.

This leap in science capability rests upon the widespread development and adaption of liquid noble TPC technology, which has a core advantage of economically scaling to the large mass necessary for measuring MeV$--$GeV $\gamma$-rays with high efficiency. To that technology basis, we add the novel GAMPix charge readout, which provides unprecedented spatial resolution over a large volume at very low power. While any new $\gamma$-ray technology will likely have other applications, including medical imaging, it is notable that our development comes at a time when the cost of launching large mass into space is dropping in a dramatic fashion. This new instrument is uniquely suited to exploit the opportunity this provides for a scientific purpose.

We would like to thank Seth Digel, Nicola Omodei, Carolyn Kierans and Andreas Zoglauer for useful discussions. This work was supported by NASA grant NNH21ZDA001N-APRA, the Kavli Institute for Particle Astrophysics and Cosmology, and SLAC.

\newpage
\appendix
\section{Detector material choice}\label{sec:appendix_LAr}
\begin{figure*}
    \centering
    \includegraphics[width=1\textwidth]{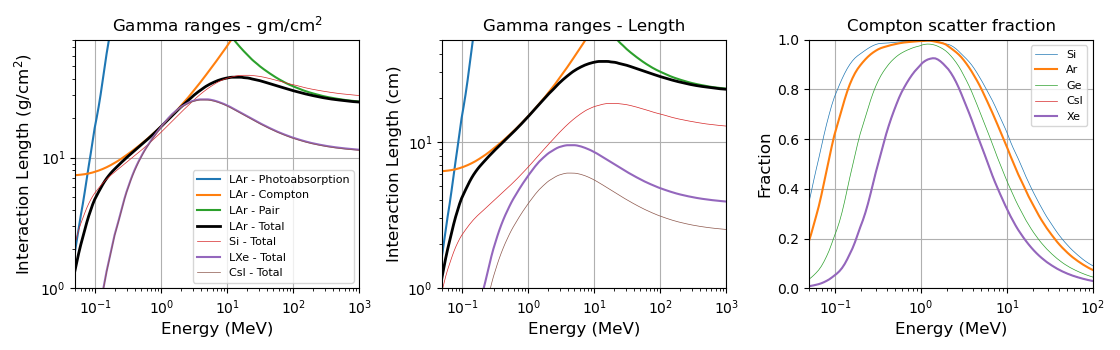}
    \caption{(Left) $\gamma$-ray interaction lengths expressed in g/cm$^2$ for various processes in LAr and for all processes in LXe and other detector materials. (Center) The corresponding physical interaction lengths. (Right) The fraction of interactions occurring via Compton scattering as a function of energy, with photoabsorption dominating at low energies and pair production at high energies. Note that the curves for CsI and LXe are nearly indistinguishable in the left and right panels.}
    \label{fig:gamma_ranges}
\end{figure*}

In this section, we explain our choice to use LAr rather than LXe for our Compton telescope. We also compare LAr with other detector materials commonly employed in $\gamma$-ray detection. This discussion parallels the analysis by \citet{Bernard2013_tpc}, who evaluated the performance of liquid and gaseous TPCs in the pair-production regime. We start by considering various material properties and parameters. The first parameter to consider is the scattering length, which we characterize by the mean free path $\lambda_{\gamma}$ (see Fig.~\ref{fig:gamma_ranges}). In the figure, $\lambda_{\gamma}$ is presented both as a physical distance and as a mass attenuation coefficient (in units of g/cm$^2$). When expressed in terms of mass attenuation, the scaling with atomic number $Z$ and atomic mass $A$ depends on the interaction process. Specifically, for photoabsorption, the scaling is approximately $\sim A/Z^{4.5}$; for Compton scattering, it scales as $\sim A/Z$; and for pair production, it follows $\sim A/\bigl[Z(Z+1)\bigr]$ \citep{Hubbell:1969xbg, pdg2022}.

Following from the discussion above, we can see that $\gamma$-rays are most penetrating in the Compton regime, where the mass attenuation $\lambda_{\gamma}$ scales approximately as $A/Z$ (roughly $\sim1/2$) and shows only a weak dependence on $Z$. This means that, for a fixed detector mass (and with thickness chosen to avoid excessive self-shielding), the stopping efficiency is largely independent of the specific material. 

However, the choice of material strongly influences the energy range over which Compton scattering dominates. At low energies, photoabsorption becomes the dominant process; while this leads to more compact events, it provides no pointing information. At high energies, pair production does enable source localization, but in the transition region between Compton scattering and pair production the angular resolution for pair events is at its worst (even though it is best for Compton events). As illustrated in Fig.~\ref{fig:gamma_ranges}, the energy span over which the Compton fraction exceeds 0.5 is roughly two orders of magnitude in $\log(E)$ for argon (a factor of $\sim 100$), compared to only about one order of magnitude (a factor of $\sim 10$) for xenon. This is a key argument in favor of using a low-$Z$ material—specifically, LAr over LXe—for the tracker.

Low-$Z$ and low-density materials offer significant advantages for spatial imaging of events. In these materials, the individual Compton scatters tend to be more spread out, which makes their geometric configuration easier to measure. Moreover, electron recoil tracks are generally longer and straighter. This not only improves the pointing accuracy in the pair-production regime but also enhances the precision in determining the interaction locations and the initial directions of the recoil electrons in the Compton regime, as can be seen in Fig~\ref{fig:position_resolution}.

On the downside, low-$Z$, low-density materials exhibit a considerable increase in the number of low-energy scatters because event-ending photoabsorption is strongly suppressed until much lower energies. This suppression leads to a modest loss of containment for a given detector mass (particularly for events that exit through the front, where additional mass cannot compensate) and introduces extra low-energy scatters that can complicate the reconstruction process.

To quantify these arguments, we begin with the simplifying assumption (ansatz) that the instrumental spatial resolution, $\sigma_{xyz}$—which is determined by the sensor pitch—is independent of the detector material. Although this is not strictly true in every case, the sensors in GammaTPC, AMEGO-X, and ASTROGAM all have the same pitch and this ansatz enables straightforward comparisons; adjustments can be made later if different $\sigma_{xyz}$ values need to be considered.

In Table~\ref{tab:materials} we list the physical interaction length $\lambda_{\gamma}$ at 1\,MeV for a variety of materials along with other key properties. One important property is the size of the recoiling electron tracks, which are easier to image when they are larger. Because these tracks are not perfectly straight—and tend to be particularly curled at low energies—a useful measure of their size is provided by the continuous slowing down approximation (CSDA) length. Denoted as $l_e$, the CSDA length is calculated by $l_e = \int dE/(dE/dx)$ which represents the “unrolled” length of the track and is typically larger than the actual spatial extent by a factor that depends on how much the track curls.\footnote{More sophisticated measures can be obtained from tracks simulated in detail with PENELOPE, such as the largest length from a principal component analysis. However, we find that those results scale with energy and material in a similar way to CSDA.} 

We also include a measure of diffusion, which is critical for TPCs and should be minimized. For comparison, we evaluate the rms spread due to diffusion, $\sigma_D(\lambda_{\gamma})$, since the detector thickness is chosen to scale with the physical length $\lambda_{\gamma}$ for a given Compton efficiency.

Next, we define the following unitless benchmark ratios that ideally should be minimized:
\begin{itemize}
    \item $\sigma_{xyz}/\lambda_{\gamma}$, which governs the precision in measuring the various vectors used for kinematic reconstruction in the Compton regime.
    \item $\sigma_{xyz}/l_e$, which quantifies the effect of the instrumental spatial resolution on imaging electron tracks.
    \item $\sigma_D(\lambda_{\gamma})/l_e$, which measures the impact of diffusion on electron track imaging.
\end{itemize}

To compare different materials, we compute each ratio for a given material and then form a “ratio of ratios” by dividing by the corresponding value for LAr (with lower values indicating better performance). As shown in Table~\ref{tab:materials}, LAr exhibits superior performance compared to the other materials considered.

A final geometric measure is the straightness of the electron recoil tracks in the face of ``multiple scattering'', which is important for electron recoil directions in the Compton regime (see Sec.~\ref{sec:angular_res}, and appendix~\ref{sec:appendix_ARM}), and directly determines angular resolution in the pair regime. Multiple scattering is described by Moliére theory~\citep{pdg2022}, which tells us that over a distance $x$, the track path is altered by an angle $\theta_o \sim p_o/(\beta c p) \sqrt{x/X_o}$, where $p_o = 13.6\,\mathrm{MeV}/c$, $p$ is the electron momentum, $\beta$ is its relativistic velocity, and $X_o$ is the material’s radiation length (a parameter also important for bremsstrahlung and pair production and listed in Table~\ref{tab:materials}). A key issue is choosing the appropriate distance $x$. Here we proceed with the follow heuristic argument. With a spatial resolution of $\sigma_{xyz}$\footnote{This $\sigma_{xyz}$ is not in general the same as that defined above, rather it is a measure of how well the lateral location of the ends of a segment of track can be measured. It arises from the centroid measurement of the track lateral location from a set of pixel samples at a given point along the length of the track. However, following the ansatz about position resolution used in his appendix, we presume this $\sigma_{xyz}$ is linearly proportional to the value used elsewhere in this section and is material independent.}, the error on the measurement of the track direction over a distance $x$ is approximately 
$\Delta \theta_m = \frac{\sigma_{xyz}}{x}$. We then choose $x$ so that $\Delta \theta_m=\theta_o$, that is, we sample a distance along the track over which the measurement error is equal to the track deviation. Eliminating x, we have $\Delta \theta \sim (p_o/\beta c p)^{2/3} (\sigma_{xyz}/X_o)^{1/3}$. A similar treatment by \citet{Bernard2013_tpc}, following \citet{Innes1993}, finds a modestly different dependence on $X_o$ and $p$: $\Delta \theta \propto (1/p)^{3/4}(1/X_o)^{3/8}$, while \emph{Fermi} found a scaling of $E^{-0.78}$ at high energies.

The ratio of $\Delta \theta$ for each material, which should be minimized, to that for LAr is again shown in Table\,\ref{tab:materials}, and once again favors LAr. We include tungsten ($W$) in the table because it was the main conversion material in \emph{Fermi}, and is clearly poor in this respect. In the pair regime, this is an incomplete comparison with detectors like \emph{Fermi} or AMEGO-X that have gaps between readout/absorber layers. These gaps provide a roughly 10:1 lever arm to aid in the geometric measurement, but also boost the effect of multiple scattering, as the local value of $\theta_o$ is greater than the value of $\theta_o$ averaged over a distance of travel. We do not attempt a more rigorous analysis of these effects.

\begin{table*}
\centering
    \caption{Material Properties and Performance Benchmarks}
    \begin{tabular}{lccccccccccc}
        \toprule
        & & $A$ & $\rho$ & $\lambda_{\gamma}$ & $l_e$ & $\chi_o$ & $\sigma_D(\lambda_{\gamma})$ & \multicolumn{4}{c}{Benchmark values} \\
        Material & $Z$ & (g/mole) & (g/cm$^3$) & (cm) & (mm) & (cm) & ($\mu$m) & $\frac{\sigma_{xyz}}{\lambda_{\gamma}}$ & $\frac{\sigma_{xyz}}{l_e}$ & $\frac{\sigma_D(\lambda_{\gamma})}{l_e}$ & $\Delta\theta$ \\        
        \hline
        Ar  & 18  & 40.0 & 1.16 & 14.9 & 5.1 & 16.8 & 500 & 1.0 & 1.0 & 1.0 & 1.0 \\
        Xe  & 54  & 131  & 2.94 & 5.9  & 2.4 & 2.9  & 470 & 2.5 & 2.0 & 1.9 & 1.8 \\
        Si  & 14  & 28.1 & 2.33 & 6.7  & 2.3 & 9.4  & ... & 2.2 & 2.2 & ... & 1.2 \\
        Ge  & 32  & 72.6 & 5.35 & 3.3  & 1.2 & 2.3  & ... & 4.5 & 4.1 & ... & 1.9 \\
        CsI & ... & ... & 4.51 & 3.8  & 1.6 & 1.9  & ... & 3.9 & 6.9 & ... & 2.1 \\
        W   & 74 & 184  & 19.3 & ...  & ... & 0.4  & ... & ... & ... & ... & 3.6 \\
        \hline
    \end{tabular}
\vspace{2em} \\ 
\textbf{Notes:} The performance benchmark values $\sigma_{xyz}/\lambda_{\gamma}$, $\sigma_{xyz}/l_e$, $\sigma_D(\lambda_{\gamma})/l_e$ and $\Delta\theta$ are relative to LAr, with smaller values being better. Assumed are LAr/LXe=120/165\,K, $l_e$=1\,MeV, and $\sigma_D(\lambda_{\gamma})$ evaluated at $E_\gamma$=1\,MeV and $\boldsymbol{E}$=0.5\,kV/cm. We assume, crudely, that $\sigma_{xyz}$ is the same for all materials, which should be kept in mind when interpreting this table.
\label{tab:materials}
\end{table*}

In the Compton regime, energy resolution is also important for the kinematic reconstruction. As this is a complicated topic, depending on the quanta of signal generated, the Fano factor, signal transport and readout noise, we settle for a brief discussion. LAr and LXe can be compared, ignoring Fano factors and assuming equivalent light and charge collection for each, by noting that energy resolution scales as $\delta\,E/E,\propto\,\sqrt{N}\,\propto\,\sqrt{1/W}$ where $N$ is quanta of light and charge generated. By this metric, LAr with $W$ measured either as \,19.5\,eV for scintillation or 23.6\,eV for charge is only modestly worse than LXe with $W$ measured as 13.7\,eV\citep{Dahl:2009nta} or 11.5\,eV~\citep{Anton2020}.\footnote{We have a combined $W$ value (see Sec.~\ref{sec:charge_light}) for the better-studied LXe system, but only separate measurements of $W$ for scintillation and charge for LAr.} Ge and Si detectors have smaller $W$ values of 2.9\,eV and 3.6\,eV respectively, and Ge detectors have much better energy resolution than LAr (or LXe), and is a major advantage for COSI. Other factors limit the resolution Si detectors, and the projected energy resolution for Amego-X and ASTROGAM is similar to our predictions for LAr (Sec.~\ref{sec:energy_resolution}).

Related to energy resolution is Doppler broadening~\citep{Zoglauer2003}, whereby the velocities of the bound atomic electrons involved in Compton scattering give an additional broadening term for and set the fundamental limit to the angular resolution in a Compton telescope. This effect most strongly favors the lower columns in the period table and disfavors the noble elements, but also favors low $Z$ over high $Z$. Thus, Ar is favored over Xe, but Si is more strongly favored over Ar.

The geometric considerations above that favor low $Z$ and low density raise the question: why not use LHe or LNe, or gaseous He, Ne, or Ar? As noted in Sec.~\ref{sec:instrument_concept} gaseous detectors have superb imaging down to low energies~\citep{Bernard_2022}, but have substantially less stopping power than liquids. LHe and LNe have a number of practical challenges such as very high energy scintillation light, and much larger demands on the cryogenics, but a more fundamental reason rules these elements out. In both, free electrons form "bubbles" of excluded space around them and acquire drift velocities similar to ions~\citep[see, e.g.,][]{Maris2008, Bruschi1972}, rendering TPCs with either completely impractically given the event rates in space. For completeness we note that while a LKr TPC is otherwise well behaved, Kr is contaminated by radioactive $^{85}$Kr at a prohibitive level of 1 MBq of decays per kg Kr~\citep{Bollhofer2019}. A final consideration between LXe and LAr is cost, which on average is 1K-2K\,USD/kg or more for Xe, and negligible for Ar. This Xe cost is a nuisance in a laboratory development setting, but not a major factor for a space borne mission.

\section{Modified Energy Minimization CKD}\label{sec:appendix_minCKD}
This appendix details the novel energy minimization approach introduced for the CKD reconstruction process. The total energy of an incoming $\gamma$-ray, $E_{\mathrm{total}}$, is assumed to be the sum of individual energy depositions across various locations in the detector:
\begin{equation}
E_{\mathrm{total}} = \sum_{i=1}^{N_0} E_i,
\end{equation}
where $E_i$ denotes the energy deposited at the $i$-th interaction point, and $N_0$ is the total number of scattering events.

For a given interaction sequence, the outgoing $\gamma$-ray energy after the $k$-th scatter is computed as:
\begin{equation}
E_{\mathrm{out}} = E_{\mathrm{total}} - \sum_{i=1}^{k} E_i,
\end{equation}
and the corresponding incoming energy, $E_{\mathrm{in}}$, is given by the outgoing energy prior to the $k$-th scatter.

The validity of a sequence is evaluated by comparing the observed outgoing energy, $E_{\mathrm{out}}$, with the predicted outgoing energy, $E_{\mathrm{out, pred}}$, which is calculated using the Compton scattering formula:
\begin{equation}
E_{\mathrm{out, pred}} = \frac{E_{\mathrm{in}}}{1 + \left(E_{\mathrm{in}}/(m_e c^2)\right)(1 - \cos \theta)},
\end{equation}
where $m_e$ is the electron mass, $c$ is the speed of light, and the scattering angle $\theta$ is determined geometrically from the interaction points:
\begin{equation}
\cos \theta = \frac{(\mathbf{r}_1 - \mathbf{r}_2) \cdot (\mathbf{r}_2 - \mathbf{r}_3)}{|\mathbf{r}_1 - \mathbf{r}_2| |\mathbf{r}_2 - \mathbf{r}_3|}.
\end{equation}

The error for a given sequence is quantified as the summed squared difference between the observed and predicted outgoing energies:
\begin{equation}
\mathrm{error} = \sum_{j=1}^{N} \left( E_{\mathrm{out, pred}}^{(j)} - E_{\mathrm{out}}^{(j)} \right)^2,
\end{equation}
where $N$ represents the number of scattering events in the sequence taken into account. A similar approach is used in ~\citet{Takada_2011} to discriminate events (with N=1).

Attempts to normalize this error metric using the sum of squared errors of measured and calculated energies did not yield significant improvements. To enhance robustness, we incorporate a regularization term (inspired by machine learning loss functions) where the error metric is augmented by a penalty term based on the Klein-Nishina probability, $P_{\mathrm{K-N}}^{(j)}$, which represents the physical likelihood of a given scattering event:
\begin{equation}
\mathrm{error_2} = \sum_{j=1}^{N} \left[ \left( E_{\mathrm{out, pred}}^{(j)} - E_{\mathrm{out}}^{(j)} \right)^2 - \lambda^2 \log P_{\mathrm{K-N}}^{(j)} \right],
\label{eq:sum_error_0}
\end{equation}
where $\lambda$ (set empirically to 1 eV) controls the influence of the penalty term. This term penalizes physically improbable events by assigning larger errors to configurations with low $P_{\mathrm{K-N}}^{(j)}$ values.

For sequences where the initial electron recoil direction, $\mathbf{R}$, can be estimated (requiring a scatter energy of more than $\sim$300 keV), a geometric constraint is imposed. Specifically, $\mathbf{R}$ is required to lie close to the plane defined by the first three scattering points $\mathbf{r}_1$, $\mathbf{r}_2$, and $\mathbf{r}_3$. The mismatch parameter, $\Delta\Phi$, is defined as:
\begin{equation}
\Delta\Phi = \frac{|\mathbf{R} \cdot \mathbf{v}|}{\|\mathbf{R}\| \|\mathbf{v}\|},
\end{equation}
where $\mathbf{v}$ is the cross product of the vectors spanning the plane:
\begin{equation}
\mathbf{v} = (\mathbf{r}_2 - \mathbf{r}_1) \times (\mathbf{r}_3 - \mathbf{r}_1).
\end{equation}
If $\mathbf{R}$ were to lie exactly in the plane, $\Delta\Phi$ would be zero.

This geometric constraint is incorporated into the total error metric via an additional penalty term:
\begin{equation}
\alpha^2 \frac{\Delta\Phi}{\Delta \beta},
\label{eg:electron_track_penalty}
\end{equation}
where $\Delta \beta$ accounts for the resolution limitations of $\mathbf{R}$ and depends on the energy and depth of the scatter, as discussed in Sec.~\ref{sec:gampix}. The parameter $\alpha^2$ (with units of eV$^2$) is chosen empirically to maintain dimensional consistency with other terms.

The complete error metric, incorporating both the Klein-Nishina penalty and the geometric constraint, is expressed as:

\begin{align}
\mathrm{total\ error} = \sum_{j=1}^{N} \bigg[ & \left( E_{\mathrm{out, pred}}^{(j)} - E_{\mathrm{out}}^{(j)} \right)^2 \nonumber \\
& - \lambda^2 \log P_{\mathrm{K-N}}^{(j)} + \alpha^2 \frac{\Delta\Phi^{(j)}}{\Delta \beta^{(j)}} \bigg].
\label{eq:sum_error}
\end{align}

To identify the most probable interaction sequence, one would ideally evaluate all possible permutations of interaction orders, selecting the sequence with the lowest total error as defined in Equation~\ref{eq:sum_error}. This exhaustive approach is the most accurate but computationally prohibitive due to the factorial growth in permutations, scaling as $N_0!$.

An alternative, less computationally demanding approach is to focus only on identifying the initial three scatters, disregarding the order of subsequent events. This reduces the problem to selecting the optimal triplet of scatters, scaling as $N_0! / (N_0 - 3)!$, but is not as reliable due to the lack of consideration for the full interaction sequence.

A practical middle ground is to evaluate the order of the first $N$ scatters, typically setting $N$ to 5 or 6. This approach, scaling as $N_0! / (N_0 - N)!$, balances computational feasibility and reliability, as searching deeper into the sequence increases the robustness of the reconstruction. Overall, the Modified Energy Minimization CKD ensures that the reconstructed sequence is both physically plausible and consistent with the observed data while remaining computationally manageable.

\section{Angular Resolution of Gamma-Ray Telescopes}
\label{sec:appendix_ARM}
The angular resolution of a $\gamma$-ray Compton telescope quantifies its ability to reconstruct the direction of incoming gamma rays. Traditionally, this is measured using the Angular Resolution Measure (ARM), defined as the Full Width at Half Maximum (FWHM) of the distribution of angular differences between the true and reconstructed photon directions. This standard ARM calculation effectively evaluates the instrument's precision but assumes no additional constraints beyond the cone geometry derived from Compton scattering.

Electron tracking allows the determination of the recoil electron trajectory, reducing the ambiguity in the reconstructed photon direction. This transforms the allowable photon directions from a ring to an arc. Consequently, the traditional ARM definition becomes less suitable in such scenarios.

\begin{figure}
    \centering
    \includegraphics[width=\linewidth]{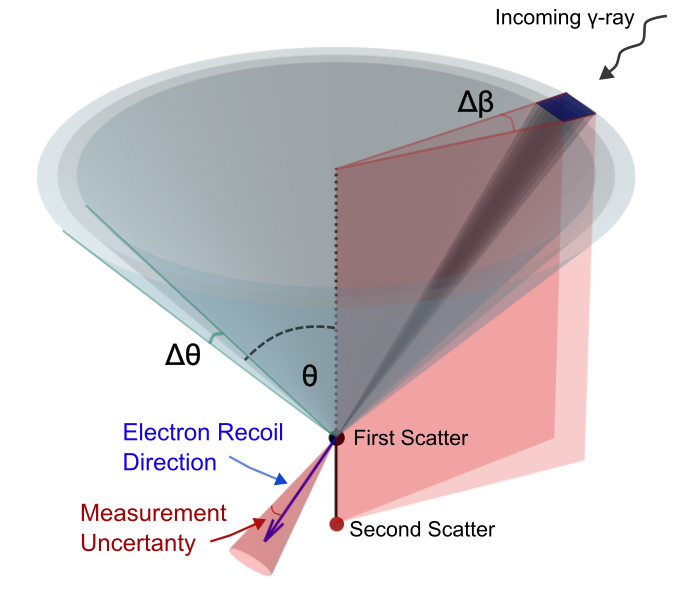}
    \caption{Geometric interpretation of angular constraints in Compton imaging. The incoming $\gamma$-ray undergoes a Compton scattering interaction at the first scatter point, resulting in a photon with a new trajectory and a recoiling electron. The angle $\theta$ is the true Compton scattering angle, while $\Delta \theta$ represents the angular deviation between the reconstructed scattering angle, $\theta_C$, and the true angle. The electron recoil direction provides an additional geometric constraint by defining a plane that includes both the recoiling electron's trajectory and the scattered photon path. $\Delta \beta$ quantifies the deviation between the reconstructed and true electron recoil planes, referred to as the Scatter Plane Deviation. Together, these angular deviations constrain the possible incoming direction of the $\gamma$-ray, reducing the source localization from a conical surface to a smaller arc. The measurement uncertainty of the electron recoil direction directly translates to the $\Delta \beta$ uncertainty, which corresponds to the projection of the cone opening angle onto the plane.} 
    \label{fig:lorentzian_definitions}
\end{figure}

\paragraph{Modified Definition of ARM}
To address this limitation, we propose a modified ARM definition that incorporates the additional constraints provided by electron tracking. The traditional ARM is calculated as:
\begin{equation} 
\text{ARM} = \text{FWHM} \left( P(\Delta \theta) \right), 
\end{equation} 
where \( \theta_C \) is defined as the reconstructed scattering angle relative to the line connecting the first and second scattering events, \( \theta_0 \) represents the true scattering angle, and \( \Delta \theta = \theta_C - \theta_0 \) denotes the angular deviation (illustrated in Fig.~\ref{fig:lorentzian_definitions}). The symbol \( P \) represents the distribution of angular deviations across multiple scattering events.

Electron track detection enhances angular resolution by providing an additional geometric constraint. When the Compton electron’s path is accurately reconstructed, it defines a plane containing both the scattered photon and the electron’s trajectory. This restricts the possible incoming direction of the $\gamma$-ray photon, reducing it from a conical distribution to a smaller arc. To reflect this constraint, we introduce a secondary angular deviation, \( \Delta \beta \), which quantifies the difference between the actual and measured electron recoil track planes. As shown in Fig.~\ref{fig:lorentzian_definitions}, \( \Delta \beta \) corresponds to the opening angle of the Scatter Plane Deviation introduced in \citet{Tanimori2015}.


To optimally account for both measurements, we redefine the ARM as:
\begin{equation} 
\text{ARM}_{\text{mod}} = \text{FWHM} \left( P\left(\min (\Delta \theta, \Delta \beta) \right)\right). 
\end{equation} 
This modified definition ensures that the smaller angular deviation, whether from the Compton scattering (\( \Delta \theta \)) or from the electron recoil plane (\( \Delta \beta \)), is utilized for each event. The motivation for this approach lies in the observation that, for certain events, the electron track reconstruction offers greater accuracy than the photon scattering angle measurement, thereby providing a more reliable constraint on the $\gamma$-ray's incoming direction.

\paragraph{Comparison of the Modified ARM with 2D Lorentzian Fitting}

To assess the accuracy of the modified ARM definition, we compare it with the "gold standard" method: fitting a two-dimensional Lorentzian function to an image of the source. For this validation, an image is generated by projecting the arcs (or cones in the absence of electron tracking) created by individual events onto a map. The intensity peak at the intersection of these arcs corresponds to the location of the event source. The two-dimensional Lorentzian function used for fitting is defined as:

\begin{equation} 
f(x, y) = \frac{I}{(x - x_0)^2/\gamma_x^2 + (y - y_0)^2/\gamma_y^2 + 1} + r,
\end{equation} 
where $(x_0, y_0)$ represents the source position, $\gamma_x$ and $\gamma_y$ are the widths of the distribution along the $x$ and $y$ axes, $I$ is the intensity at the peak, and $r$ accounts for a uniform background offset. From this fit, the ARM is determined using the relationship:

\begin{equation} 
\text{ARM}_{2d} = \frac{1}{\pi} \left(\frac{\gamma_x + \gamma_y}{2}\right)^{3/2}. 
\end{equation}

\begin{figure}
    \centering
    \includegraphics[width=\linewidth]{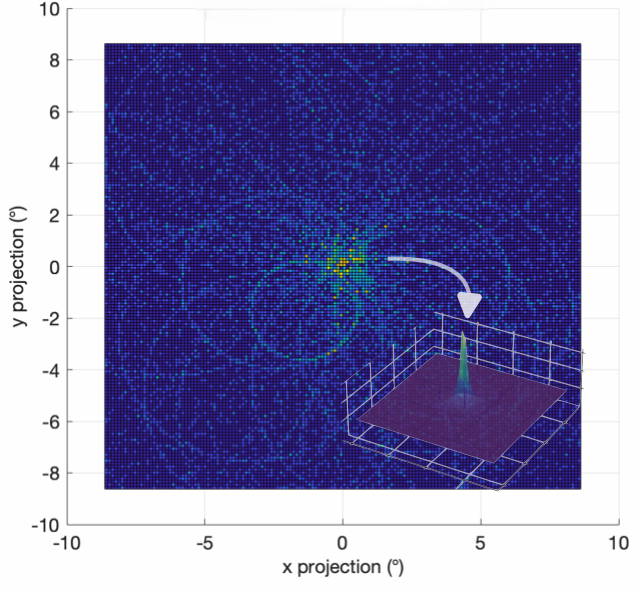}
    \caption{Image of a source generated by overlapping rings from individual Compton events. The rings intersect at the source location, forming a bright spot. A 2D Lorentzian fit to this image provides a highly accurate measure of angular resolution.}
    \label{fig:image_from_rings}
\end{figure}

An example of such an image is shown in Fig.~\ref{fig:image_from_rings}, where the overlapping rings correspond to individual Compton events. The intersection of these rings at the source location produces a bright spot. Although fitting a 2D Lorentzian function provides a precise measure of angular resolution, it is computationally expensive. 

\paragraph{Evaluation of the Modified ARM Through Simulations}
Simulations were conducted to evaluate the performance of the modified ARM definition and compare it to both the traditional ARM and the ARM derived from a 2D Lorentzian fit. The simulation involved generating circular scattering patterns on a spatial grid, with each event characterized by an angular deviation ($\Delta \theta$) and a random radial distance $r$. Additionally, Scatter Plane Deviations ($\Delta \beta$) were introduced with varying distribution widths, tailored to different simulation scenarios to account for varying electron track readout accuracy.

For each simulated event, the minimum angular separation, $\min(\Delta \theta, \Delta \beta)$, was computed and analyzed using the modified ARM definition. The traditional ARM was obtained by fitting a Lorentzian distribution to the $\Delta \theta$ values, while the 2D Lorentzian ARM was determined through a fit to the spatial density distribution of events on the grid.

\begin{figure}
    \centering 
    \includegraphics[width=\linewidth]{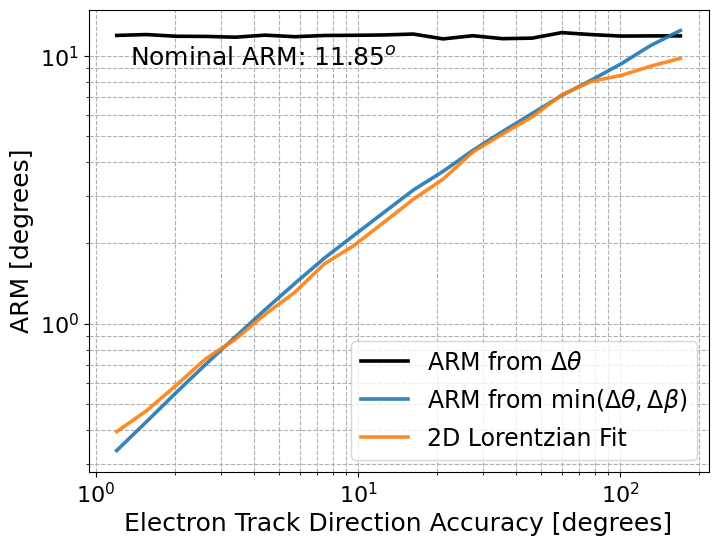} 
    \caption{Comparison of the modified ARM and the ARM derived from the 2D Lorentzian fit, demonstrating strong agreement across different simulation scenarios. The results highlight the robustness and efficiency of the modified ARM definition for quantifying angular resolution.} 
    \label{fig:arm_comparison} 
\end{figure}

Fig.~\ref{fig:arm_comparison} presents the simulation results, demonstrating strong agreement between the modified ARM and the 2D Lorentzian ARM across a range of electron track readout accuracy scenarios, defined by the maximum $\Delta \beta$. The simulations validate that the modified ARM effectively incorporates the geometric constraints provided by electron tracking, resulting in a robust and computationally efficient metric for angular resolution. By closely matching the results of the computationally intensive 2D Lorentzian fitting, the modified ARM offers a practical and precise alternative for performance evaluation in $\gamma$-ray Compton telescopes.

\paragraph{Impact of Electron Tracking on Angular Resolution}
The uncertainty in Compton scattering spreads the possible source direction over a solid angle proportional to \( 2\pi \Delta \theta \). Electron tracking provides an orthogonal constraint by defining the electron recoil plane, reducing the uncertainty area to \( \Delta \theta \cdot \Delta \beta \). This interplay of constraints rapidly enhances angular resolution, as shown in Fig.~\ref{fig:arm_comparison}. Even for poor electron track accuracy (e.g., \( \sim 100^\circ \)), where the ARM remains dominated by \( \Delta \theta \), the inclusion of \( \Delta \beta \) reduces the effective uncertainty by approximately a factor of 5, consistent with findings of \citet{Tanimori2015}. At higher electron track accuracy (e.g., \( < 10^\circ \)), the modified ARM converges closely with the 2D Lorentzian fit, demonstrating the power of combining these orthogonal constraints to significantly improve angular resolution.


\begin{thebibliography}{}
\expandafter\ifx\csname natexlab\endcsname\relax\def\natexlab#1{#1}\fi
\providecommand{\url}[1]{\href{#1}{#1}}
\providecommand{\dodoi}[1]{doi:~\href{http://doi.org/#1}{\nolinkurl{#1}}}
\providecommand{\doeprint}[1]{\href{http://ascl.net/#1}{\nolinkurl{http://ascl.net/#1}}}
\providecommand{\doarXiv}[1]{\href{https://arxiv.org/abs/#1}{\nolinkurl{https://arxiv.org/abs/#1}}}

\bibitem[{Aalbers {et~al.}(2023)}]{LZ2023}
Aalbers, J., {et~al.} 2023, Phys. Rev. Lett., 131, 041002,
  \dodoi{10.1103/PhysRevLett.131.041002}

\bibitem[{Aalbers {et~al.}(2024)}]{aalbers2024}
---. 2024, Dark Matter Search Results from 4.2 Tonne-Years of Exposure of the
  LUX-ZEPLIN (LZ) Experiment.
\newblock \doarXiv{2410.17036}

\bibitem[{Aalseth {et~al.}(2018)}]{Aalseth:2017fik}
Aalseth, C.~E., {et~al.} 2018, Eur. Phys. J. Plus, 133, 131,
  \dodoi{10.1140/epjp/i2018-11973-4}

\bibitem[{Aartsen {et~al.}(2018)}]{IceCube:2018dnn}
Aartsen, M., {et~al.} 2018, Science, 361, eaat1378,
  \dodoi{10.1126/science.aat1378}

\bibitem[{Abbott {et~al.}(2017{\natexlab{a}})}]{Monitor:2017mdv}
Abbott, B., {et~al.} 2017{\natexlab{a}}, Astrophys. J. Lett., 848, L13,
  \dodoi{10.3847/2041-8213/aa920c}

\bibitem[{Abbott {et~al.}(2017{\natexlab{b}})}]{GBM:2017lvd}
---. 2017{\natexlab{b}}, Astrophys. J. Lett., 848, L12,
  \dodoi{10.3847/2041-8213/aa91c9}

\bibitem[{Abbott {et~al.}(2017{\natexlab{c}})}]{Abbott:2017dke}
Abbott, B.~P., {et~al.} 2017{\natexlab{c}}, Astrophys. J. Lett., 851, L16,
  \dodoi{10.3847/2041-8213/aa9a35}

\bibitem[{Abed \& others. DUNE~Collaboration(2021)}]{protodune_2021}
Abed, A., \& others. DUNE~Collaboration. 2021, Design, construction and
  operation of the ProtoDUNE-SP Liquid Argon TPC.
\newblock \doarXiv{2108.01902}

\bibitem[{Acerbi {et~al.}(2023)Acerbi, Altamura, {Di Ruzza}, Merzi, Spinnato,
  \& Gola}]{Acerbi2023}
Acerbi, F., Altamura, A., {Di Ruzza}, B., {et~al.} 2023, Nuclear Instruments
  and Methods in Physics Research Section A: Accelerators, Spectrometers,
  Detectors and Associated Equipment, 1045, 167502,
  \dodoi{https://doi.org/10.1016/j.nima.2022.167502}

\bibitem[{Adamowski {et~al.}(2014)Adamowski, Carls, Dvorak, Hahn, Jaskierny,
  Johnson, Jostlein, Kendziora, Lockwitz, Pahlka, Plunkett, Pordes, Rebel,
  Schmitt, Stancari, Tope, Voirin, \& Yang}]{Adamowski_2014}
Adamowski, M., Carls, B., Dvorak, E., {et~al.} 2014, Journal of
  Instrumentation, 9, P07005, \dodoi{10.1088/1748-0221/9/07/p07005}

\bibitem[{Adams {et~al.}(2020)Adams, Bass, Bishai, Bromberg, Calcutt, Chen,
  Fried, Furic, Gao, Gastler, Hugon, Joshi, Kirby, Liu, Mahn, Mooney, Morris,
  Pereyra, Pons, Radeka, Raguzin, Shooltz, Spanu, Timilsina, Tufanli, Tzanov,
  Viren, Gu, Williams, Wood, Worcester, Worcester, Yang, \& Zhang}]{Adams_2020}
Adams, D., Bass, M., Bishai, M., {et~al.} 2020, Journal of Instrumentation, 15,
  P06017, \dodoi{10.1088/1748-0221/15/06/p06017}

\bibitem[{Agazie {et~al.}(2023)}]{agazie2023nanograv}
Agazie, G., {et~al.} 2023, The NANOGrav 15-year Data Set: Constraints on
  Supermassive Black Hole Binaries from the Gravitational Wave Background.
\newblock \doarXiv{2306.16220}

\bibitem[{Agency(2019)}]{penelope}
Agency, N.~E. 2019, PENELOPE 2018: A code system for Monte Carlo simulation of
  electron and photon transport,
  \dodoi{https://doi.org/https://doi.org/10.1787/32da5043-en}

\bibitem[{Agnes {et~al.}(2015)Agnes, Alexander, Alton, Arisaka, Back, Baldin,
  Biery, Bonfini, Bossa, Brigatti, Brodsky, Budano, Cadonati, Calaprice, Canci,
  Candela, Cao, Cariello, Cavalcante, Chavarria, Chepurnov, Cocco, Crippa,
  D'Angelo, D'Incecco, Davini, {De Deo}, Derbin, Devoto, {Di Eusanio}, {Di
  Pietro}, Edkins, Empl, Fan, Fiorillo, Fomenko, Forster, Franco, Gabriele,
  Galbiati, Goretti, Grandi, Gromov, Guan, Guardincerri, Hackett, Herner,
  Hungerford, Ianni, Ianni, Jollet, Keeter, Kendziora, Kidner, Kobychev, Koh,
  Korablev, Korga, Kurlej, Li, Loer, Lombardi, Love, Ludhova, Luitz, Ma,
  Machulin, Mandarano, Mari, Maricic, Marini, Martoff, Meregaglia, Meroni,
  Meyers, Milincic, Montanari, Monte, Montuschi, Monzani, Mosteiro, Mount,
  Muratova, Musico, Nelson, Odrowski, Okounkova, Orsini, Ortica, Pagani,
  Pallavicini, Pantic, Papp, Parmeggiano, Parsells, Pelczar, Pelliccia,
  Perasso, Pocar, Pordes, Pugachev, Qian, Randle, Ranucci, Razeto, Reinhold,
  Renshaw, Romani, Rossi, Rossi, Rountree, Sablone, Saggese, Saldanha, Sands,
  Sangiorgio, Segreto, Semenov, Shields, Skorokhvatov, Smirnov, Sotnikov,
  Stanford, Suvorov, Tartaglia, Tatarowicz, Testera, Tonazzo, Unzhakov,
  Vogelaar, Wada, Walker, Wang, Wang, Watson, Westerdale, Wojcik, Wright,
  Xiang, Xu, Yang, Yoo, Zavatarelli, Zec, Zhu, \& Zuzel}]{AGNES2015456}
Agnes, P., Alexander, T., Alton, A., {et~al.} 2015, Physics Letters B, 743,
  456, \dodoi{https://doi.org/10.1016/j.physletb.2015.03.012}

\bibitem[{Agnes {et~al.}(2018)Agnes, Dawson, De~Cecco, Fan, Fiorillo, Franco,
  Galbiati, Giganti, Johnson, Korga, Kryn, Lebois, Mandarano, Martoff,
  Navrer-Agasson, Pantic, Qi, Razeto, Renshaw, Riffard, Rossi, Savarese,
  Schlitzer, Suvorov, Tonazzo, Wang, Wang, Watson, \& Wilson}]{Agnes2018}
Agnes, P., Dawson, J., De~Cecco, S., {et~al.} 2018, Phys. Rev. D, 97, 112005,
  \dodoi{10.1103/PhysRevD.97.112005}

\bibitem[{{Ait-Ouamer} {et~al.}(1992){Ait-Ouamer}, {Kerrick}, {O'Neill},
  {Tumer}, {Zych}, \& {White}}]{SN1987A}
{Ait-Ouamer}, F., {Kerrick}, A.~D., {O'Neill}, T.~J., {et~al.} 1992, Astrophys.
  J., 386, 715, \dodoi{10.1086/171052}

\bibitem[{Ajaj {et~al.}(2019)}]{Ajaj:2019imk}
Ajaj, R., {et~al.} 2019, Phys. Rev., D100, 022004,
  \dodoi{10.1103/PhysRevD.100.022004}

\bibitem[{Ajello {et~al.}(2009)}]{Ajello2009}
Ajello, M., {et~al.} 2009, Astrophys. J., 697, 1071,
  \dodoi{10.1088/0004-637X/697/2/1071}

\bibitem[{Ajello {et~al.}(2016)}]{Ajello:2015kwa}
---. 2016, Astrophys. J., 819, 44, \dodoi{10.3847/0004-637X/819/1/44}

\bibitem[{Akerib \& otherz(2020)}]{Akerib_2020}
Akerib, D., \& otherz. 2020, Nuclear Instruments and Methods in Physics
  Research Section A: Accelerators, Spectrometers, Detectors and Associated
  Equipment, 953, 163047, \dodoi{10.1016/j.nima.2019.163047}

\bibitem[{Akerib {et~al.}(2019)}]{Akerib_2019}
Akerib, D., {et~al.} 2019, Physical Review D, 100,
  \dodoi{10.1103/physrevd.100.022002}

\bibitem[{Alner {et~al.}(2007)Alner, Araújo, Bewick, Bungau, Camanzi, Carson,
  Cashmore, Chagani, Chepel, Cline, Davidge, Davies, Daw, Dawson, Durkin,
  Edwards, Gamble, Gao, Ghag, Howard, Jones, Joshi, Korolkova, Kudryavtsev,
  Lawson, Lebedenko, Lewin, Lightfoot, Lindote, Liubarsky, Lopes, Lüscher,
  Majewski, Mavrokoridis, McMillan, Morgan, Muna, Murphy, Neves, Nicklin, Ooi,
  Paling, Pinto~da Cunha, Plank, Preece, Quenby, Robinson, Salinas,
  Sergiampietri, Silva, Solovov, Smith, Smith, Spooner, Sumner, Thorne, Tovey,
  Tziaferi, Walker, Wang, White, \& Wolfs}]{Alner_2007}
Alner, G., Araújo, H., Bewick, A., {et~al.} 2007, Physics Letters B, 653,
  161–166, \dodoi{10.1016/j.physletb.2007.08.030}

\bibitem[{Altamura {et~al.}(2023)Altamura, Acerbi, {Di Ruzza}, Verroi, Merzi,
  Mazzi, \& Gola}]{Altamura2023}
Altamura, A.~R., Acerbi, F., {Di Ruzza}, B., {et~al.} 2023, Nuclear Instruments
  and Methods in Physics Research Section A: Accelerators, Spectrometers,
  Detectors and Associated Equipment, 1045, 167488,
  \dodoi{https://doi.org/10.1016/j.nima.2022.167488}

\bibitem[{Angle {et~al.}(2008)Angle, Aprile, Arneodo, Baudis, Bernstein,
  Bolozdynya, Brusov, Coelho, Dahl, DeViveiros, Ferella, Fernandes, Fiorucci,
  Gaitskell, Giboni, Gomez, Hasty, Kastens, Kwong, Lopes, Madden, Manalaysay,
  Manzur, McKinsey, Monzani, Ni, Oberlack, Orboeck, Plante, Santorelli, dos
  Santos, Shagin, Shutt, Sorensen, Schulte, Winant, \& Yamashita}]{Angle_2008}
Angle, J., Aprile, E., Arneodo, F., {et~al.} 2008, Physical Review Letters,
  100, \dodoi{10.1103/physrevlett.100.021303}

\bibitem[{Anton {et~al.}(2020)Anton, Badhrees, Barbeau, Beck, Belov, Bhatta,
  Breidenbach, Brunner, Cao, Cen, Chambers, Cleveland, Coon, Craycraft,
  Daniels, Darroch, Daugherty, Davis, Delaquis, Der Mesrobian-Kabakian, DeVoe,
  Dilling, Dolgolenko, Dolinski, Echevers, Fairbank, Fairbank, Farine,
  Feyzbakhsh, Fierlinger, Fudenberg, Gautam, Gornea, Gratta, Hall, Hansen,
  Hoessl, Hufschmidt, Hughes, Iverson, Jamil, Jessiman, Jewell, Johnson,
  Karelin, Kaufman, Koffas, Kr\"ucken, Kuchenkov, Kumar, Lan, Larson, Lenardo,
  Leonard, Li, Li, Li, Licciardi, Lin, MacLellan, McElroy, Michel, Mong, Moore,
  Murray, Neilson, Njoya, Nusair, Odian, Ostrovskiy, Piepke, Pocar, Reti\`ere,
  Robinson, Rowson, Runge, Schmidt, Sinclair, Soma, Stekhanov, Tarka, Todd,
  Tolba, Tosi, Totev, Veenstra, Veeraraghavan, Vuilleumier, Wagenpfeil,
  Watkins, Weber, Wen, Wichoski, Wrede, Wu, Xia, Yahne, Yang, Yen, Zeldovich,
  \& Ziegler}]{Anton2020}
Anton, G., Badhrees, I., Barbeau, P.~S., {et~al.} 2020, Phys. Rev. C, 101,
  065501, \dodoi{10.1103/PhysRevC.101.065501}

\bibitem[{Aoyama {et~al.}(2022)Aoyama, Tanaka, Kimura, \& Yorita}]{Aoyama_2022}
Aoyama, K., Tanaka, M., Kimura, M., \& Yorita, K. 2022, Progress of Theoretical
  and Experimental Physics, 2022, \dodoi{10.1093/ptep/ptac064}

\bibitem[{Aprile {et~al.}(1993)Aprile, Bolotnikov, Chen, \&
  Mukherjee}]{Aprile1993}
Aprile, E., Bolotnikov, A., Chen, D., \& Mukherjee, R. 1993, Nuclear
  Instruments and Methods in Physics Research Section A: Accelerators,
  Spectrometers, Detectors and Associated Equipment, 327, 216,
  \dodoi{https://doi.org/10.1016/0168-9002(93)91446-T}

\bibitem[{Aprile {et~al.}(1987)Aprile, Ku, Park, \& Schwartz}]{Aprile_1987}
Aprile, E., Ku, W. H.-M., Park, J., \& Schwartz, H. 1987, Nuclear Instruments
  and Methods in Physics Research Section A: Accelerators, Spectrometers,
  Detectors and Associated Equipment, 261, 519,
  \dodoi{https://doi.org/10.1016/0168-9002(87)90362-7}

\bibitem[{Aprile {et~al.}(2003)Aprile, Curioni, Giboni, Kobayashi, Oberlack,
  Chupp, Dunphy, Doke, Kikuchi, \& Ventura}]{Aprile_2003}
Aprile, E., Curioni, A., Giboni, K.-L., {et~al.} 2003, in X-Ray and Gamma-Ray
  Telescopes and Instruments for Astronomy, ed. J.~E. Truemper \& H.~D.
  Tananbaum, Vol. 4851 (SPIE), 1196, \dodoi{10.1117/12.461567}

\bibitem[{Aprile {et~al.}(2006)Aprile, Dahl, de~Viveiros, Gaitskell, Giboni,
  Kwong, Majewski, Ni, Shutt, \& Yamashita}]{Aprile_2006}
Aprile, E., Dahl, C.~E., de~Viveiros, L., {et~al.} 2006, Physical Review
  Letters, 97, \dodoi{10.1103/physrevlett.97.081302}

\bibitem[{Aprile {et~al.}(2020)}]{XENON2020}
Aprile, E., {et~al.} 2020, Eur. Phys. J. C, 80, 785,
  \dodoi{10.1140/epjc/s10052-020-8284-0}

\bibitem[{Aprile {et~al.}(2023)}]{xenonnt2023}
---. 2023, Physical Review Letters, 131, \dodoi{10.1103/physrevlett.131.041003}

\bibitem[{Aramaki {et~al.}(2020)Aramaki, Adrian, Karagiorgi, \&
  Odaka}]{Aramaki_2020}
Aramaki, T., Adrian, P. O.~H., Karagiorgi, G., \& Odaka, H. 2020, Astroparticle
  Physics, 114, 107–114, \dodoi{10.1016/j.astropartphys.2019.07.002}

\bibitem[{Arnold {et~al.}(2009)Arnold, Christiansen, Davis, Hyde, Lear, Liou,
  Lyons, Prior, Studor, Ratliff, Ryan, Giovane, \& Corsaro}]{Arnold_2009}
Arnold, J., Christiansen, E., Davis, A., {et~al.} 2009, Handbook for Designing
  MMOD Protection, https://ntrs.nasa.gov/citations/20090010053

\bibitem[{Atwood {et~al.}(2009)}]{Atwood_2009}
Atwood, W.~B., {et~al.} 2009, The Astrophysical Journal, 697, 1071–1102,
  \dodoi{10.1088/0004-637x/697/2/1071}

\bibitem[{Back \& Ramachers(2008)}]{Back_2008}
Back, J., \& Ramachers, Y. 2008, Nuclear Instruments and Methods in Physics
  Research Section A: Accelerators, Spectrometers, Detectors and Associated
  Equipment, 586, 286–294, \dodoi{10.1016/j.nima.2007.12.008}

\bibitem[{Baldini {et~al.}(2018)Baldini, Baracchini, Bemporad, Berg, Biasotti,
  Boca, Cattaneo, Cavoto, Cei, Chiappini, Chiarello, Chiri, Cocciolo,
  Corvaglia, de~Bari, De~Gerone, D’Onofrio, Francesconi, Fujii, Galli, Gatti,
  Grancagnolo, Grassi, Grigoriev, Hildebrandt, Hodge, Ieki, Ignatov, Iwai,
  Iwamoto, Kaneko, Kasami, Kettle, Khazin, Khomutov, Korenchenko, Kravchuk,
  Libeiro, Maki, Matsuzawa, Mihara, Milgie, Molzon, Mori, Morsani,
  Mtchedilishvili, Nakao, Nakaura, Nicolò, Nishiguchi, Nishimura, Ogawa,
  Ootani, Panareo, Papa, Pepino, Piredda, Popov, Raffaelli, Renga, Ripiccini,
  Ritt, Rossella, Rutar, Sawada, Signorelli, Simonetta, Tassielli, Uchiyama,
  Usami, Venturini, Voena, Yoshida, Yudin, \& Zhang}]{Baldini_2018}
Baldini, A.~M., Baracchini, E., Bemporad, C., {et~al.} 2018, The European
  Physical Journal C, 78, \dodoi{10.1140/epjc/s10052-018-5845-6}

\bibitem[{Berlin {et~al.}(2023)Berlin, Krnjaic, \& Pinetti}]{berlin2023}
Berlin, A., Krnjaic, G., \& Pinetti, E. 2023, Reviving MeV-GeV Indirect
  Detection with Inelastic Dark Matter.
\newblock \doarXiv{2311.00032}

\bibitem[{Bernard(2013{\natexlab{a}})}]{Bernard_2013_pol}
Bernard, D. 2013{\natexlab{a}}, Nuclear Instruments and Methods in Physics
  Research Section A: Accelerators, Spectrometers, Detectors and Associated
  Equipment, 729, 765–780, \dodoi{10.1016/j.nima.2013.07.047}

\bibitem[{Bernard(2013{\natexlab{b}})}]{Bernard2013_tpc}
---. 2013{\natexlab{b}}, Nuclear Instruments and Methods in Physics Research
  Section A: Accelerators, Spectrometers, Detectors and Associated Equipment,
  701, 225, \dodoi{https://doi.org/10.1016/j.nima.2012.11.023}

\bibitem[{Bernard {et~al.}(2022{\natexlab{a}})Bernard, Chattopadhyay, Kislat,
  \& Produit}]{Bernard_2022_pol}
Bernard, D., Chattopadhyay, T., Kislat, F., \& Produit, N. 2022{\natexlab{a}},
  Gamma-Ray Polarimetry (Springer Nature Singapore), 1–42,
  \dodoi{10.1007/978-981-16-4544-0_52-1}

\bibitem[{Bernard {et~al.}(2022{\natexlab{b}})Bernard, Hunter, \&
  Tanimori}]{Bernard_2022}
Bernard, D., Hunter, S.~D., \& Tanimori, T. 2022{\natexlab{b}}, Time Projection
  Chambers for Gamma-Ray Astronomy (Springer Nature Singapore), 1–50,
  \dodoi{10.1007/978-981-16-4544-0_50-1}

\bibitem[{Berner {et~al.}(2019)}]{Berner:2019uvt}
Berner, R., {et~al.} 2019, Instruments, 3, 28,
  \dodoi{10.3390/instruments3020028}

\bibitem[{Blandford {et~al.}(2019)Blandford, Meier, \&
  Readhead}]{Blandford:2018iot}
Blandford, R., Meier, D., \& Readhead, A. 2019, Ann. Rev. Astron. Astrophys.,
  57, 467, \dodoi{10.1146/annurev-astro-081817-051948}

\bibitem[{{Bloemen} {et~al.}(1995){Bloemen}, {Bennett}, {Blom}, {Collmar},
  {Hermsen}, {Lichti}, {Morris}, {Schoenfelder}, {Stacy}, {Strong}, \&
  {Winkler}}]{Bloemen1995}
{Bloemen}, H., {Bennett}, K., {Blom}, J.~J., {et~al.} 1995, Astronomy and
  Astrophysics, 293, L1

\bibitem[{Blum {et~al.}(2008)Blum, Rolandi, \& Riegler}]{Blum2008}
Blum, W., Rolandi, L., \& Riegler, W. 2008, {Particle detection with drift
  chambers}, Particle Acceleration and Detection (Springer),
  \dodoi{10.1007/978-3-540-76684-1}

\bibitem[{{Boggs, S. E.} \& {Jean, P.}(2000)}]{Boggs_2000}
{Boggs, S. E.}, \& {Jean, P.} 2000, Astron. Astrophys. Suppl. Ser., 145, 311,
  \dodoi{10.1051/aas:2000107}

\bibitem[{Bollhöfer {et~al.}(2019)Bollhöfer, Schlosser, Schmid, Konrad,
  Purtschert, \& Krais}]{Bollhofer2019}
Bollhöfer, A., Schlosser, C., Schmid, S., {et~al.} 2019, Journal of
  Environmental Radioactivity, 205-206, 7,
  \dodoi{https://doi.org/10.1016/j.jenvrad.2019.04.014}

\bibitem[{Bruschi {et~al.}(1972)Bruschi, Mazzi, \& Santini}]{Bruschi1972}
Bruschi, L., Mazzi, G., \& Santini, M. 1972, Phys. Rev. Lett., 28, 1504,
  \dodoi{10.1103/PhysRevLett.28.1504}

\bibitem[{Burns {et~al.}(2019)}]{Burns:2019zzo}
Burns, E., {et~al.} 2019, Astro2020 White Paper.
\newblock \doarXiv{1903.04461}

\bibitem[{Buuck {et~al.}(2023)Buuck, Mishra, Charles, Lalla, Hitchcock,
  Monzani, Omodei, \& Shutt}]{Buuck_2023}
Buuck, M., Mishra, A., Charles, E., {et~al.} 2023, The Astrophysical Journal,
  942, 77, \dodoi{10.3847/1538-4357/aca329}

\bibitem[{Cao {et~al.}(2015)Cao, Alexander, Aprahamian, Avetisyan, Back, Cocco,
  DeJongh, Fiorillo, Galbiati, Grandi, Guardincerri, Kendziora, Lippincott,
  Love, Lyons, Manenti, Martoff, Meng, Montanari, Mosteiro, Olvitt, Pordes,
  Qian, Rossi, Saldanha, Sangiorgio, Siegl, Strauss, Tan, Tatarowicz, Walker,
  Wang, Watson, Westerdale, \& Yoo}]{Cao_2015}
Cao, H., Alexander, T., Aprahamian, A., {et~al.} 2015, Physical Review D, 91,
  \dodoi{10.1103/physrevd.91.092007}

\bibitem[{Caputo {et~al.}(2019)}]{McEnery:2019tcm}
Caputo, R., {et~al.} 2019, Astro2020 White Paper.
\newblock \doarXiv{1907.07558}

\bibitem[{Clark {et~al.}(2018)Clark, Nadeau, Hills, Dujardin, \& {Di
  Stefano}}]{Clark2018}
Clark, M., Nadeau, P., Hills, S., Dujardin, C., \& {Di Stefano}, P. 2018,
  Nuclear Instruments and Methods in Physics Research Section A: Accelerators,
  Spectrometers, Detectors and Associated Equipment, 901, 6,
  \dodoi{https://doi.org/10.1016/j.nima.2018.05.066}

\bibitem[{Conti {et~al.}(2003)Conti, DeVoe, Gratta, Koffas, Waldman, Wodin,
  Akimov, Bower, Breidenbach, Conley, \& et~al.}]{Conti_2003}
Conti, E., DeVoe, R., Gratta, G., {et~al.} 2003, Physical Review B, 68,
  \dodoi{10.1103/physrevb.68.054201}

\bibitem[{Coogan {et~al.}(2021{\natexlab{a}})Coogan, Morrison, \&
  Profumo}]{Coogan_2021}
Coogan, A., Morrison, L., \& Profumo, S. 2021{\natexlab{a}}, Journal of
  Cosmology and Astroparticle Physics, 2021, 044,
  \dodoi{10.1088/1475-7516/2021/08/044}

\bibitem[{Coogan {et~al.}(2021{\natexlab{b}})Coogan, Morrison, \&
  Profumo}]{Coogan_2021b}
---. 2021{\natexlab{b}}, Phys. Rev. Lett., 126, 171101,
  \dodoi{10.1103/PhysRevLett.126.171101}

\bibitem[{Cumani {et~al.}(2019)Cumani, Hernanz, Kiener, Tatischeff, \&
  Zoglauer}]{Cumani_2019}
Cumani, P., Hernanz, M., Kiener, J., Tatischeff, V., \& Zoglauer, A. 2019,
  Experimental Astronomy, 47, 273–302, \dodoi{10.1007/s10686-019-09624-0}

\bibitem[{Dahl(2009)}]{Dahl:2009nta}
Dahl, C.~E. 2009, PhD thesis, Princeton U.
\newblock \url{https://www.princeton.edu/physics/graduate-program
  /theses/theses-from-2009 /E.Dahlthesis.pdf}

\bibitem[{De~Angelis {et~al.}(2017)De~Angelis, Tatischeff, Tavani, Oberlack,
  Grenier, Hanlon, Walter, Argan, von Ballmoos, \& et~al.}]{De_Angelis_2017}
De~Angelis, A., Tatischeff, V., Tavani, M., {et~al.} 2017, Experimental
  Astronomy, 44, 25–82, \dodoi{10.1007/s10686-017-9533-6}

\bibitem[{{De Geronimo} \& {O'Connor}(2000)}]{OConnor2000}
{De Geronimo}, G., \& {O'Connor}, P. 2000, IEEE Transactions on Nuclear
  Science, 47, 1458, \dodoi{10.1109/23.872996}

\bibitem[{Deng {et~al.}(2018)Deng, He, Liu, Liu, Li, \& Yue}]{Deng_2018}
Deng, Z., He, L., Liu, F., {et~al.} 2018, Journal of Instrumentation, 13,
  P08019, \dodoi{10.1088/1748-0221/13/08/p08019}

\bibitem[{Diehl {et~al.}(2024)Diehl, Feindt, Hansen, Lachnit, Poblotzki,
  Rastorguev, Spannagel, Vanat, \& Vignola}]{diehl2024digital}
Diehl, I., Feindt, F., Hansen, K., {et~al.} 2024, A Digital Silicon
  Photomultiplier.
\newblock \doarXiv{2402.12305}

\bibitem[{Ding {et~al.}(2020)Ding, Pershey, Chernyak, \& Liu}]{ding2020}
Ding, K., Pershey, D., Chernyak, D., \& Liu, J. 2020, Prospect of undoped
  inorganic scintillators at 77 Kelvin for the detection of non-standard
  neutrino interactions at the Spallation Neutron Source.
\newblock \doarXiv{2008.00939}

\bibitem[{Dobi(2014)}]{Dobi2014}
Dobi, A. 2014, PhD thesis, Maryland U., College Park, \dodoi{10.13016/M24P5P}

\bibitem[{Doke {et~al.}(1988)Doke, Crawford, Hitachi, Kikuchi, Lindstrom,
  Masuda, Shibamura, \& Takahashi}]{DOKE_1988}
Doke, T., Crawford, H.~J., Hitachi, A., {et~al.} 1988, Nuclear Instruments and
  Methods in Physics Research Section A: Accelerators, Spectrometers, Detectors
  and Associated Equipment, 269, 291,
  \dodoi{https://doi.org/10.1016/0168-9002(88)90892-3}

\bibitem[{Doke {et~al.}(2002)Doke, Hitachi, Kikuchi, Masuda, Okada, \&
  Shibamura}]{Doke_2002}
Doke, T., Hitachi, A., Kikuchi, J., {et~al.} 2002, Japanese Journal of Applied
  Physics, 41, 1538, \dodoi{10.1143/JJAP.41.1538}

\bibitem[{{DUNE Collaboration}(2023)}]{dunefd2_2023}
{DUNE Collaboration}. 2023, The DUNE Far Detector Vertical Drift Technology,
  Technical Design Report.
\newblock \doarXiv{2312.03130}

\bibitem[{Dwyer {et~al.}(2018)Dwyer, Garcia-Sciveres, Gnani, Grace, Kohn,
  Kramer, Krieger, Lin, Luk, Madigan, Marshall, Steiner, \&
  Stezelberger}]{Dwyer_2018}
Dwyer, D., Garcia-Sciveres, M., Gnani, D., {et~al.} 2018, Journal of
  Instrumentation, 13, P10007, \dodoi{10.1088/1748-0221/13/10/p10007}

\bibitem[{Dzhatdoev \& Podlesnyi(2019)}]{Dzhatdoev_2019}
Dzhatdoev, T., \& Podlesnyi, E. 2019, Astroparticle Physics, 112, 1–7,
  \dodoi{10.1016/j.astropartphys.2019.04.004}

\bibitem[{Fischer {et~al.}(2022)Fischer, Zimmermann, \& Maisano}]{Fischer2022}
Fischer, P., Zimmermann, R.~K., \& Maisano, B. 2022, Nuclear Instruments and
  Methods in Physics Research Section A: Accelerators, Spectrometers, Detectors
  and Associated Equipment, 1040, 167033,
  \dodoi{https://doi.org/10.1016/j.nima.2022.167033}

\bibitem[{{Fleischhack}(2021)}]{Fleischhack_2021}
{Fleischhack}, H. 2021, arXiv e-prints, arXiv:2108.02860.
\newblock \doarXiv{2108.02860}

\bibitem[{Frach {et~al.}(2009)Frach, Prescher, Degenhardt, de~Gruyter, Schmitz,
  \& Ballizany}]{Frach2009}
Frach, T., Prescher, G., Degenhardt, C., {et~al.} 2009, in 2009 IEEE Nuclear
  Science Symposium Conference Record (NSS/MIC), 1959--1965,
  \dodoi{10.1109/NSSMIC.2009.5402143}

\bibitem[{Funk(2017)}]{Funk2017}
Funk, S. 2017, High-Energy Gamma Rays from Supernova Remnants, ed. A.~W.
  Alsabti \& P.~Murdin (Cham: Springer International Publishing), 1737--1750,
  \dodoi{10.1007/978-3-319-21846-5_12}

\bibitem[{Gallina(2024)}]{Gallina2024}
Gallina, G. 2024, Performance of novel Silicon Photo-Multipliers for the nEXO
  and Darkside-20k experiments.,  Presnted at the Nagoay Workshop on Technology
  and Instrumentation in Future Liquid Noble Gas Detectors.
\newblock \url{https://indico.kmi.nagoya-u.ac.jp/event/6/contributions/3/}

\bibitem[{Gallina {et~al.}(2022)Gallina, Guan, Retiere, Cao, Bolotnikov, Kotov,
  Rescia, Soma, Tsang, Darroch, {et~al.}}]{Gallina_2022}
Gallina, G., Guan, Y., Retiere, F., {et~al.} 2022, European Physical Journal.
  C, Particles and Fields (Online), 82, \dodoi{10.1140/epjc/s10052-022-11072-8}

\bibitem[{Garutti \& Musienko(2019)}]{Garutti2019}
Garutti, E., \& Musienko, Y. 2019, Nuclear Instruments and Methods in Physics
  Research Section A: Accelerators, Spectrometers, Detectors and Associated
  Equipment, 926, 69, \dodoi{https://doi.org/10.1016/j.nima.2018.10.191}

\bibitem[{Gibbons {et~al.}(2024)Gibbons, Chen, Haselschwardt, Xia, \&
  Sorensen}]{Gibbons_2024}
Gibbons, R., Chen, H., Haselschwardt, S., Xia, Q., \& Sorensen, P. 2024,
  Journal of Instrumentation, 19, P01013,
  \dodoi{10.1088/1748-0221/19/01/p01013}

\bibitem[{Gros {et~al.}(2018)Gros, Amano, Attié, Baron, Baudin, Bernard,
  Bruel, Calvet, Colas, Daté, Delbart, Frotin, Geerebaert, Giebels, Götz,
  Hashimoto, Horan, Kotaka, Louzir, Magniette, Minamiyama, Miyamoto, Ohkuma,
  Poilleux, Semeniouk, Sizun, Takemoto, Yamaguchi, Yonamine, \&
  Wang}]{GROS2018}
Gros, P., Amano, S., Attié, D., {et~al.} 2018, Astroparticle Physics, 97, 10,
  \dodoi{https://doi.org/10.1016/j.astropartphys.2017.10.008}

\bibitem[{{Gupta} {et~al.}(2020){Gupta}, {Pena-Perez}, {Markovic}, {Doering},
  {Reese}, {Tamma}, {Ali}, {Caragiulo}, {Petrignani}, {Rota}, {Kamath}, {Xu},
  {Abu-Nimeh}, \& {Dragone}}]{CryoAsic2}
{Gupta}, A., {Pena-Perez}, A., {Markovic}, B., {et~al.} 2020, in 2020 IEEE 63rd
  International Midwest Symposium on Circuits and Systems (MWSCAS), 611--614,
  \dodoi{10.1109/MWSCAS48704.2020.9184452}

\bibitem[{Hartwig(2016)}]{Hartwig2016}
Hartwig, J. 2016, in Proceedings of the 52nd AIAA/SAE/ASEE Joint Propulsion
  Conference, \dodoi{10.2514/6.2016-4772}

\bibitem[{Holladay {et~al.}(2004)Holladay, Day, \& Gill}]{Holladay_2004}
Holladay, J., Day, G., \& Gill, L. 2004, in 34th International Conference on
  Environmental Systems (ICES)

\bibitem[{Hubbell(1969)}]{Hubbell:1969xbg}
Hubbell, J.~H. 1969, Photon cross sections, attenuation coefficients, and
  energy absorption coefficients from 10 keV to 100 GeV:,  National Institute
  of Standards and Technology, Gaithersburg, MD,
  \dodoi{https://doi.org/10.6028/NBS.NSRDS.29}

\bibitem[{Hunter {et~al.}(2014)Hunter, Bloser, Depaola, Dion, DeNolfo, Hanu,
  Iparraguirre, Legere, Longo, McConnell, Nowicki, Ryan, Son, \&
  Stecker}]{HUNTER2014}
Hunter, S.~D., Bloser, P.~F., Depaola, G.~O., {et~al.} 2014, Astroparticle
  Physics, 59, 18, \dodoi{https://doi.org/10.1016/j.astropartphys.2014.04.002}

\bibitem[{Innes(1993)}]{Innes1993}
Innes, W.~R. 1993, Nuclear Instruments and Methods in Physics Research Section
  A: Accelerators, Spectrometers, Detectors and Associated Equipment, 329, 238,
  \dodoi{https://doi.org/10.1016/0168-9002(93)90942-B}

\bibitem[{Johnson {et~al.}(1997)Johnson, Kinzer, Kurfess, Strickman, Purcell,
  Grabelsky, Ulmer, Hillis, Jung, \& Cameron}]{OSSE:1997}
Johnson, W., Kinzer, R., Kurfess, J., {et~al.} 1997, The Astrophysical Journal
  Supplement Series, 86, \dodoi{10.1086/191795}

\bibitem[{Khek {et~al.}(2022)Khek, Mishra, Buuck, \& Shutt}]{Khek_2022}
Khek, B., Mishra, A., Buuck, M., \& Shutt, T. 2022, AI, 3, 975,
  \dodoi{10.3390/ai3040058}

\bibitem[{Kierans {et~al.}(2022)Kierans, Takahashi, \& Kanbach}]{Kierans2022}
Kierans, C., Takahashi, T., \& Kanbach, G. 2022, Compton Telescopes for
  Gamma-Ray Astrophysics (Singapore: Springer Nature Singapore), 1--72,
  \dodoi{10.1007/978-981-16-4544-0_46-1}

\bibitem[{Kimura {et~al.}(2019)Kimura, Tanaka, Washimi, \& Yorita}]{Kimura2019}
Kimura, M., Tanaka, M., Washimi, T., \& Yorita, K. 2019, Phys. Rev. D, 100,
  032002, \dodoi{10.1103/PhysRevD.100.032002}

\bibitem[{Kuźniak \& Szelc(2021)}]{Kuzniak2021}
Kuźniak, M., \& Szelc, A.~M. 2021, Instruments, 5,
  \dodoi{10.3390/instruments5010004}

\bibitem[{Lebedenko {et~al.}(2009)Lebedenko, Araújo, Barnes, Bewick, Cashmore,
  Chepel, Currie, Davidge, Dawson, Durkin, Edwards, Ghag, Horn, Howard, Hughes,
  Jones, Joshi, Kalmus, Kovalenko, Lindote, Liubarsky, Lopes, Lüscher,
  Majewski, Murphy, Neves, Pinto~da Cunha, Preece, Quenby, Scovell, Silva,
  Solovov, Smith, Smith, Stekhanov, Sumner, Thorne, \& Walker}]{Lebedenko_2009}
Lebedenko, V.~N., Araújo, H.~M., Barnes, E.~J., {et~al.} 2009, Physical Review
  D, 80, \dodoi{10.1103/physrevd.80.052010}

\bibitem[{Li {et~al.}(2023)}]{pandax2023}
Li, S., {et~al.} 2023, Physical Review Letters, 130,
  \dodoi{10.1103/physrevlett.130.261001}

\bibitem[{MacMullin {et~al.}(2012)MacMullin, Boswell, Devlin, Elliott,
  Fotiades, Guiseppe, Henning, Kawano, LaRoque, Nelson, \&
  O'Donnell}]{MacMullin2014}
MacMullin, S., Boswell, M., Devlin, M., {et~al.} 2012, Phys. Rev. C, 85,
  064614, \dodoi{10.1103/PhysRevC.85.064614}

\bibitem[{Maris(2008)}]{Maris2008}
Maris, H.~J. 2008, Journal of the Physical Society of Japan, 77, 111008,
  \dodoi{10.1143/JPSJ.77.111008}

\bibitem[{McCooey(2015)}]{McCooey}
McCooey, D. 2015, Dual Geodesic Icosahedron Pattern 19.
\newblock \url{http://dmccooey.com/polyhedra/index.html}

\bibitem[{McGrath {et~al.}(2022)McGrath, D’Orazio, \&
  Creighton}]{McGrath_2022}
McGrath, C., D’Orazio, D.~J., \& Creighton, J. 2022, Monthly Notices of the
  Royal Astronomical Society, 517, 1242–1263, \dodoi{10.1093/mnras/stac2593}

\bibitem[{Mianowski {et~al.}(2023)}]{POLAR-2:2022tiw}
Mianowski, S., {et~al.} 2023, Exper. Astron., 55, 343,
  \dodoi{10.1007/s10686-022-09873-6}

\bibitem[{Mikhailik {et~al.}(2015)Mikhailik, Kapustyanyk, Tsybulskyi, Rudyk, \&
  Kraus}]{Mikhailik_2015}
Mikhailik, V.~B., Kapustyanyk, V., Tsybulskyi, V., Rudyk, V., \& Kraus, H.
  2015, physica status solidi (b), 252, 804–810,
  \dodoi{10.1002/pssb.201451464}

\bibitem[{Mooney(2015)}]{mooney_2015}
Mooney, M. 2015, The MicroBooNE Experiment and the Impact of Space Charge
  Effects.
\newblock \doarXiv{1511.01563}

\bibitem[{Mrozinski \& DiNicola(2017)}]{NICMCryocooler_2017}
Mrozinski, J., \& DiNicola, M. 2017, Nasa Instrument Cost Model: Cryocooler,
  https://www.nasa.gov/wp-content/uploads/2023/06/14-nicmcryocooler-costsymposium-2017-urs-final-tagged.pdf

\bibitem[{{National Academies of Sciences, Engineering, and
  Medicine}(2023)}]{Decadal2020}
{National Academies of Sciences, Engineering, and Medicine}. 2023, Pathways to
  Discovery in Astronomy and Astrophysics for the 2020s (Washington, DC: The
  National Academies Press), \dodoi{10.17226/26141}

\bibitem[{{nEXO Collaboration}(2018)}]{nexo_2018}
{nEXO Collaboration}. 2018, {nEXO Pre-Conceptual Design Report}.
\newblock \doarXiv{1805.11142}

\bibitem[{{Pena-Perez} {et~al.}(2022){Pena-Perez}, {Gupta}, {Markovic}, {Rota},
  {Doering}, {Convery}, \& {Dragone}}]{CryoAsic3}
{Pena-Perez}, A., {Gupta}, A., {Markovic}, B., {et~al.} 2022, in 2022 IEEE
  Nuclear Science Symposium and Medical Imaging Conference Record (NSS/MIC) -
  Accepted for publication, 1--2, \dodoi{10.1109/NSS/MIC42677.2020.9507812}

\bibitem[{{Pena-Perez} {et~al.}(2020){Pena-Perez}, {Doering}, {Gupta}, {Tamma},
  {Markovic}, H.~{Ali}, {Rota}, {Kamath}, {Petrignani}, {Xu}, {Abu-Nimeh},
  {(Sander) Breur}, {Tsang}, {Convery}, \& {Dragone}}]{CryoAsic1}
{Pena-Perez}, A., {Doering}, D., {Gupta}, A., {et~al.} 2020, in 2020 IEEE
  Nuclear Science Symposium and Medical Imaging Conference Record (NSS/MIC),
  1--2, \dodoi{10.1109/NSS/MIC42677.2020.9507812}

\bibitem[{Plachta {et~al.}(2018)Plachta, Stephens, Johnson, \&
  Zagarola}]{Plachata2018}
Plachta, D., Stephens, J., Johnson, W., \& Zagarola, M. 2018, Cryogenics, 94,
  95, \dodoi{https://doi.org/10.1016/j.cryogenics.2018.07.005}

\bibitem[{Plante {et~al.}(2022)Plante, Aprile, Howlett, \& Zhang}]{Plante_2022}
Plante, G., Aprile, E., Howlett, J., \& Zhang, Y. 2022, The European Physical
  Journal C, 82, \dodoi{10.1140/epjc/s10052-022-10832-w}

\bibitem[{Radeka(2003)}]{Radeka}
Radeka, V. 2003, Annual Review of Nuclear and Particle Science, 38, 217,
  \dodoi{10.1146/annurev.ns.38.120188.001245}

\bibitem[{Retiere(2017)}]{Retiere2017}
Retiere, F. 2017, 3Dimensionally integrated Digital SiPM,  Presented at the
  International Conference on Technology and Instrumentation in Particle
  Physics 2017(TIPP2017).
\newblock \url{https://indico.ihep.ac.cn/event/6387/contributions/83436/}

\bibitem[{{Rybicki} \& {Lightman}(1986)}]{RybickiLightman}
{Rybicki}, G.~B., \& {Lightman}, A.~P. 1986, {Radiative Processes in
  Astrophysics} (John Wiley \& Sons, Ltd)

\bibitem[{Schoenfelder {et~al.}(1993)}]{COMPTEL:1993}
Schoenfelder, V., {et~al.} 1993, Astrophys. J. Suppl., 86, 657,
  \dodoi{10.1086/191794}

\bibitem[{{Sch{\"o}nfelder} {et~al.}(2000){Sch{\"o}nfelder}, {Bennett}, {Blom},
  {Bloemen}, {Collmar}, {Connors}, {Diehl}, {Hermsen}, {Iyudin}, {Kippen},
  {Kn{\"o}dlseder}, {Kuiper}, {Lichti}, {McConnell}, {Morris}, {Much},
  {Oberlack}, {Ryan}, {Stacy}, {Steinle}, {Strong}, {Suleiman}, {van Dijk},
  {Varendorff}, {Winkler}, \& {Williams}}]{Schoenfelder:2000bu}
{Sch{\"o}nfelder}, V., {Bennett}, K., {Blom}, J.~J., {et~al.} 2000, Astron.
  Astrophys. Suppl., 143, 145, \dodoi{10.1051/aas:2000101}

\bibitem[{Shutt {et~al.}(2007)Shutt, Dahl, Kwong, Bolozdynya, \&
  Brusov}]{Shutt2007}
Shutt, T., Dahl, C., Kwong, J., Bolozdynya, A., \& Brusov, P. 2007, Nuclear
  Physics B - Proceedings Supplements, 173, 160,
  \dodoi{https://doi.org/10.1016/j.nuclphysbps.2007.08.140}

\bibitem[{Shutt {et~al.}(2024)Shutt, Trbalic, Pena-Perez, {et~al.}}]{Gampix}
Shutt, T., Trbalic, B., Pena-Perez, A., {et~al.} 2024, tbd.
\newblock \doarXiv{tbd}

\bibitem[{Smith {et~al.}(2023)}]{Smith_2023}
Smith, D.~A., {et~al.} 2023, The Astrophysical Journal, 958, 191,
  \dodoi{10.3847/1538-4357/acee67}

\bibitem[{Szydagis {et~al.}(2011)Szydagis, Barry, Kazkaz, Mock, Stolp, Sweany,
  Tripathi, Uvarov, Walsh, \& Woods}]{Szydagis_2011}
Szydagis, M., Barry, N., Kazkaz, K., {et~al.} 2011, Journal of Instrumentation,
  6, P10002, \dodoi{10.1088/1748-0221/6/10/p10002}

\bibitem[{Szydagis {et~al.}(2021)Szydagis, Block, Farquhar, Flesher, Kozlova,
  Levy, Mangus, Mooney, Mueller, Rischbieter, \& Schwartz}]{Szydagis_2021}
Szydagis, M., Block, G.~A., Farquhar, C., {et~al.} 2021, Instruments, 5, 13,
  \dodoi{10.3390/instruments5010013}

\bibitem[{Takada {et~al.}(2011)Takada, Kubo, Nishimura, Ueno, Hattori, Kabuki,
  Kurosawa, Miuchi, Mizuta, Nagayoshi, Nonaka, Okada, Orito, Sekiya, Takeda, \&
  Tanimori}]{Takada_2011}
Takada, A., Kubo, H., Nishimura, H., {et~al.} 2011, The Astrophysical Journal,
  733, 13 (15pp), \dodoi{10.1088/0004-637X/733/1/13}

\bibitem[{Takada {et~al.}(2022)Takada, Takemura, Yoshikawa, Mizumura, Ikeda,
  Nakamura, Onozaka, Abe, Hamaguchi, Kubo, Kurosawa, Miuchi, Saito, Sawano, \&
  Tanimori}]{Takada_2022}
Takada, A., Takemura, T., Yoshikawa, K., {et~al.} 2022, The Astrophysical
  Journal, 930, 6, \dodoi{10.3847/1538-4357/ac6103}

\bibitem[{Tanimori {et~al.}(2015)}]{Tanimori2015}
Tanimori, T., {et~al.} 2015, Astrophys. J., 810, 28,
  \dodoi{10.1088/0004-637X/810/1/28}

\bibitem[{Tanimori {et~al.}(2017)}]{Tanimori2017}
---. 2017, Sci. Rep., 7, 41511, \dodoi{10.1038/srep41511}

\bibitem[{Tomsick {et~al.}(2019{\natexlab{a}})Tomsick, Zoglauer, Sleator,
  Lazar, Beechert, Boggs, Roberts, Siegert, Lowell, Wulf, Grove, Phlips,
  Brandt, Smale, Kierans, Burns, Hartmann, Leising, Ajello, Fryer, Amman,
  Chang, Jean, \& von Ballmoos}]{cosi_2019}
Tomsick, J.~A., Zoglauer, A., Sleator, C., {et~al.} 2019{\natexlab{a}}, The
  Compton Spectrometer and Imager,  arXiv, \dodoi{10.48550/ARXIV.1908.04334}

\bibitem[{Tomsick {et~al.}(2019{\natexlab{b}})Tomsick, Zoglauer, Sleator,
  Lazar, Beechert, Boggs, Roberts, Siegert, Lowell, Wulf, Grove, Phlips,
  Brandt, Smale, Kierans, Burns, Hartmann, Leising, Ajello, Fryer, Amman,
  Chang, Jean, \& von Ballmoos}]{tomsick2019}
---. 2019{\natexlab{b}}, The Compton Spectrometer and Imager.
\newblock \doarXiv{1908.04334}

\bibitem[{Washimi {et~al.}(2018)Washimi, Kimura, Tanaka, \&
  Yorita}]{Washimi2018}
Washimi, T., Kimura, M., Tanaka, M., \& Yorita, K. 2018, Nuclear Instruments
  and Methods in Physics Research Section A: Accelerators, Spectrometers,
  Detectors and Associated Equipment, 910, 22,
  \dodoi{https://doi.org/10.1016/j.nima.2018.09.019}

\bibitem[{Weber {et~al.}(2000)Weber, Stover, Gilbert, Nevitt, \&
  Ouderkirk}]{Weber2000}
Weber, M.~F., Stover, C.~A., Gilbert, L.~R., Nevitt, T.~J., \& Ouderkirk, A.~J.
  2000, Science, 287, 2451, \dodoi{10.1126/science.287.5462.2451}

\bibitem[{Wilkens \& (on behalf~ofthe ATLAS
  LArg~Collaboration)(2009)}]{Wilkens_2009}
Wilkens, H., \& (on behalf~ofthe ATLAS LArg~Collaboration). 2009, Journal of
  Physics: Conference Series, 160, 012043,
  \dodoi{10.1088/1742-6596/160/1/012043}

\bibitem[{Workman {et~al.}(2022)}]{pdg2022}
Workman, R.~L., {et~al.} 2022, PTEP, 2022, 083C01, \dodoi{10.1093/ptep/ptac097}

\bibitem[{Zheng {et~al.}(2022)Zheng, Gao, Wen, Zeng, Pan, Xu, Liu, Zhang, Peng,
  Jiang, Long, Lu, Yang, Feng, Zeng, Cang, \& Tian}]{Zheng2022}
Zheng, X., Gao, H., Wen, J., {et~al.} 2022, Nuclear Instruments and Methods in
  Physics Research Section A: Accelerators, Spectrometers, Detectors and
  Associated Equipment, 1044, 167510,
  \dodoi{https://doi.org/10.1016/j.nima.2022.167510}

\bibitem[{{Zoglauer}(2019)}]{MEGALIB}
{Zoglauer}, A. 2019, {MEGAlib: Medium Energy Gamma-ray Astronomy library}.
\newblock \doeprint{1906.018}

\bibitem[{Zoglauer \& Boggs(2007)}]{NN_CKD}
Zoglauer, A., \& Boggs, S. 2007, IEEE Nuclear Science Symposium Conference
  Record, 6, 4436 , \dodoi{10.1109/NSSMIC.2007.4437096}

\bibitem[{Zoglauer \& Kanbach(2003)}]{Zoglauer2003}
Zoglauer, A., \& Kanbach, G. 2003, in X-Ray and Gamma-Ray Telescopes and
  Instruments for Astronomy, ed. J.~E. Truemper \& H.~D. Tananbaum, Vol. 4851,
  International Society for Optics and Photonics (SPIE), 1302 -- 1309,
  \dodoi{10.1117/12.461177}

\end{thebibliography}

\end{document}